\documentclass[nofootinbib,notitlepage,superscriptaddress,twocolumn]{revtex4-1}
\pdfoutput=1

\usepackage{aas_macros,amsmath,amssymb,esint,graphicx,overpic,mathrsfs,stackengine,rotating,scalerel,array,rotating,tabularray}
\usepackage[table]{xcolor}
\usepackage[bottom]{footmisc}
\usepackage[colorlinks=true]{hyperref}
\usepackage[T1]{fontenc}
\definecolor{red}{RGB}{228,26,28}
\definecolor{green}{RGB}{77,175,74}
\definecolor{blue}{RGB}{55,126,184}
\definecolor{purple}{RGB}{152,78,163}

\definecolor{Red}{RGB}{252, 119, 119}
\definecolor{Green}{RGB}{226,239,217,}
\definecolor{Blue}{RGB}{222,236,247}

\let\originalleft\left
\let\originalright\right
\renewcommand{\left}{\mathopen{}\mathclose\bgroup\originalleft}
\renewcommand{\right}{\aftergroup\egroup\originalright}

\newcommand{\br}[1]{\left[#1\right]}
\newcommand{\cu}[1]{\left\{#1\right\}}
\newcommand{\pa}[1]{\left(#1\right)}
\newcommand{\abs}[1]{\left\vert#1\right\vert}
\newcommand{\ed}{\mathop{}\!\mathrm{d}}

\renewcommand{\O}[1]{\mathcal{O}\pa{#1}}

\DeclareMathOperator\sign{sign}

\newcommand{\ovcirc}[1]{\overset{\circ}{#1}}

\newcommand{\CutD}{{\resizebox{2ex}{!}{$\cap$}}}
\newcommand{\CutU}{{\resizebox{2ex}{!}{$\cup$}}}
\newcommand{\vertvert}{\vert\ \vert}

\newcommand{\gmor}{g_{\scaleto{\mathrm{MOR}\mathstrut}{5pt}}}
\newcommand{\gmob}{g_{\scaleto{\mathrm{MOB}\mathstrut}{5pt}}}
\newcommand{\gamor}{g_{\scaleto{\mathrm{AMOR}\mathstrut}{5pt}}}
\newcommand{\gamob}{g_{\scaleto{\mathrm{AMOB}\mathstrut}{5pt}}}
\newcommand{\risco}{r_{\scaleto{\mathrm{ISCO}\mathstrut}{5pt}}}

\newcommand{\rffp}{r_{\scaleto{\bullet\mathstrut}{5pt}}}
\newcommand{\gffp}{g_{\scaleto{\bullet\mathstrut}{5pt}}}

\newcommand{\gmorhat}{\hat{g}_{\scaleto{\mathrm{MOR}\mathstrut}{5pt}}}
\newcommand{\gmobhat}{\hat{g}_{\scaleto{\mathrm{MOB}\mathstrut}{5pt}}}

\newcommand{\Cmobhat}{\hat{\mathcal{C}}_{\scaleto{\mathrm{MOB}\mathstrut}{5pt}}}
\newcommand{\Cmorhat}{\hat{\mathcal{C}}_{\scaleto{\mathrm{MOR}\mathstrut}{5pt}}} 
\newcommand{\Rmobhat}{\hat{\mathcal{R}}_{\scaleto{\mathrm{MOB}\mathstrut}{5pt}}}
\newcommand{\Rmorhat}{\hat{\mathcal{R}}_{\scaleto{\mathrm{MOR}\mathstrut}{5pt}}}

\newcommand{\VF}[1]{\scalebox{-1}[1]{#1}}

\newcolumntype{P}[1]{>{\hspace{5pt}}p{#1}<{\hspace{5pt}}}

\usepackage{comment}

\begin{document}

\title{Morphology of Relativistically Broadened Line Emission \texorpdfstring{\\}{} from Axisymmetric Equatorial Accretion Disks}

\author{Delilah E.~A. Gates}
\email{delilah.gates@cfa.harvard.edu}
\affiliation{Center for Astrophysics $\arrowvert$ Harvard \& Smithsonian, 60 Garden Street, Cambridge, MA 02138, USA}
\affiliation{Black Hole Initiative at Harvard University, 20 Garden Street, Cambridge, MA 02138, USA}

\author{Chau Truong}
\affiliation{Department of Physics, Princeton University, Princeton, New Jersey 08544, USA}

\author{Amrita Sahu}
\affiliation{Department of Astrophysical Sciences, Princeton University, Princeton, New Jersey 08544, USA}

\author{Alejandro C\'ardenas-Avenda\~no}
\affiliation{Computational Physics and Methods (CCS-2), Center for Nonlinear Studies (CNLS) \& Center for Theoretical Astrophysics (CTA), Los Alamos National Laboratory, Los Alamos NM 87545, USA}

\begin{abstract}
Single-frequency emission from an accretion disk around a black hole is broadened into a line profile due to gravitational redshift and the motion of the disk's particles relative to the observer. The ensemble of relativistically broadened emission frequencies from the disk elements forms the spectrum viewed by an observer. Over the past decades, the broadened spectra of accreting systems have been used to constrain the spin of the black hole, the observer's inclination, and the astrophysical model parameters of the system. These inferences are usually made under the assumption that the accretion disk consists of particles orbiting around the black hole on stable circular orbits in the equatorial plane. Under this Standard disk model, in this work, we revisit line profile morphology, i.e., its extent, kinks, and fall-off. We provide a unified analytical explanation for these line profile morphological features, which encode the black hole spin, viewing inclination, and locations of the disk's inner and outer edges. We then show that these features, however, are model-dependent, by parametrically relaxing some of the astrophysical assumptions. In particular, we explore how allowing the disk particles to deviate from stable circular orbits rapidly degenerates the characteristic features of the line profile under the Standard disk model. Our results further demonstrate how sensitive our understanding of black hole and system properties can be to assumptions we make when interpreting these types of measurements.
\end{abstract}

\maketitle

\section{Introduction}
\label{Sec:Introduction}

Before the launch of space-based gravitational wave detectors, such as the Laser Interferometer Space Antenna (LISA)~\cite{LISA:2022kgy}, photons will continue to be the best messengers for measuring the spins of black holes (BHs) with masses across several orders of magnitude, from stellar-mass BHs to supermassive BHs. Earth-based gravitational wave detectors, like the LIGO-Virgo-KAGRA (LVK) gravitational wave network, detect the merger of stellar-mass compact objects (BHs and neutron stars), and are more sensitive to a combination of the spins for the merging objects, but not to their individual spins~\cite{Purrer:2015nkh}. The spin of the resulting, post-merger BH, on the other hand, is better estimated, but its uncertainty is still large~\cite{LIGOScientific:2020kqk,KAGRA:2021duu}.

In the electromagnetic sector, there are many sources and several methods to infer the spin. In the radio, for example, using Very Long Baseline Interferometry (VLBI), spins measurements of two supermassive BHs, Sgr A* and M87*~\cite{EventHorizonTelescope:2019pgp, Nemmen:2019idv,EventHorizonTelescope:2022exc} are not yet strongly inferred, but are expected to improve with future space-based VLBI capabilities~\cite{Johnson:2024ttr,Lupsasca:2024xhq}. At higher frequencies, thermal continuum fitting, X-ray polarimetry, high-frequency quasi-periodic oscillations, and inner disk reflection modeling are the most popular methods to infer the spins of BHs~\cite{Zhang:1997dy,Fabian:2000nu,Brenneman:2013oba,Reynolds2019,Reynolds:2020jwt}. Included in the latter is the X-ray reflection method, which studies the features of the broadening of spectral lines emitted from the accretion disks surrounding these BHs. The characteristics of this broadening are the primary focus of this study.

As the material in the disk spirals around the central massive object, it experiences intense gravitational redshift and Doppler broadening, resulting in distinctive signatures in the emitted X-ray spectra. These signatures, broadening and skewness, of the emission lines, were predicted more than three decades ago in Ref.~\cite{Fabian:1989ej}. When the model proposed in Ref.~\cite{Fabian:1989ej} was applied to observations of MCG-6-30-15, it provided a remarkable agreement~\cite{Tanaka:1995en}. Since then, inferring the spin from spectroscopic observations has been routinely performed~\cite{Reynolds2019,Reynolds:2020jwt}.

However, the spin inference is not straightforward due to the intrinsic degeneracies of the models and the variability of sources. In particular, breaking the degeneracies between the BH's parameters and the accretion disk's properties has proven to be extremely challenging (see, e.g., the discussions in Refs.~\cite{Krolik:2002ae,2004ApJS..153..205D,Zdziarski:2023zuh}), and inferences from X-ray measurements have almost always been approached with great caution.

The most commonly used model to interpret the data assumes that the accretion disk is described by the Novikov--Thorne (infinitely thin) disk model~\cite{Novikov1973}, in which particles move on stable circular equatorial orbits around the BH, ending at the innermost stable circular orbit (ISCO) or at a larger radius. The underlying assumptions of this model are very strong, and may not be applicable to some systems~\cite{Lasota:2024lcl}. Moreover, the variability of the sources makes the inference scheme even more challenging.  For example, the initial seminal observation reported in Ref.~\cite{Tanaka:1995en} was well-fitted by a non-rotating (Schwarzschild) BH. However, subsequent observations and analyses revealed that the source exhibited a state with very low continuum emission. For those states, the studied broadened line's width increased significantly, and a Schwarzschild BH with a disk ending at the ISCO could no longer account for the observed line profile~\cite{Iwasawa:1996uh, Brenneman2006}.

The profile of the broadened lines has always been understood as the combination of Newtonian Doppler effects (causing double peaks), special relativity contributions (transverse Doppler shifts and the relativistic beaming), and pure general relativity effects (the gravitational redshifting)~\cite{Fabian:2000nu}. In this work, we first revisit this relativistic broadening, under the Standard thin disk model~\cite{Novikov1973}, and analytically explain the morphology of the line profile—its extent, kinks, and fall-off—in a unified framework. We also provide a complete classification and characterization of the morphology. 

While all these features encode information about the BH's spin, viewing inclination, and the location of the disk's inner and outer edges, they are inherently model-dependent~\cite{Krolik:2002ae,2004ApJS..153..205D,Zdziarski:2023zuh}. Furthermore, creating an accretion model that accurately captures all the physical processes is far from trivial. For instance, for high accretion rates, one may need to include ``slim disk'' modifications~\cite{1988ApJ...332..646A,2009ApJS..183..171S,Abdikamalov:2020oci}, which, among other effects, allow particles to move on (slightly) sub-Keplerian trajectories in most of the disk, push the disk's inner edge interior to the ISCO radius, or even make some parts of the disk not visible due to the finite thickness of the disk. Even when stationarity and axisymmetry are assumed and some effects are still neglected---such as the angular momentum flux taken away by radiation or the radial flux of radiation---these models are computationally expensive, and one may still lack the sufficient coverage of the parameter space needed for accurate parameter estimation~\cite{2009ApJS..183..171S}. 

As a way to model modifications of some of the astrophysical assumptions, we allow the particles in the disk to move on generic equatorial orbits following the phenomenological parametrization presented in Refs.~\cite{Penna:2013zga,Cardenas-Avendano:2022csp}. This parametrization allows for smooth interpolation of the radial (azimuthal) velocity from that corresponding to radial infall starting from rest at infinity to the sub-Keplerian value. It includes the Keplerian disk model~\cite{Cunningham1975} and purely radial infalling motion as special cases. Under this phenomenological approach, we show that deviations from Keplerian orbits can lead to systematic biases in the interpreted features. While some features in the broadened emission survive after changes in the disk's dynamics, their locations are shifted. Our results underscore the sensitivity of BH property inferences to the underlying astrophysical assumptions. 

Overall, the findings of this work have at least the following three main implications for inferring parameters of BH-disk systems:

\emph{\underline{Implication 1}: The procedure for mapping line profile morphological features that we have developed can be applied to any disk model. Therefore, as new analytic approximations for the disk motion are obtained, that can account for more physical effects, our analysis can be used to enhance parameter inferences for BH-disk systems (Sec.~\ref{sec:CriticalRedshifts}).}

\emph{\underline{Implication 2}: In regimes where the deviations of disk motion from that of the Standard disk model are negligible, our results explicitly provide the mapping from the line profile features to the parameters of the BH-disk system in the disk region where the emissivity of the emission line has support (Sec.~\ref{sec:StandardDiskMorphologies}). Furthermore, constraints can be made in the case of both under-resolved and resolved lines. For under-resolved lines, bounds on the spin and inclination can be made from the amount of broadening. Further, when sharp features can be resolved in the line profile (e.g., with high-resolution spectra from current and future X-ray missions expected in the near future~\cite{eXTP:2018anb,Barret:2016ett,2020SPIE11444E..22T,Reynolds:2023vvf}), their location can be used to get tighter parameter inferences (Sec.~\ref{sec:MeasureSpin&Inclination}).}

\emph{\underline{Implication 3}: The location of sharp features in the line profile is controlled by the inner and outer disk radii on which the fluorescent line emission has support, as well as the black hole spin parameters. In contrast, the \emph{shape} of the line profiles is controlled by the functional form of the emissivity profile. As such, there may be methods to fit line profiles for the emissivity separately from fitting for BH spin and observer inclination (Sec.~\ref{sec:Emissitivty})}.

\emph{\underline{Implication 4}: Given that current BH-disk constraints via X-ray spectroscopy typically do not take into account deviations from the Standard disk model, the identification provided in this work for the sharp features that arise in the line profile of a given disk model can be used to construct a budget for that model's systematics, as one can systematically track how changing the model's parameters impact the location of these features. Therefore, one can start estimating the uncertainty in the model introduces to the BH-disk parameters inference derived from line emission (Sec.~\ref{sec:NonStandard}).}

\begin{figure*}
    {\resizebox{\linewidth}{!}{
    \begin{tabular}{c}
         \scalebox{2}{\Huge Image of the Disk} \\
         \includegraphics[height=\textheight]{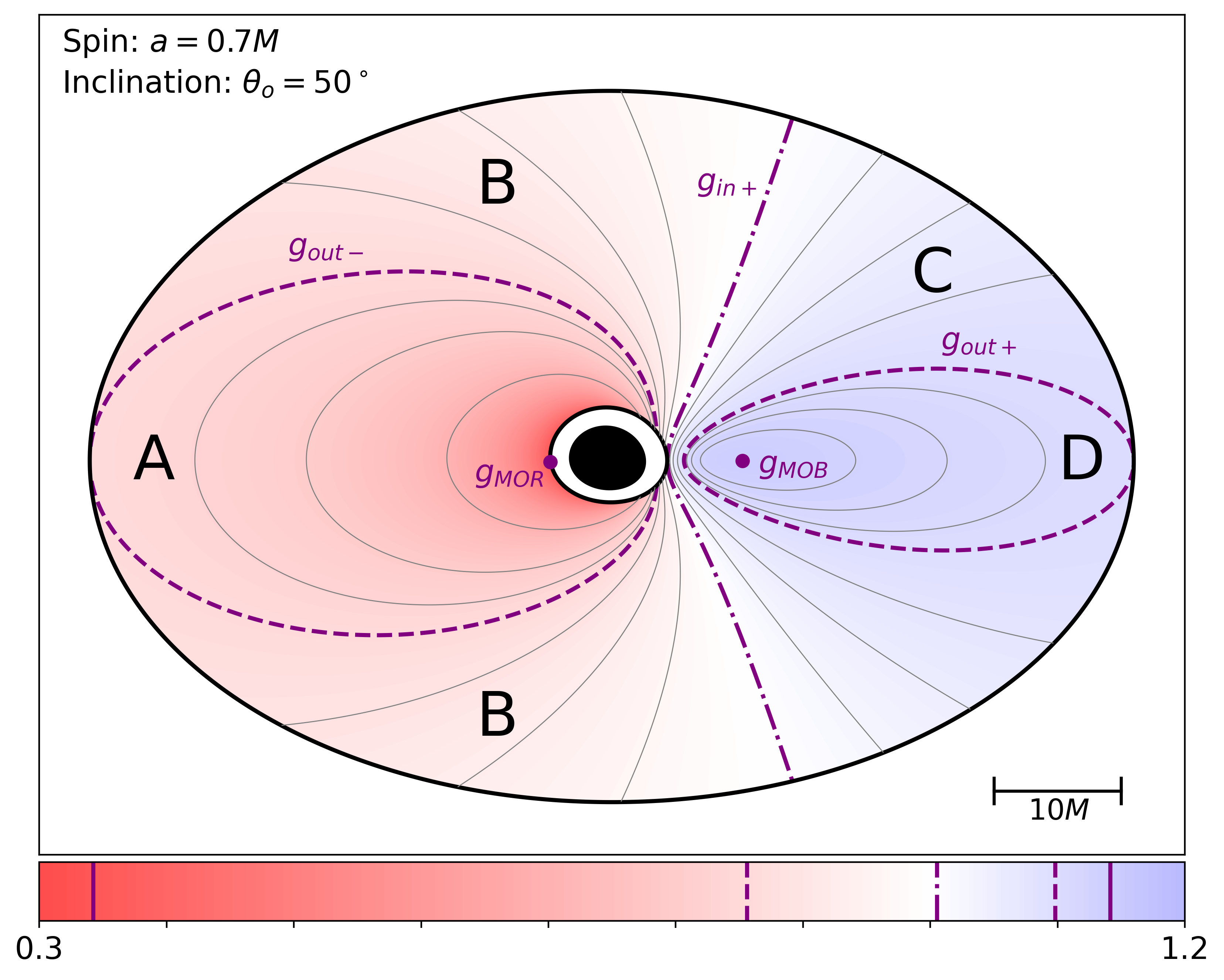}  \\
         \scalebox{1.25}{\Huge Redshift factor, $g$} 
    \end{tabular} \ 
    \begin{tabular}{c}
        \scalebox{2}{\Huge Line Profile} \\
         \includegraphics[height=\textheight]{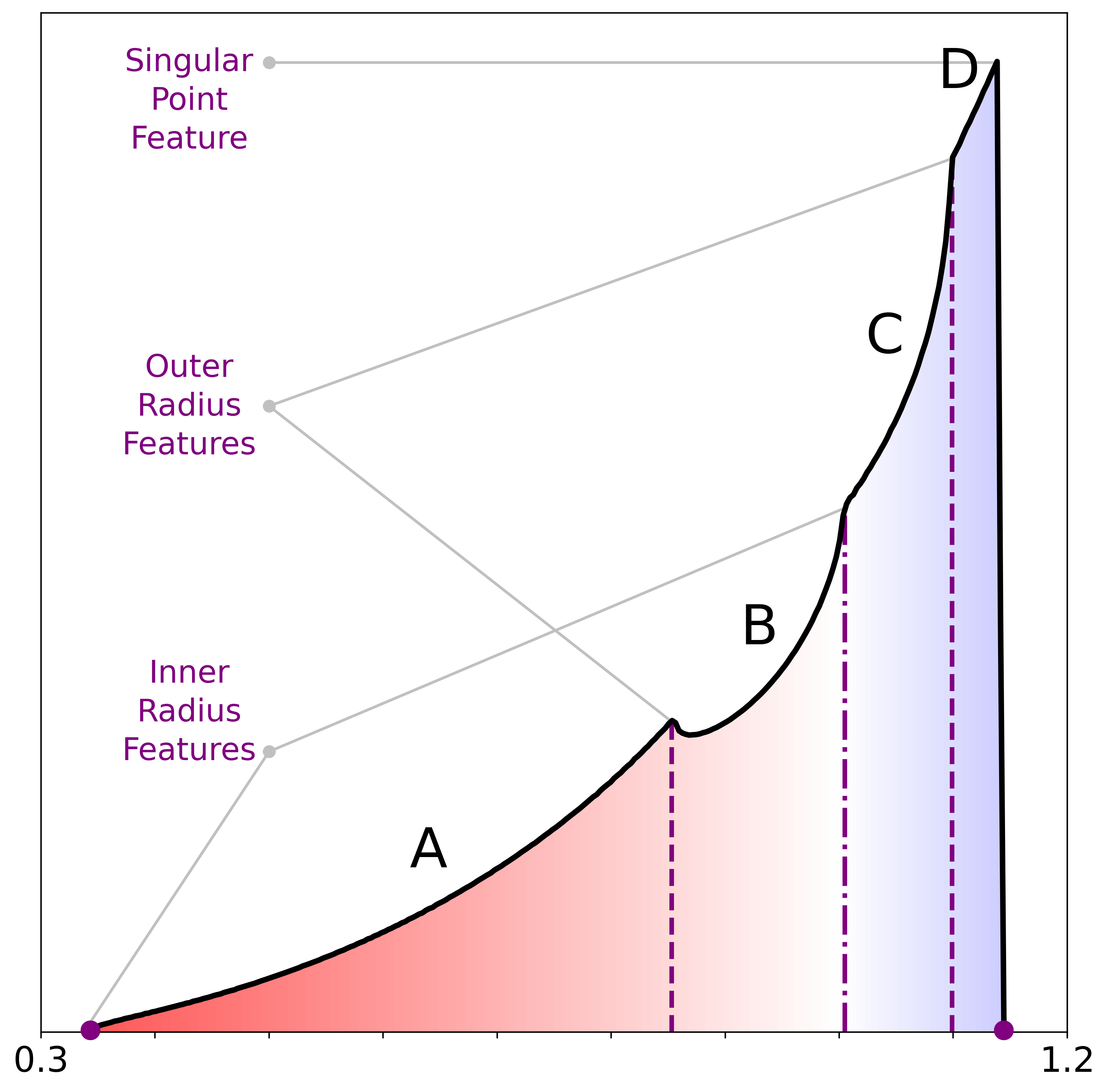}  \\
         \scalebox{1.25}{\Huge Redshift factor, $g$} 
      \end{tabular} }}
    \caption{Topology of the constant redshift factor curves on the resolved image of the accretion disk seed features in the observed spectrum for line emission. 
    \textbf{Left}: Image of an accretion disk, with inner radius $r_{\rm in}=\risco$ and outer radius $r_{\rm out}=40M$, composed of particles traveling on prograde stable circular orbits around a rotating Kerr BH with spin $a=0.7M$, observed at an inclination $\theta_{\rm o}=50^\circ$. (Both the BH and disk rotate clockwise.) The region between the contours of $r_{\rm in}$ and $r_{\rm out}$ (black curves) comprises the flux contributing region (FCR) of the observer's screen and is colored by the value of the redshift factor, $g$. The disk has five critical redshift values denoted by purple curves and points: the extremal redshift of the inner radius $g_{{\rm in}\pm}$ (where $g_{\rm in_-}=g_{\rm MOR}$ is the maximum observable redshift), the extremal redshift of the outer radius $g_{{\rm out}\pm}$, and the maximum observable blueshift $\gmob$. The critical redshift values divide the disk's image into four regions: $g\in\pa{g_{\rm in -},g_{\rm out -}}$ (A), $g\in\pa{g_{\rm out -},g_{\rm in +}}$ (B), $g\in\pa{g_{\rm in +},g_{\rm out +}}$ (C), and $g\in\pa{g_{\rm out +},\gmob}$ (D). Each region contains generic redshift contours (gray solid curves) of distinct topologies: individual open curves (A and C), pairs of open curves (B), and individual closed curves (D). The black region corresponds to the  apparent image of the BH horizon radius.
    \textbf{Right}: The observed spectrum for an emission line, i.e., line profile, assuming a disk emissivity $I_{\rm s} \propto r^{-3}$. The critical redshift values and the segments that correspond to emission from the disk's regions A to D, as described in the left panel, are highlighted. The line profile is normalized by its peak value.}
    \label{fig:ExampleDisk&Line}
\end{figure*}
\section{Summary}
\label{sec:LineProfileMorpology}

To guide the reader, we will now present a high-level, self-contained overview of the remaining sections of the paper.

The observed flux from an accretion disk can be calculated by integrating the specific intensity at each point on the observer's screen $I_{\rm o}(E_{\rm o},\bf{x})$ over the differential solid angle ($\ed \Omega (\bf{x})$) of the disk's image~\footnote{The ``screen'' is the image plane of sky as seen by the observer whereon screen coordinates correspond to directions in the sky.}~\cite{Cunningham1975}. Assuming the spacetime is described by the Kerr geometry, the integration can be performed on either the source or screen coordinates, as points in the disk $(r_{\rm s},\phi_{\rm s})$ can be mapped to points on the observer's screen $\bf{x}$, via the rays that connect the observer to points on the disk.

The screen's specific intensity can be written in terms of the disk's specific intensity, $I_{\rm s}(E_{\rm s},{\bf x})$, using the Liouville's theorem, $E_{\rm o}^{-3}I_{\rm o}(E_{\rm o},{\bf x})=E_{\rm s}^{-3}I_{\rm s}(E_{\rm s},{\bf x})$. The last ingredient to compute the flux is the four-velocity, $u$, of the emitting particles in the disk, as it provides the redshift factor
\begin{equation}
g=\frac{u_{\rm o}\cdot k}{u_{\rm s}\cdot k}=\frac{E_{\rm o}}{E_{\rm s}},
\end{equation}
where $k$ is the four-momentum of the photon. The redshift factor can be expressed as a function of the observer's screen coordinates, $g(\bf{x})$, because frequency-shifting effects, i.e., gravity and the relative disk motion, affect all emission frequencies equally. Putting everything together, the observed flux is given by~\cite{Cunningham1975}
\begin{equation}
    \label{eq:LineProfileEo}
    F_{E_{\rm o}} (E_{\rm o})=\int \ed \Omega ({\bf x})\ g^3({\bf x}) I_{\rm s}(E_{\rm s}=E_{\rm o}/g(\bf{x}),{\bf x}).
\end{equation}

If one considers a rest frame emission of energy $E_{\rm s}$, for which the observed energy $E_{\rm o}$ is one-to-one with the redshift factor $g$, we can define the rest-energy-rescaled specific energy flux $F_g(g)\equiv F_{E_{\rm o}}(E_{\rm o}=g E_{\rm s}) / E_{\rm s}$, normalized such that the total observed flux is
\begin{equation}
    \label{eq:TotalFlux}
    \mathcal{F}_{\rm o}= \int \ed E_{\rm o} F_{E_{\rm o}}(E_{\rm o})=E_{\rm s}^2 \int \ed g \ F_g(g).
\end{equation}
The function $F_g$ is more convenient to work with, as it does not require specifying an emission energy $E_{\rm s}$. Focusing on the line emission, which is the subject of this work, we take the specific intensity to be proportional to the line-emissivity of the disk $I_s({\bf x})$. We refer to the specific flux, $F_{E_{\rm o}}$ or $F_g$, from the line emission as a \emph{line profile}.

To analytically solve for the line profile, the function $F_g(g)$ would need to be explicitly written.
To do so, one would need to invert the redshift factor $g({\bf x})$, and a second function $z_s({\bf x})$ (e.g., either source radius $r_{\rm s}({\bf x})$ or angle $\phi_s({\bf x})$) for the screen coordinates ${\bf x}(g,z_s)$.
The injective map of ${\bf x} \to (g,z_s)$ may not be bijective, and in such cases the inversion must be performed on multiple patches of the observer's screen. For instance, for axisymmetric and equatorial accretion flows like those discussed herein, the source radius $r
_{\rm s}$ is the usual choice of $z_{\rm s}$ when the disk is viewed off the BH's spin axis, requiring two patches on the observer's screen: one patch covering roughly the upper-half of the screen in which $g$ increases with $\phi_{\rm s}$ at fixed $r_{\rm s}$, and the other covering roughly the lower half of the screen in which $g$ decreases with $\phi_{\rm s}$ at fixed $r_{\rm s}$. For on-axis viewing, where $g$ and $r_{\rm s}$ are functions of the magnitude of screen location $|{\bf x}|$, $\phi_{\rm s}$ is the usual choice of $z_{\rm s}$, and a single patch covering the entire observer's screen can be used.

After this inversion, the line profile can then be written as
\begin{subequations}
\label{eq:LineProfileG}
\begin{align}
    F_g(g)&=\sum_{m=1}^\text{\# of patches} F_m(g),\\
    F_m(g)&= {g^3} \int_{z_s^-(g)}^{z_s^+(g)} \ed z_s \abs{\frac{\partial \Omega ({\bf x})}{\partial g \partial z_s}} \  I_{\rm s}(g,z_s).
\end{align}  
\end{subequations}
Unfortunately, the inversion of $(g,z_s)$ for $({\bf x})$ is generally not analytical~\footnote{The inversion is analytic in flat space and presented in App.~\ref{app:NewtonianDisk}.}. 
That is why the observed spectrum, Eq.~\eqref{eq:LineProfileEo}, is typically calculated numerically~\footnote{The code used for the simulations presented in this work is publicly available at~\url{https://github.com/iAART/LineAART}}. (See App.~\ref{sec:Discretize} for details of the method used in this work.) Nonetheless, we can understand its extent and the locations of the transitions in the line profile's behavior, by examining the topology of the contours of constant redshift factor in the region of the screen where there is a contribution to the observed flux. 

In the right panel of Fig.~\ref{fig:ExampleDisk&Line}, we show an example of the line profile morphology under the Standard disk model. The Standard model assumes a geometrically thin, optically thick, equatorial disk  that terminates at the ISCO, made of emitters traveling on Keplerian circular orbits. In the left panel of Fig.~\ref{fig:ExampleDisk&Line}, we show the corresponding resolved image of the accretion disk, highlighting the critical redshifts that indicate the extent of the line profile and transitions in its behavior. As we will explain, the number and order of critical redshift factors entirely determine the associated line profile's morphological features.

In this work, we explicitly show how the line profile morphology encodes information about the disk's extent, BH's spin, and observer's inclination via the critical redshift factors under different disk models for the accretion flow. 
In Sec.~\ref{sec:CriticalRedshifts}, we analytically develop the mechanism by which the accretion disk gives rise to key features in the line profile via critical redshift values determined by the locations of the disk's inner and outer radii.
In Sec.~\ref{sec:StandardDiskMorphologies}, we develop a classification of the line profile morphologies, based on the ordering of the critical redshift values under the Standard disk. In Sec.~\ref{sec:MeasureSpinIncRout}, we discuss how to discriminate between different line profile morphologies, and how these line profile features provide a precise way of constraining the BH's spin, as well as the disk's inclination and extent, without having to specify an emissivity profile. In Sec.~\ref{sec:NonStandard}, we explore how the line profile morphology changes under two additional disk models---i) the Cunningham model, which generalizes the Standard disk model by including emission from sources interior to the ISCO (the so-called plunging region), and ii) a model allowing for non-Keplerian circular accretion flows---pointing out how sensitive interpretation of the BH-disk parameters is on the assumed disk model. A discussion on prospects for future work is presented in Sec.~\ref{Sec:Discussion}. The appendices contain most of the technical details of the numerical implementation for computing the lines and the disk models. 

Throughout the paper, we use geometric units in which $G_{\rm N}=1=c$, the $(-,+,+,+)$ metric signature, and assume the spacetime to be described by the Kerr geometry (see, e.g., Ref.~\cite{Bambi2024} for a review on beyond-Kerr spacetimes). Additionally, we will parameterize the observer inclination using
\begin{align}
    x_{\rm o}=\sin \theta_o.
\end{align}
In resolved disk images, we align the projection of the BH spin axis along the vertical direction~\footnote{Reversing the direction of the BH's rotation results in a resolved disk image being reflected about the projected BH spin axis on the observer's screen. Similarly changing the alignment of the projected BH spin axis on the observer's screen results in a resolved disk image that is rotated. The results we derive for line profiles are invariant under these changes in a resolved disk image, as such changes do not affect the observed spectra.}. For illustrative purposes, we will show resolved images of BHs rotating clockwise. We also restrict our analysis to prograde disks. 

\section{The Critical Redshift Values \& the Line Profile Morphologies}
\label{sec:CriticalRedshifts}

In this section, we discuss how the emitting region in the spacetime imprints key features on the line profile morphology, by studying the behavior of all constituent functions of the line profile, Eq.~\eqref{eq:LineProfileG}, on the observer's screen. Without making strong assumptions about the disk emissivity, we will determine for the line profile 1) the extent and the location of sharp features, i.e., kinks, 2) the behavior near the edges and kinks, and 3) the behavior between kinks. All these features are linked to critical values given by the local maxima and minima of the redshift factor of the disk on the observer's screen. Studying the link between the resolved image of emitting region and line profile will allow us to develop a general procedure for identifying key line profile features, which we will apply to the line profiles produced from an equatorial accretion disk around a Kerr BH.

Mapping the emission region in the spacetime to the observer's screen, we can identify the set of (connected) points that mark boundary curves from which the observer receives flux. These boundaries enclose the flux contributing region (FCR). If the emissivity is a positive and continuous function of the emitter location in the emission region in the spacetime, then the emissivity as a function of screen coordinates will also be positive and continuous in the FCR. Likewise, if the accretion flow is continuous in the emission region, then the redshift factor on the observer's screen will also be continuous in the FCR. Under these conditions, critical values in the line profiles can be found by studying only the redshift factor in the FCR (as exemplified in the right panel of Fig.~\ref{fig:ExampleDisk&Line}). 

\emph{Line profile extent and kinks}---Examining Eq.~\eqref{eq:LineProfileG}, the line profile must be a piecewise function changing functional form whenever any of its constituent functions changes form. Thus, with $(g,z_s)$ being inversions of screen coordinates ${\bf x}$, these transitions in the line profile $F_g(g)$ will occur at the local maxima and minima of $g({\bf x})$ in the FCR. The local maxima and minima of the critical redshift values correspond to the edges and kinks in the line profile.

The line profile edges, the maximum observable redshift (MOR) and the maximum observable blueshift (MOB)~\cite{Gates2020} are given by the global minimum and maximum values of the critical redshift factors, i.e., 
\begin{subequations}
\label{eq:MORMOB}
\begin{align}
    \gmor&=\min\cu{g({\bf x})| {\bf x}\in \text{FCR}},\\
    \gmob&=\max\cu{g({\bf x})| {\bf x}\in \text{FCR}}.
\end{align}   
\end{subequations}

Taking into account all local minima and maxima, a disk with $N$ critical values $\cu{g^c_i|i\in\cu{0,1,...,N-1}}$ in ascending order, has MOR and MOB given by $\gmor=g_{0}^c$ and $\gmob=g_{N-1}^c$, respectively.
All other critical values $g\in\cu{g^c_i} \setminus \cu{\gmor,\gmob}=\cu{g_i^c|i\in\cu{1,2,...,N-2}}$ mark transitions in the line profile behavior. Hence, the line profile exhibits two fewer kinks ($n=N-2$) than the number of critical values, and has $N-1$ intervals $(g_{i-1}^c,g_{i}^c)$ on which it is smooth.

Even without access to the explicit form of the line profile, we can understand and identify these critical values, which mark behavior changes in the line profile, by examining the topology of the redshift factor level sets across the FCR.
The topology of the on-screen redshift contours in the FCR determines the behavior of the line profile. These redshift factor levels can be points, closed curves, or open curves. We distinguish the level set topologies in terms of the kind of contour (point, closed curve, or open curve) and which boundaries of the FCR, if any, they touch. Topology changes in redshift contours only occur at critical values given by local minima and maxima of $g({\bf x})$, both interior and at the boundaries of the FCR.  
We can use the critical value contours to divide the FCR into $N-1$ regions $\mathcal{G}_i$ wherein $g({\bf x})\in(g^c_{i-1},g^c_{i})$ (as exemplified in the left panel of Fig.~\ref{fig:ExampleDisk&Line}). We find that within an FCR subset $\mathcal{G}_i$, the number of contours of each topology at each generic, i.e. non-critical, redshift factor value is constant. 

\emph{Behavior near critical values}---Additionally, we find that the behavior of the line profile near a critical value is determined by whether the critical value is associated with local maxima or minima of $g({\bf x})$. 
When the line profile is discontinuous at the critical value $g_i^c$, the discrepancy between the line profile's value as we approach $g_i^c$ from the left and right is positive (negative) when $g_i^c$ is a maximum (minimum),
\begin{align}
    \label{eq:RelativeSteepness}
    F_g(g_i^{c-}) - F_g(g_i^{c+}) \gtrless 0,
\end{align}
e.g., the line profile at $\gmob$ in the left panel of Fig.~\ref{fig:ExampleDisk&Line}.
When the line profile is continuous at critical value $g_i^c$, the discrepancy in the line profile's steepness, i.e., magnitude of the line profile's derivative, as we approach $g_i^c$ from the left and right is positive (negative) when $g_i^c$ is a maximum (minimum),
\begin{align}
    \abs{F_g'(g_i^{c-})} -\abs{ F_g'(g_i^{c+})} \gtrless 0,
\end{align}
e.g., the line profile at $\cu{g_{{\rm in}\pm},g_{{\rm out}\pm}}$ in left panel of Fig.~\ref{fig:ExampleDisk&Line}.

These statements follow from examining the neighborhood around critical redshift contours on the observer's screen. Consider a contour $g_i^c$ for which the function $g({\bf x})$ is maximized/minimized at every point along the $g_i^c$. The neighborhood of $g_i^c$ contributes to the flux at $g\lessgtr g_i^c$ but not $g\gtrless g_i^c$. 
As such, $F_g(g_i^{c\pm})=F_g(g_i^{c\pm})=0$. 
Therefore, $F_g(g_i^c)$ is either discontinuous or the limit from the left/right is decreasing to reach zero and $\abs{F_g(g_i^{c\pm})}<0$. (See $g_{in-}$ and $\gmob$ in the left panel of Fig.~\ref{fig:ExampleDisk&Line} as an example.) Consider a contour $g_i^c$ which touches a boundary curve at a point ${\bf x}_b$ and extends into the interior of the FCR, such that in a neighborhood of ${\bf x}_b$, the only local maximum/minium of $g({\bf x})$ along $g_i^c$ is at ${\bf x}_b$, i.e., $g_i^c=g({\bf x}_b)$. Examining the flux contribution from a neighborhood of ${\bf x}_b$, if we  deform the boundary such that ${\bf x}_b$ becomes a point on the interior of the FCR, the flux contribution from the $g_i\lessgtr g({\bf x}_b)$ would increase while the contribution from the $g\gtrless g({\bf x}_b)$ would remain fixed. Further, this deformation would cause the line profile to become smooth at $g({\bf x}_b)$. Therefore, when ${\bf x}_b$ is a local maximum/minimum of the boundary of the FCR, the line profile must be steeper on the left/right side of $g_i^c=g({\bf x}_b)$, $\abs{F_g(g_i^{c+})}\lessgtr \abs{F_g(g_i^{c-})}$.
(For an example consider the points where $g_{\rm in+}$ and $g_{{\rm out}\pm}$ meet the $r_{\rm in,out}$ in left Fig.~\ref{fig:ExampleDisk&Line} and imagine deforming $r_{\rm in,out}$ inward/outward.)

If a critical value represents both a local minimum and a local maximum, then the relative value and steepness of the line profile at $g_i^c$ can take any value. This can occur, for example, when a $g_i^c$ contour is an open curve connecting two points ${\bf x}_{b_1}$ and ${\bf x}_{b_2}$, each on a different FCR boundary curve, with $g({\bf x})$ minimized at ${\bf x}_{b_1}$ and maximized at ${\bf x}_{b_2}$.

\emph{Behavior between critical values}---Finally, the functional form of the line profile is subject to the topology of the redshift contours. Thus, if two FCR subsets, $\mathcal{G}_i$ and $\mathcal{G}_j$, have the generic redshift contours of the same (differing) topologies, then the line profile will have the same functional forms (differing functional forms) on the corresponding intervals, $g\in(g^c_{i-1},g^c_{i})$ and $g\in(g^c_{j-1},g^c_{j})$.

Having developed an understanding of which (emissivity independent) features the disk imprints on the line profile, we now outline the procedure to obtain these key features: 
\begin{itemize}
    \item Find the FCR on the observer's screen by finding the region in which the integrand of Eq.~\eqref{eq:LineProfileG} is positive.
    \item Determine the critical values by finding the local minima and maxima of the redshift factor over the FCR; or, by foliating the FCR with contours of constant redshift factor and identifying redshift factor values where the the contours change topology.
    \item Assess the line profile's functional form in the generic redshift factor intervals, i.e., continuous intervals extending between consecutive critical values, by dividing the FCR into regions corresponding to the generic redshift factor intervals and examining the topology of the generic redshift contour therein.
\end{itemize}
Now we apply this procedure to BHs with axisymmetric equatorial accretion disks.

\emph{The FCR}---On the observer's screen, the constant radii contours are always concentric closed curves, so the FCR is the continuous screen region between the on-screen contours of the inner disk and outer disk radii, $r_{\rm in}$ and $r_{\rm out}$, respectively. The radii $r_{\rm in,out}$ are the miniminum/maximum disk radii at which the emitted frequency $E_{\rm s}$ has support. These values will, naturally, depend on the specific astrophysical details of the BH-disk system (see, e.g., Refs.~\cite{2009ApJ...707L..87T,2010A&A...521A..15A,Fabian:2014tda} for analyses where the inner radius is not at the ISCO).

\emph{The critical values}---The on-screen contour of constant radius $r_{\rm s}$ is a double cover of redshift factors $g\in(g_{\rm s -},g_{\rm s +})$ where~\footnote{This is generically true for infinitely thin axisymmetric accretion flows where the velocity profiles have radial and angular components which are monotonic functions of the source radius.} 
\begin{subequations}
\label{eq:ExtremalResdhiftAtFixedRadius}
\begin{align} 
    g_{\rm s -}(a,x_{\rm o},r_{\rm s})&\equiv \min \cu{g(a,x_{\rm o},{\textbf{x}})\vert r_{\rm s} (a,x_{\rm o},\textbf{x}) \text{ fixed}},\\
    g_{\rm s+}(a,x_{\rm o},r_{\rm s})&\equiv \max \cu{g(a,x_{\rm o},{\textbf{x}})\vert r_{\rm s} (a,x_{\rm o},\textbf{x}) \text{ fixed}}.
\end{align}    
\end{subequations}
As such, the critical redshift factors from the boundary of the FCR are the minima and maxima of the disk's inner and outer radii, 
\begin{subequations}
\label{eq:ExtremalResdhiftAtDiskEdges}
\begin{align}
    g_{\rm in \pm}(a,x_{\rm o})&\equiv g_{\rm s \pm}(a,x_{\rm o},r_{\rm in}), \\
    g_{\rm out \pm}(a,x_{\rm o})&\equiv g_{\rm s \pm}(a,x_{\rm o},r_{\rm out}).
\end{align}   
\end{subequations}

Further, for the accretion flows studied herein, $g_{\rm s-}$ is monotonically decreasing with radius; thus, the MOR is always sourced by the disk's inner radius $\gmor=g_{\rm in -}$. The radius that sources the MOB, on the other hand, depends on the BH-disk parameters (viewing inclination, spin, and the disk's inner and outer radii). When the disk is viewed down the spin axis ($x_{\rm o}=0$), the redshift factor at fixed source radius is constant, and the MOB is sourced by the disk's outer radius and is given by $g_{\rm out}$. In this case, there are two critical redshift values, $\cu{g_{\rm in},g_{\rm out}}$, and there are no transitions in the line profile's behavior.

When the disk is viewed off the spin axis ($x_{\rm o}>0$), the MOB may be sourced at any radius on the disk: either the disk's inner radius, given by $g_{\rm in +}$; the disk's outer radius, given by $g_{\rm out +}$; or at a finite radius,
\begin{align}
\label{eq:FFPRadius}
    \rffp(a,x_{\rm o})\in\pa{r_{\rm in},r_{\rm out}},
\end{align}
given by
\begin{subequations}
\label{eq:FFPRedshift}
\begin{align}
   \gffp(a,x_{\rm o})&\equiv g_{\rm s+}(a,x_{\rm o},\rffp(a,x_{\rm o})),\\
   &=\max_{r_{\rm s} \in (r_{\rm in},r_{\rm out})} g_{\rm s+}(a,x_{\rm o},r_{\rm s}). 
\end{align}    
\end{subequations}
When the MOB is sourced at the boundary of the disk, the redshift factor has no extrema on the interior of the FCR. Therefore, the line profile has up to four distinct critical redshift values, $\cu{g_{{\rm in}\pm},g_{{\rm out}\pm}}$, and will have at most two kinks. On the other hand, when the MOB is not sourced on the boundary, $\gffp$ is the only extrema on the interior of the FCR. Therefore, the line profile has up to five distinct critical redshift values, $\cu{g_{{\rm in}\pm},g_{{\rm out}\pm},\gffp}$, and will have at most three kinks.

Additionally, when the MOB is at a radius $\rffp\in\pa{r_{\rm in},r_{\rm out}}$, $\gffp$ corresponds to a unique point on the observer's screen where the Jacobian between the screen coordinates ${\bf x}$ and $(g,r_{\rm s})$ becomes singular. Each point in the FCR generally provides an infinitesimal amount of flux; however, the flux at $\gffp$ is finite~\footnote{In reference to flux of the line profile throughout this work, we will use ``finite'' to mean neither zero nor infinite.}~\cite{Gates2020}. Thus, we will refer to this as the finite flux point (FFP). 

\emph{Functional form between critical values}---For the accretion flows considered in this work, there are only four topologies which the generic redshift contours exhibit. The specific topologies of the generic contours in each FCR subset $\mathcal{G}_i$ is related to the ordering of critical redshift factors values, which is subject to the BH-disk parameters. 

To aid in the discussion of line profile morphologies, let us introduce a useful notation.  
We denote the topologies at generic redshift factor values as
 \begin{align}
    \bigcirc: & \text{ closed curves touching neither } r_{\rm in,out},\nonumber\\
    \CutD: & \text{ open curves connecting }r_{\rm in}\text{ to }r_{\rm in},\nonumber\\
    \CutU: & \text{ open curves connecting }r_{\rm out}\text{ to }r_{\rm out},\nonumber\\
    \vertvert: & \text{ pairs of open curves connecting }r_{\rm in}\text{ to }r_{\rm out}.\nonumber
\end{align}
Additionally, we introduce shorthand notation which allow us to describe the disk which produces the line profile, wherein we organize the critical redshift values in ascending order and write down the topologies the generic redshift contours for values between consecutive critical values. We will call this a ``redshift factor configuration''. As an example, consider the disk shown in Fig.~\ref{fig:ExampleDisk&Line} which has five critical values (the MOR, three kinks, and the MOB). These critical values divide the image of disk into four subsets A-D in ascending order. 
We describe this disk with the configuration $\CutD~\vertvert~\CutU~\bigcirc$.

The configuration notation is also helpful as it allows one to read off the order of critical values associated with the kinks. Consider neighboring topologies in the configuration, if going from one topology to the next requires adding (removing) a segment to the contour, the critical value is a local maximum (minimum); if going from one topology to the next requires affecting the top (bottom) of the contour, the critical value is associated with the outer (inner) disk radius. For example, the critical values associated with the topology changes of the configuration $\CutD~\vertvert~\CutU~\bigcirc$ are $\cu{g_{\rm out-},g_{\rm in+},g_{\rm out+}}$, respectively. 

In the next section, we will explore the BH-disk parameters of the Standard disk model and systematically classify all the line profile morphologies allowed. For readability throughout our discussions, we will regularly refer to the quantities defined in Eqns.~\eqref{eq:ExtremalResdhiftAtFixedRadius}-\eqref{eq:FFPRedshift}, without necessarily writing all their arguments.

\section{Classification of the Morphologies of the Line Profile }
\label{sec:StandardDiskMorphologies}

The description of the critical redshift values provided in the previous section did not assume a velocity profile for the particles in the disk. In this section, we will fully characterize and classify the morphology of the line profiles produced by the Standard disk model~\cite{Novikov1973}. This model assumes that the observable emission comes from accreting disk particles that move on stable circular orbits on the equatorial plane down to the ISCO radius $\risco$. While the Standard disk model sets $r_{\rm in}=\risco$~\footnote{The inner radius of the disk is often left as a free parameter, i.e., $r_{\rm in}\geq \risco$, when performing data analyses.}, many of the features described in previous section  will hold whether the inner edge is at the ISCO or a larger radius. Therefore, we will use $r_{\rm in}$ whenever the results hold for a $r_{\rm in}\geq \risco$

To disentangle the order of the critical redshift values marking transitions caused by the disk's inner and outer radii in the associated line profile morphologies, we will begin by studying an (idealized) infinitely large disk. This idealized disk will allow us to probe the effects of the disk's inner radius on the line profile morphology. Once that relation is established, we will explore finite disks to probe the effects of the disk's outer radius on the line profile morphology. 

\subsection{An Infinitely Large Accretion Disk}
\label{sec:InfiniteDisk}

Let us examine the topology of constant redshift contours in the FCR of the observer's screen, which, for the infinitely large disk, corresponds to all the screen points outside the $r_{\rm in}$ contour. In this case, the contours of constant redshift factor can have one of two topologies, $\bigcirc$ or $\CutD$, depending on the viewing inclination and spin of the BH~\footnote{A special case is the $g=1$ contour for off-axis observers, which appears to be an open curve extending to infinity on both ends. We will consider this case to be a closed curve.}. 

There are three distinct configurations of the redshift contours on the observer's screen, each corresponding to a different set of MOB source radius values. We summarize the on-screen redshift contour configurations in Tab.~\ref{tab:InfiniteDisk} and detail them below.

\begin{table}
{\centering
\begin{tabular}{c| c| c} 
\hline
  Disk type & MOB source radius & Configuration  \\
\hline
\hline
I & $r_{\rm out}$ & $\bigcirc$  \\
\hline
II & $r_{\rm s}\in\pa{r_{\rm in},r_{\rm out}}$ & $\CutD~\bigcirc$ \\
\hline
III  & $r_{\rm in}$ & $\CutD$ \\
\hline
\end{tabular}}
\caption{Types of configurations of constant redshift factor contours on the observer's screen, for an infinitely large disk. The configuration for Type II should be read from left to right, increasing in redshift factor.}
\label{tab:InfiniteDisk}
\end{table}

\begin{figure*}
    \begin{tabular}{ccccc}
    {Type I Disk: $\bigcirc$} &\quad & {Type II Disks: $\CutD$  $\bigcirc$} &\quad & {Type III Disk: $\CutD$}\\
    {\resizebox{.19\linewidth}{!}{
    \begin{tabular}{|c|}
    \hline\\ \\[5ex]
    \includegraphics{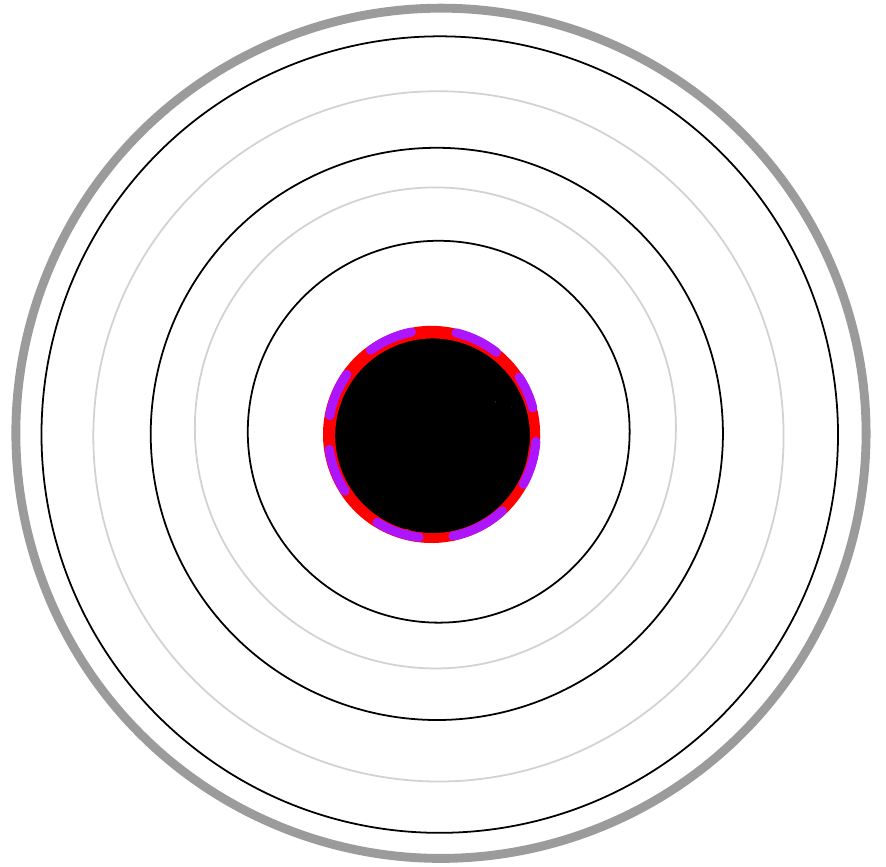}\\[5ex]
    \includegraphics[width=\linewidth]{ 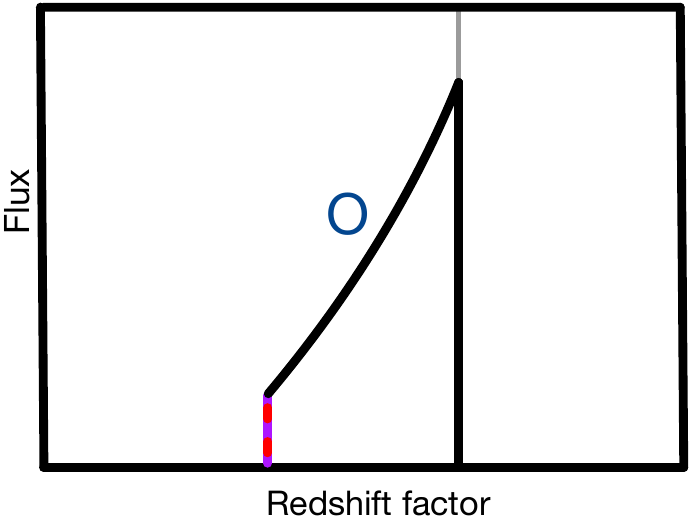}\\[5ex]
    \hline\\
    \Huge\underline{$a/M=0.7, \ \theta_{\rm o}=0^\circ$}\\
    \includegraphics[width=\linewidth]{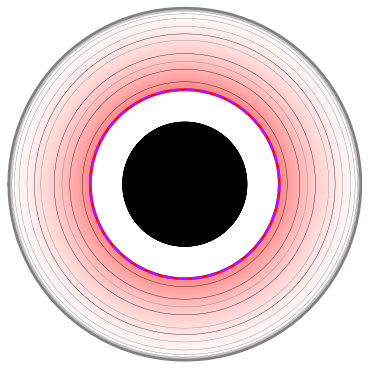}\\
    \includegraphics[width=\linewidth]{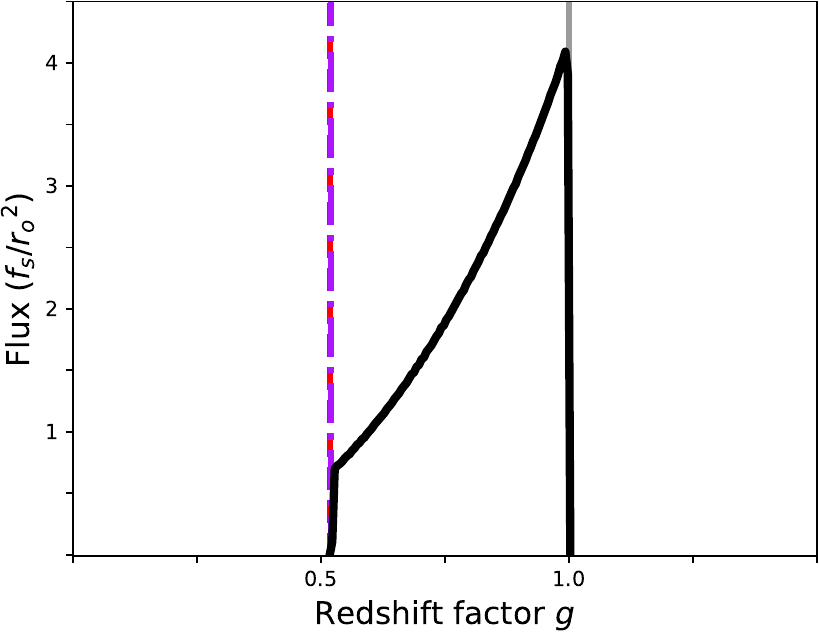}\\[5ex]
    \hline
    \end{tabular}}}
    &&
    {\resizebox{.565\linewidth}{!}{
    \begin{tabular}{|c|c|c|}
    \hline
    &&\\
    {\Huge\underline{$g_{\rm in +}<1$}}& 
    {\Huge\underline{$g_{\rm in +}=1$}}&  
    {\Huge\underline{$g_{\rm in +}>1$}}\\[5ex]
    \VF{\includegraphics{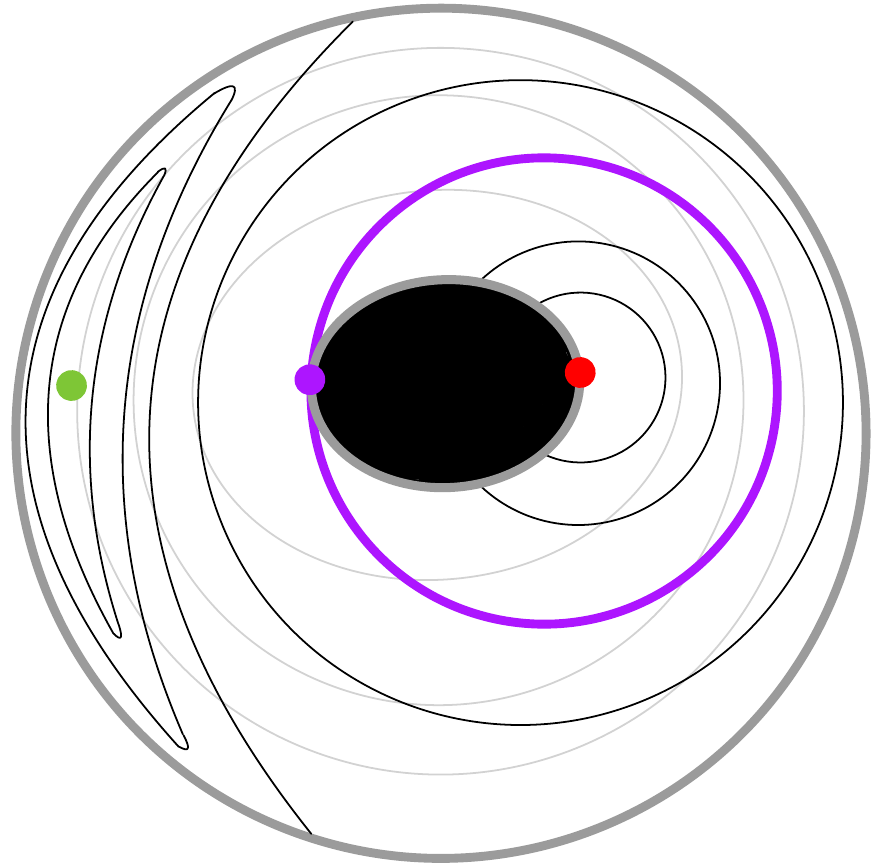}}&
    \VF{\includegraphics{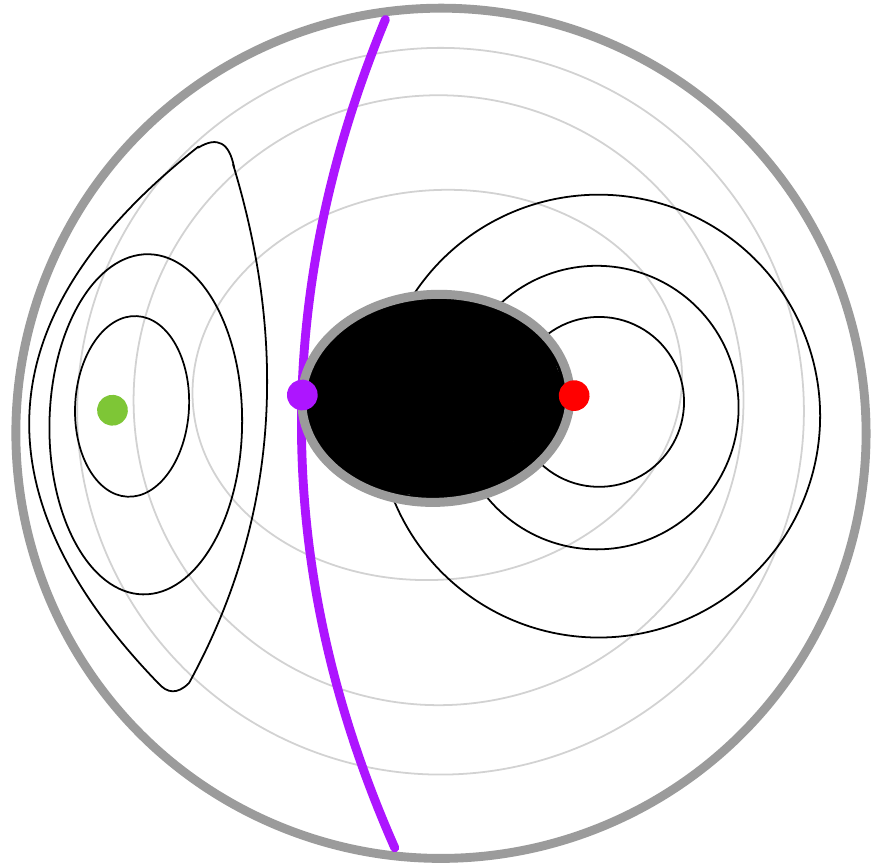}}&
    \VF{\includegraphics{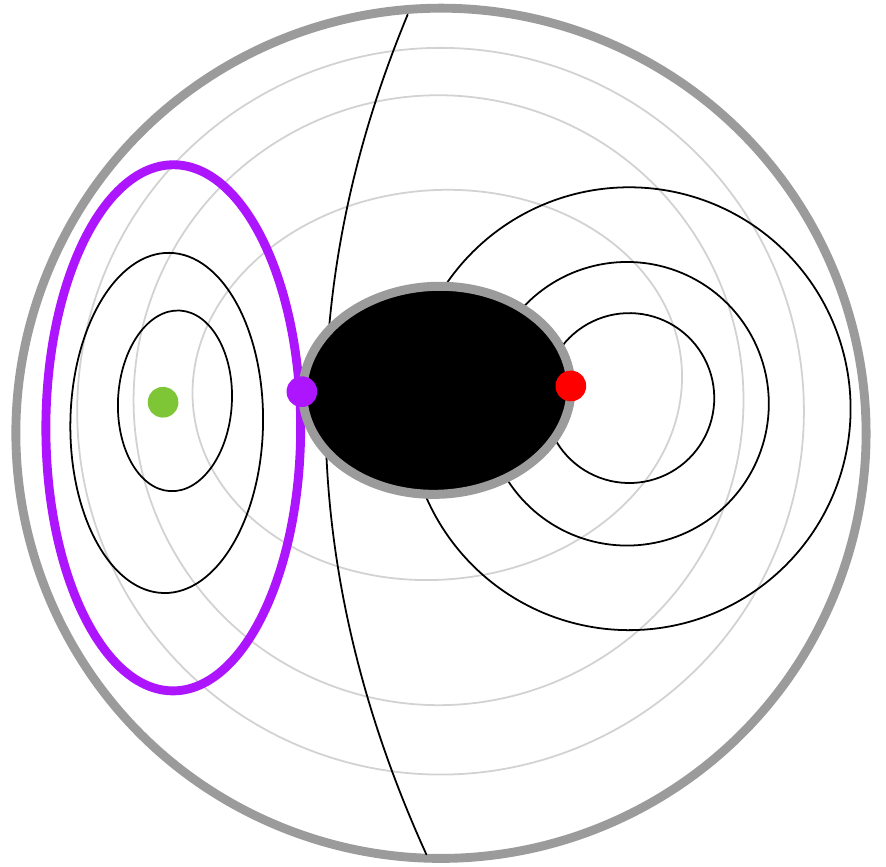}}\\[5ex]
    \includegraphics[width=\linewidth]{ 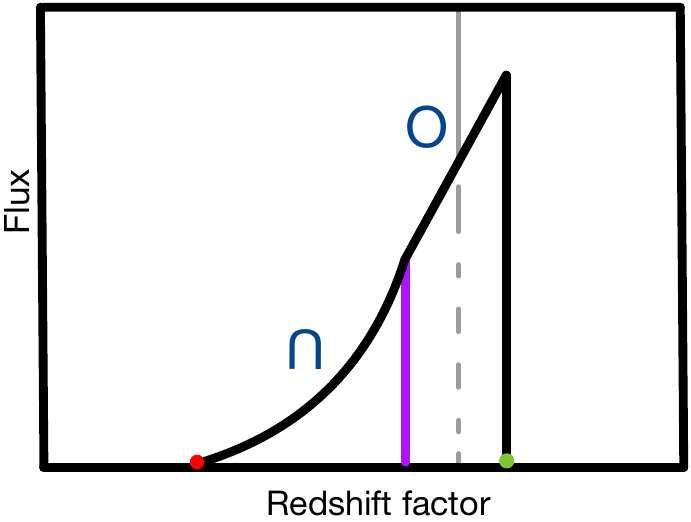}&
    \includegraphics[width=\linewidth]{ 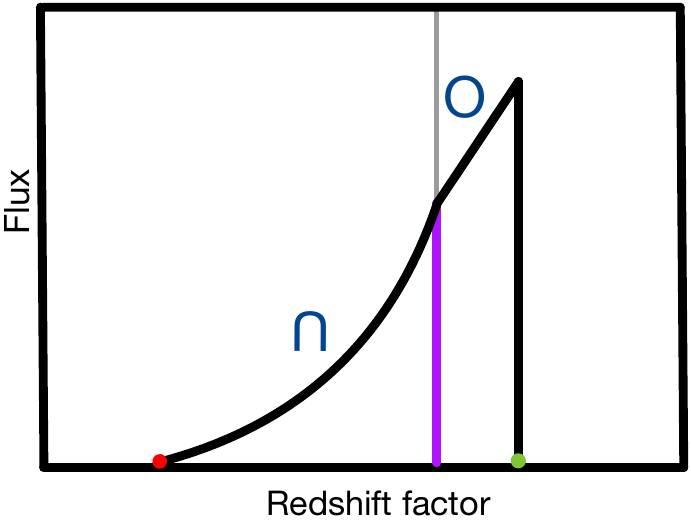}&
    \includegraphics[width=\linewidth]{ 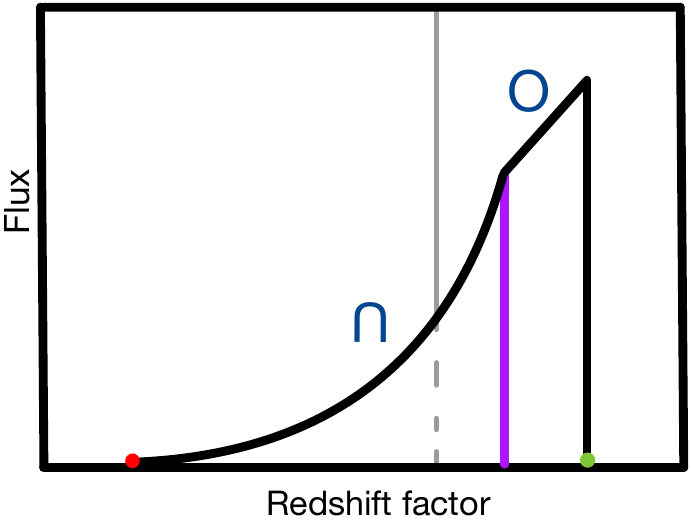}\\[5ex]
    \hline
    &&\\
    \Huge\underline{$a/M=0.7, \ \theta_{\rm o}=30^\circ$}&
    \Huge\underline{$a/M=0.7, \ \theta_{\rm o}=50^\circ$}&
    \Huge\underline{$a/M=0.7, \ \theta_{\rm o}=60^\circ$}\\
    {\includegraphics[width=\linewidth]{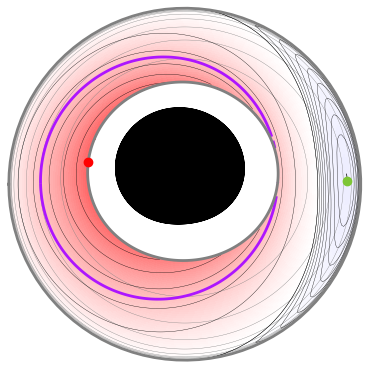}}& 
    {\includegraphics[width=\linewidth]{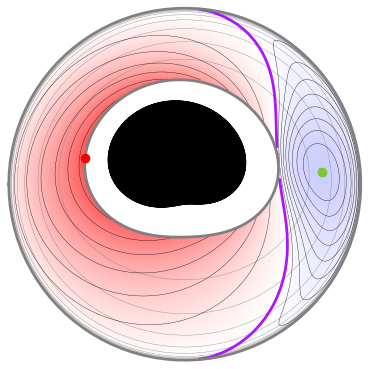}}&
    {\includegraphics[width=\linewidth]{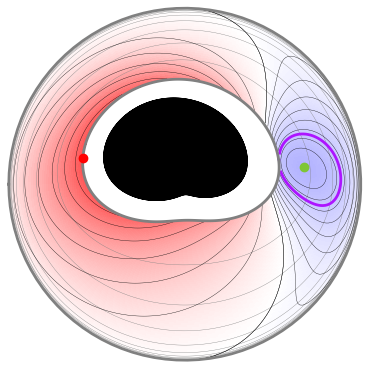}}\\
    \includegraphics[width=\linewidth]{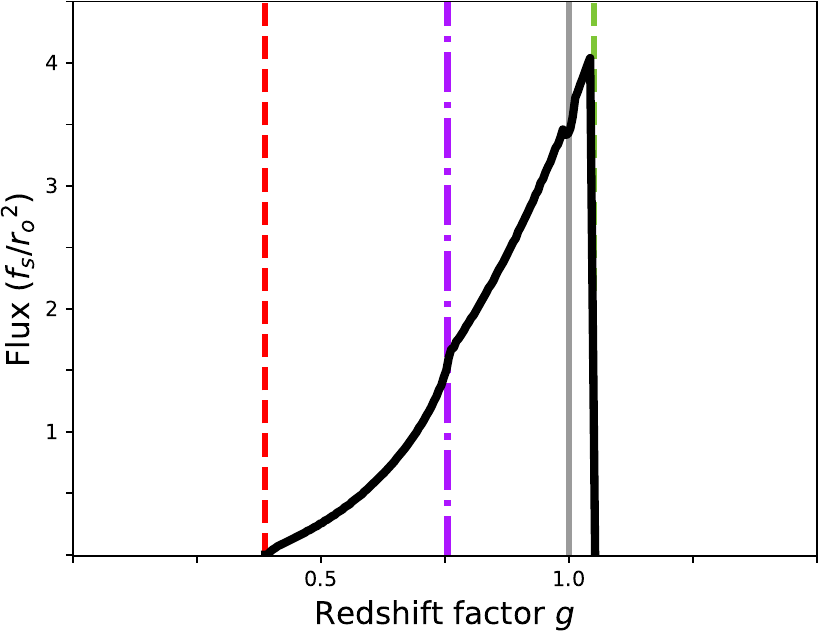}&
    \includegraphics[width=\linewidth]{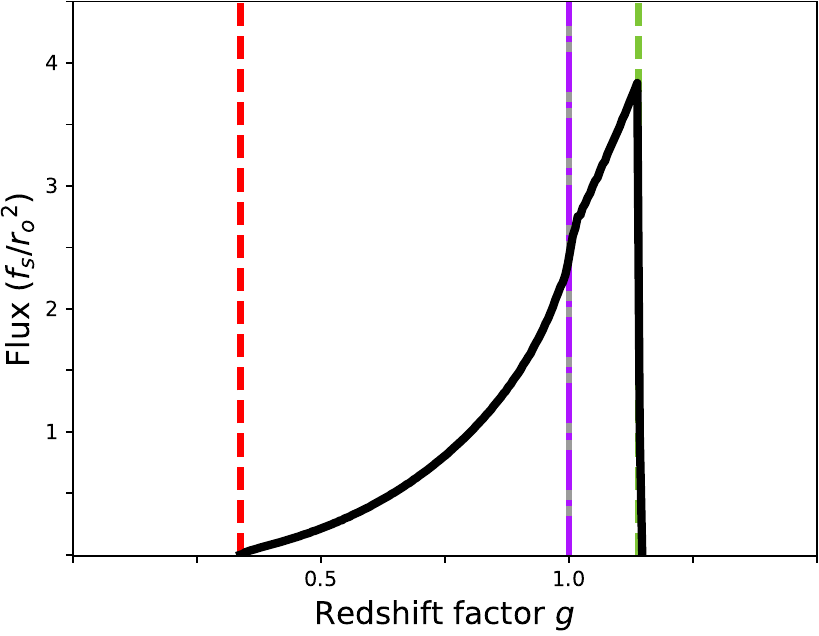}&
    \includegraphics[width=\linewidth]{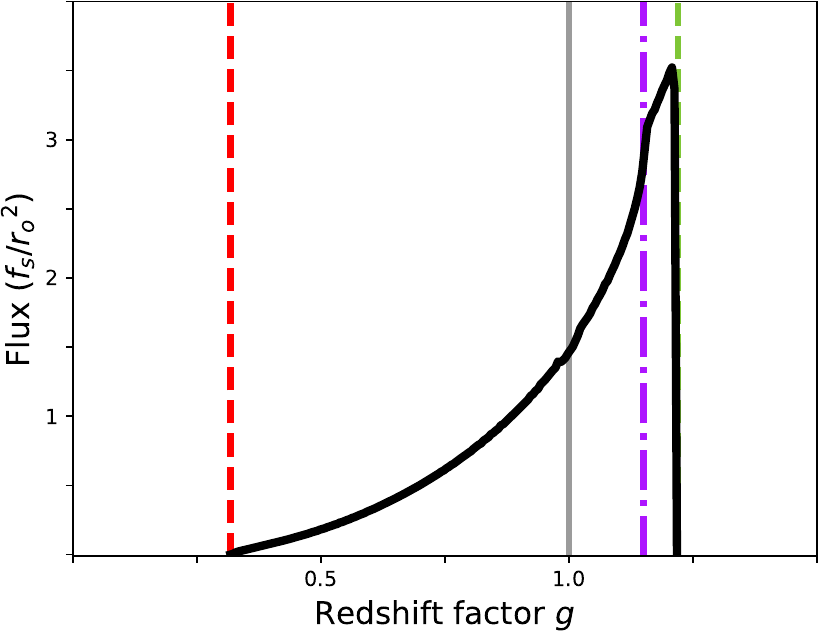}\\[5ex]
    \hline
    \end{tabular}}} 
    &&
    {\resizebox{.19\linewidth}{!}{
    \begin{tabular}{|c|}
    \hline\\ \\[5ex]
    \VF{\includegraphics{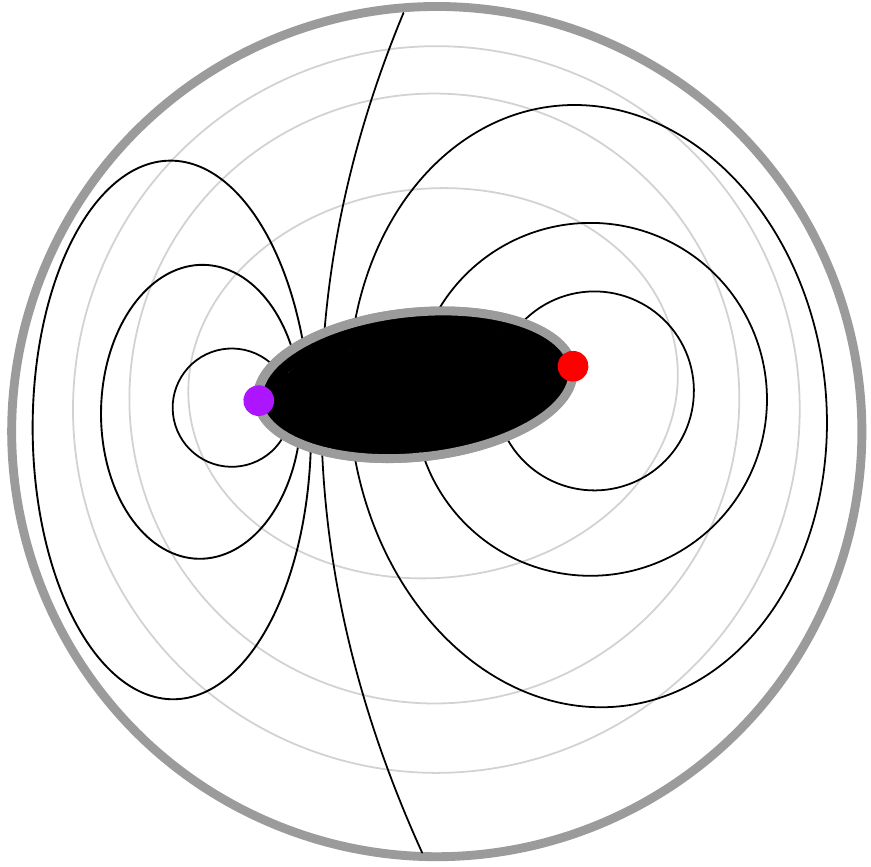}}\\[5ex]
    \includegraphics[width=\linewidth]{ 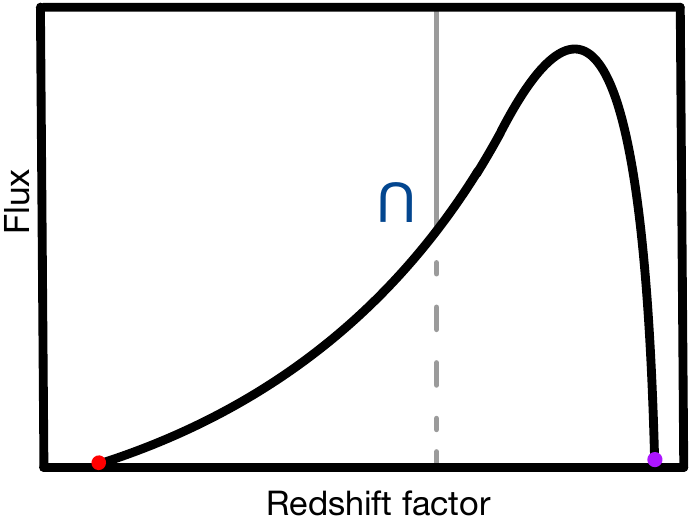}\\[5ex]
    \hline\\
    \Huge\underline{$a/M=0.7, \ \theta_{\rm o}=75^\circ$}\\
    {\includegraphics[width=\linewidth]{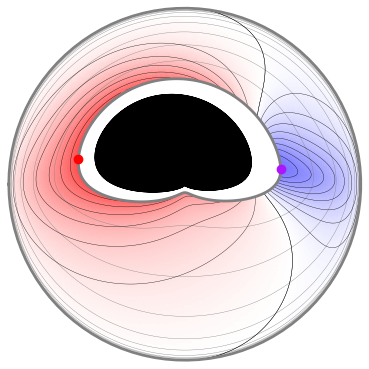}}\\
    \includegraphics[width=\linewidth]{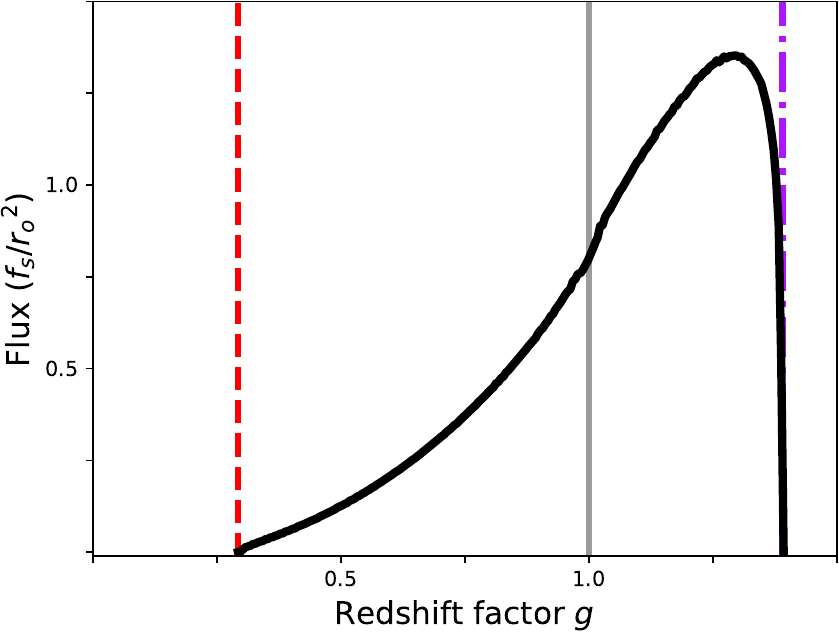} \\[5ex]
    \hline
    \end{tabular}}}
    \end{tabular}
    \caption{Line profile morphologies associated with the redshift contour configurations for infinite large disks.
    {\textbf{First row}:} Schematics for the observer's screen in compactified coordinates (Eq.~\ref{eq:CompactBardeenRadius}). On the observer's screen, the gray curves correspond to constant radii $r\in[\risco,\infty)$, and the black lines denote contours of constant redshift factor. The interior of the disk is shown with the black solid circle, and the minimum and maximum redshifts of the disk inner radius, $g_{\rm in -}$ and $g_{\rm in +}$, are shown with red and purple contours, respectively. For the Type II disk, we show the FFP redshift factor $\gffp$ with a green dot.
    {\textbf{Second row}:} Schematics for the line profiles that arise from the redshift contour configurations, where, to guide the reader, we have marked with a vertical gray line $g=1$. Suggested rephrasing: Each relevant section of the line is labeled with its corresponding configuration.
    {\textbf{Third row}:} Explicit examples of the observer's screen in compactified coordinates~\eqref{eq:CompactBardeenRadius} for spinning BH with $a=0.7M$, viewed at angles, from left to right, $\theta_{\rm o}=0^\circ,\ 30^\circ,\ 50^\circ,\ 60^\circ$ and $\ 75^\circ$. The screen is colored by redshift factor for $r>\risco$ using the same scale as in Fig.~\ref{fig:gMinMaxISCO}, and the apparent location of the horizon is shown in black.
    {\textbf{Fourth row}:} The corresponding line profiles for large disks with inner and outer disk radii $r_{\rm in}=\risco\simeq 3.4M$ and $r_{\rm out}=5000M$, respectively, at the aforementioned inclinations for disk emissivity $I_{\rm s}=f_s r_{\rm s}^{-3}$. The line profile transitions around $g=1$ for the Type II and III disks are unmarked; these transitions are addressed in the next section (Sec.~\ref{sec:FiniteDisk}).}
    \label{fig:InfiniteDisks}
\end{figure*}

\underline{\emph{Type I:}}
When the disk is viewed from the spin axis ($x_{\rm o}=0$), there is a one-to-one map between the emitter radius $r_{\rm s}$ and the redshift factor $g(r_{\rm s})$, so each radius only contributes to the flux at one redshift factor. 
(As only zero angular momentum photons reach the spin-axis from the equatorial plane, the redshift factor $g(r_{\rm s})$ is analytic and given by \eqref{eq:StandardDiskG} evaluated at $\lambda=0$.) The redshift factor $g(r_{\rm s})$ is monotonically increasing with radius. Thus, the MOR is set by the inner edge of the disk $\gmor=g(r_{\rm in})$. The MOB is sourced by $r_{\rm s}=\infty$, $\gmob=g(r_{\rm s}=\infty)=1$.  Described heuristically, radii closer to the BH experience more gravitational redshift, and the plane of motion of the disk is perpendicular to the observer's line of sight. Thus, no emission, in this case, enjoys Doppler shifting.

Further, the constant radii contours, and thus redshift contours, produce concentric circles on the observer's screen, so the whole disk image has topology $\bigcirc$. As each redshift contour is a closed curve on the observer's screen, the line profile flux \eqref{eq:LineProfileG} is finite for all redshifts $g\in[g(r_{\rm in}),1)$. The line profile flux at the MOB $\gmob=1$ is set by emissivity in the large radius limit. For emissivity of order $I_s\sim\O{r_{\rm s}^{-q}}$ when $r_{\rm s}\to\infty$, the MOB flux is vanishing for $q>3$, finite for $g=3$, or diverging for $q<3$.

\underline{\emph{Type II}:}
When viewing the disk from an angle $x_{\rm o}>0$, the constant radii contours are still closed and concentric curves but are no longer circles. A curve of fixed source radius $r_{\rm s}$ contributes to the flux via i) one screen point at each of the redshifts $g_{\rm s-}(r_{\rm s})$ and $g_{\rm s+}(r_{\rm s})$, and ii) two screen points at each $g\in\pa{g_{\rm s-}(r_{\rm s}),g_{\rm s+}(r_{\rm s})}$. Thus, the MOR and MOB are each sourced by single points in the disk. 

For certain inclinations and spin combinations (inclination sufficiently close to the spin axis or spin sufficiently large) where the maximum redshift factor as a function of the source radius, $g_{\rm s+}(r_{\rm s})$, is non-monotonic, one gets that $r_\mathrm{MOB}=\rffp\in [r_{\rm in},\infty)$. For this case, we examine the cobtours of constant redshift to determine the behavior of the line profile at and between the MOR $\gmor=g_{\rm in -}$ and MOB $\gmob=\gffp$. Typically each screen point contributes an infinitesimal amount of flux. All points in the FCR are generic points for which the Jacobian in the flux profile $\eqref{eq:LineProfileG}$ is finite for all redshift values with the exception of $\gffp$ for which the Jacobian becomes singular, i.e., the aforementioned FFP. 
  
The redshift contours are divided into two regions. For $g\in\pa{g_{\rm in -},g_{\rm in +}}$, we have topology $\CutD$ curves. The line profile flux for this region is finite and approaches zero as $g\to \gmor$, where the contours shrink to a point. We will term this region the ``inner radius wing'', as it represents the portion of the line profile affected by the inner edge of the disk. At $g_{\rm in +}$, the redshift contour closes, thus representing a change in the topology of the redshift contours, resulting in a transition in the line profile behavior. 
For $g\in (g_{\rm in +},\gffp)$, the redshift contours are closed, i.e., topology $\bigcirc$. In this region the line profile flux is finite. As $g\to \gffp$, the contours shrink to a point (the FFP) where the Jacobian \eqref{eq:LineProfileG} diverges, conspiring to produce finite flux at $\gffp$.

At fixed spin, the FFP sourcing radius $\rffp$ decreases monotonically with increasing $x_{\rm o}$, while the FFP value $\gffp$ increases monotonically with increasing $x_{\rm o}$. Described heuristically, the velocity of the orbiters is monotonically decreasing with the source radius as the orbiters approaching the BH must move faster to maintain the fixed radius; as the observer inclines away from the spin axis, we get a net blueshift when the motion of the orbiters near the BH overcomes the gravitational redshift. 

As $\rffp\to r_{\rm in}$ the topology $\bigcirc$ region shrinks~\footnote{For $r_{\rm in}$ near $\rffp$, the topology $\bigcirc$ region may be hard to numerically resolve; nonetheless, the MOB flux is finite as the FFP is still inside the FCR, though it is near the inner edge.}. A change in the redshift contour configuration occurs when $\rffp=r_{\rm in}$. These transitions happen at inclinations $\tilde{x}(a,r_{\rm in})$, defined as the inverse of $r_{\rm in}=\rffp(a,x_{\rm o})$. For sufficiently small inner disk radius and sufficiently large spin values, there does not exist a solution $\tilde{x}(a,r_{\rm in})$; that is, for sufficiently large spin values with disks that terminate sufficiently close to the BH, $r_\mathrm{MOB}=\rffp> r_{\rm in}$ for all inclinations and the disk will remain Type II for all viewing inclinations. 

\underline{\emph{Type III:}}
When the inclination is sufficiently far from the spin axis, $x_{\rm o}\in(\tilde{x}(a, r_{\rm in}),1$], and the spin is sufficiently small, $a<\tilde a (r_{\rm in})$ where $\tilde a (r_{\rm in})$ is the spin value such that $\tilde{x}(a, r_{\rm in})=1$, the maximum redshift factor as a function of radius $g_{\rm s+}(r_{\rm s})$ is monotonically increasing for $r_{\rm s}\in[r_{\rm in},r_{\rm out}]$; thus, the MOB is sourced by the inner edge of the disk $g_\mathrm{mob}=g_{\rm in +}$. All the redshift contours in $g\in \pa{g_{\rm in -},g_{\rm in +}}$ are of topology $\CutD$. Additionally, as there is no FFP in the FCR, the Jacobian of Eq.~\eqref{eq:LineProfileG} is finite everywhere in the FCR. Therefore, the line profile has vanishing flux at the MOB and MOR and finite flux everywhere in between. 

Each of the redshift contour configurations we have outlined has a distinct line profile morphology. In Fig.~\ref{fig:InfiniteDisks} we show the relationship between the redshift factor configuration and the line profile morphology. The two top rows correspond to schematics of the cases, while the last two rows show a particular example with $a=0.7M$, viewed at different angles, from left to right, $\theta_{\rm o}=0^\circ,\ 30^\circ,\ 50^\circ,\ 60^\circ$ and $\ 75^\circ$. We plot the disk images on a compactified observer's screen of radial coordinate $\bar\rho\in\br{0,1}$~\footnote{Under this compactification, the horizon radius contour may be concave, while in ordinary screen coordinates it is always convex.}, where
\begin{align}
    \label{eq:CompactBardeenRadius}
    \bar\rho=\pa{\frac{|{\bf x}|}{{|{\bf x}|+1}}}^4,
\end{align}
and use an emissivity of the form $I_{\rm s}(r_{\rm s})=f_s r_{\rm s}^{-3}$ for the line profiles.
Having examined the redshift contour patterns for an infinitely large disk, which highlights how the inner edge of the disk influences the line profile, we will now explore the impact of the disk's outer edge on the redshift contour configurations.

\subsection{A Disk with a Finite Extension}
\label{sec:FiniteDisk}

A finite outer disk radius, which is the relevant astrophysics case, can alter the topology of redshift contours displayed on-screen. The finite outer disk radius naturally reduces the FCR to the region between the contours of the inner and outer disk radii, $r_{\rm in}$ and $r_{\rm out}$, respectively. Any redshift contour that crosses the outer disk radius contour now has a portion of it amputated as compared to the infinite disk case. Topology $\bigcirc$ contours which intersect $r_{\rm out}$ are no longer closed and become topology $\CutU$, and topology $\CutD$ contours which intersect $r_{\rm out}$ get split into two disjointed line segments and become topology $\vertvert$. 
As the $g=1$ contour closes at infinity in the infinitely large disk case, the $g=1$ contour must be of topology $\CutU$ or $\vertvert$ in the case of a finite disk. Altogether, this introduces transitions to the line profile's behavior as the topology of the redshift contours changes. The redshift factor values at which the topologies change are the minimum and maximum redshift factors for the disk outer radius, $g_{\rm out -}$ and $g_{\rm out +}$, respectively. Finally as introducing a finite outer disk radius shrinks the FCR, amputating or removing redshift contours which contributed to the line profile of the infinitely large disk, the finite disk line profiles can be thought of as  arising by removing flux from the infinite disk line profiles.

\begin{figure*}   
    {\centering Type II FFP-inclusive Disks}\\
    {\resizebox{.7\linewidth}{!}{
    \begin{tabular}{|c|c|c|c|c|}
    \hline
    &&&&\\
    \Huge\underline{$\CutD~\bigcirc~\CutU~\bigcirc$ (II~${}^\bullet$inc~A)}&
    \Huge\underline{$\CutD~\CutU~\bigcirc$ (II~${}^\bullet$inc~B)}&
    \Huge\underline{$\CutD~\vertvert~\CutU~\bigcirc$ (II~${}^\bullet$inc~C)}&
    \Huge\underline{$\CutD~\vertvert~\bigcirc$ (II~${}^\bullet$inc~D)}&
    \Huge\underline{$\CutD~\vertvert~\CutD~\bigcirc$ (II~${}^\bullet$inc~E)}\\[2.5ex]
    \VF{\includegraphics[width=\linewidth]{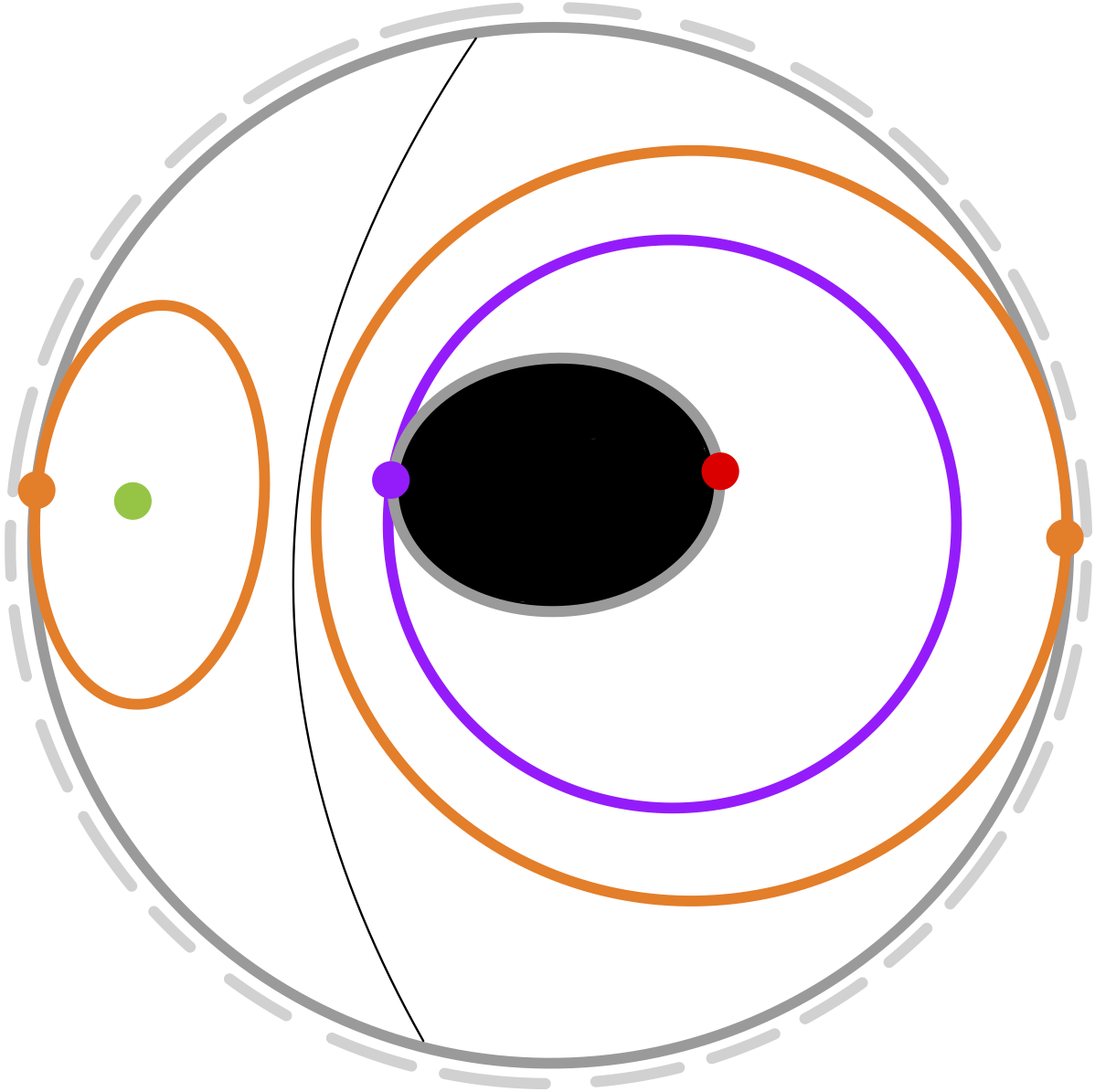}}&
    \VF{\includegraphics[width=\linewidth]{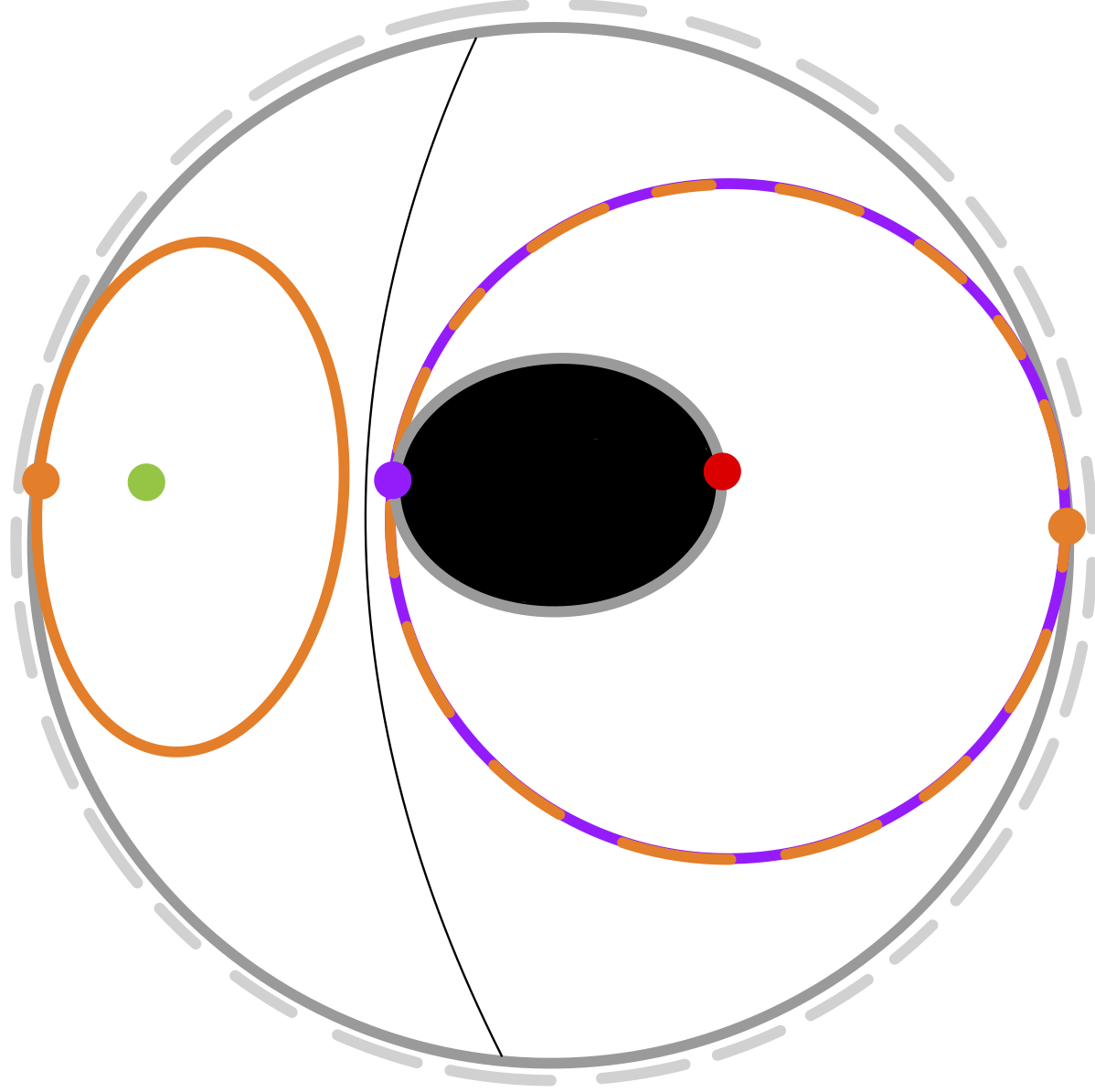}}&
    \VF{\includegraphics[width=\linewidth]{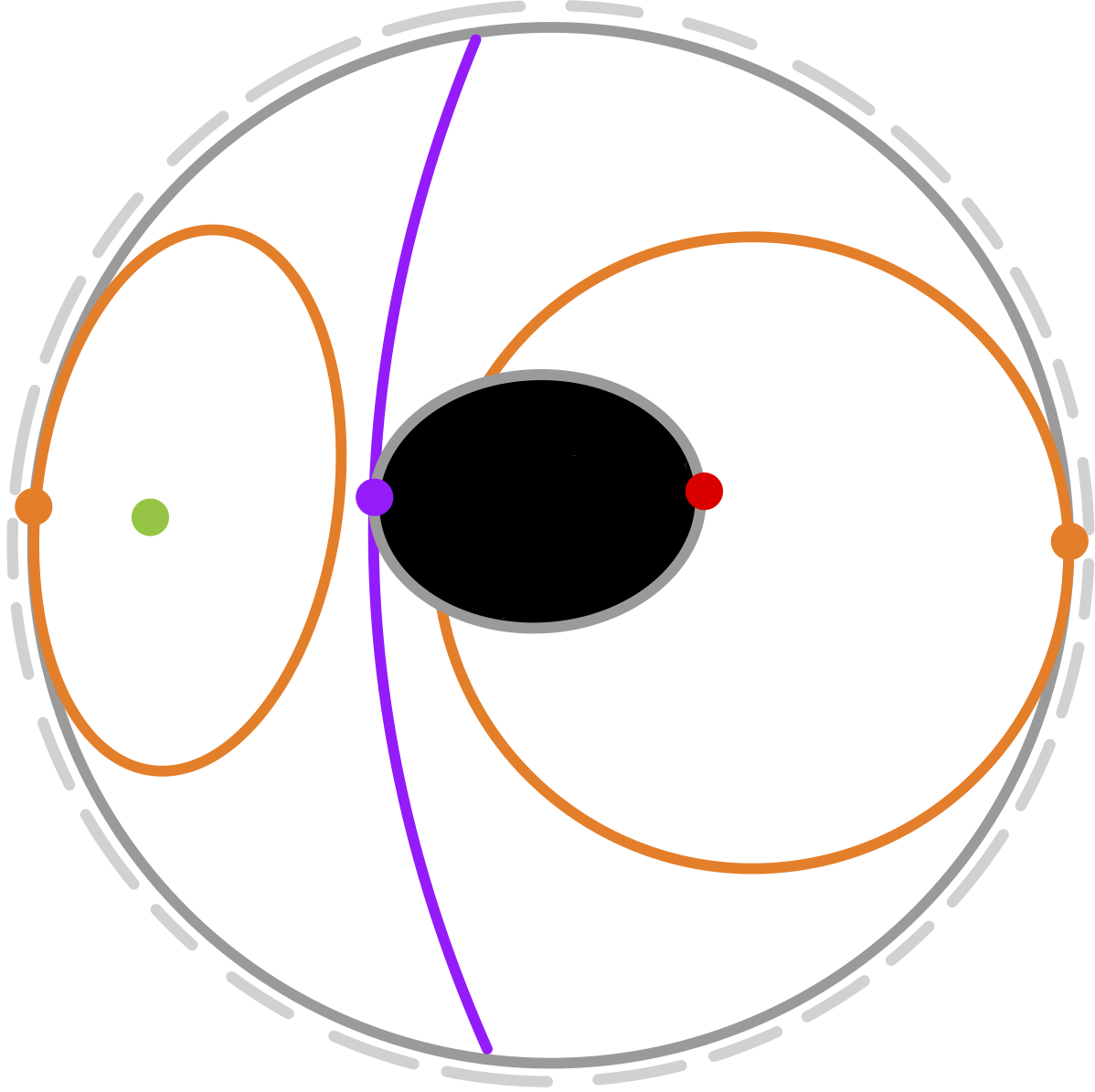}}&
    \VF{\includegraphics[width=\linewidth]{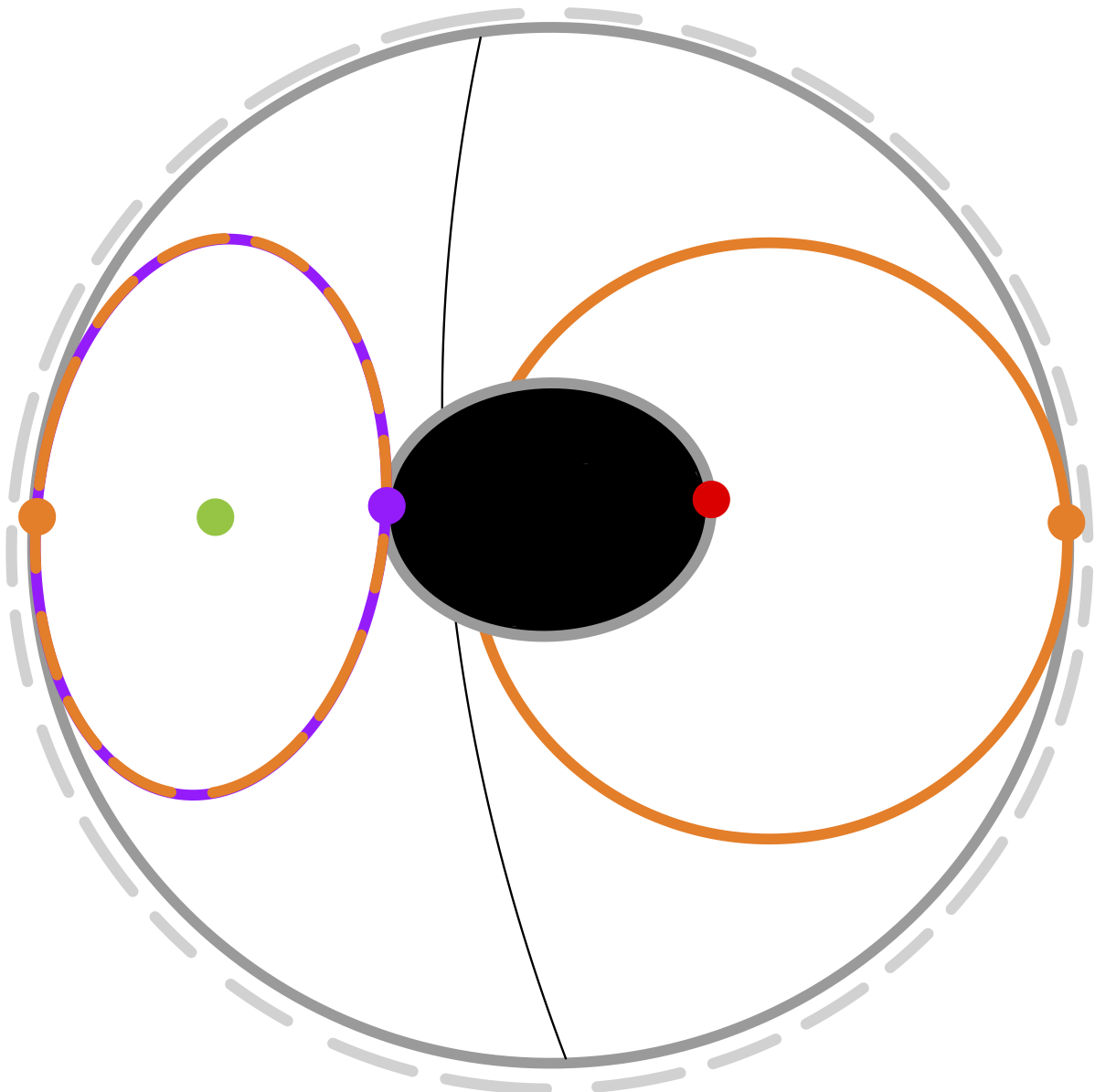}}&
    \VF{\includegraphics[width=\linewidth]{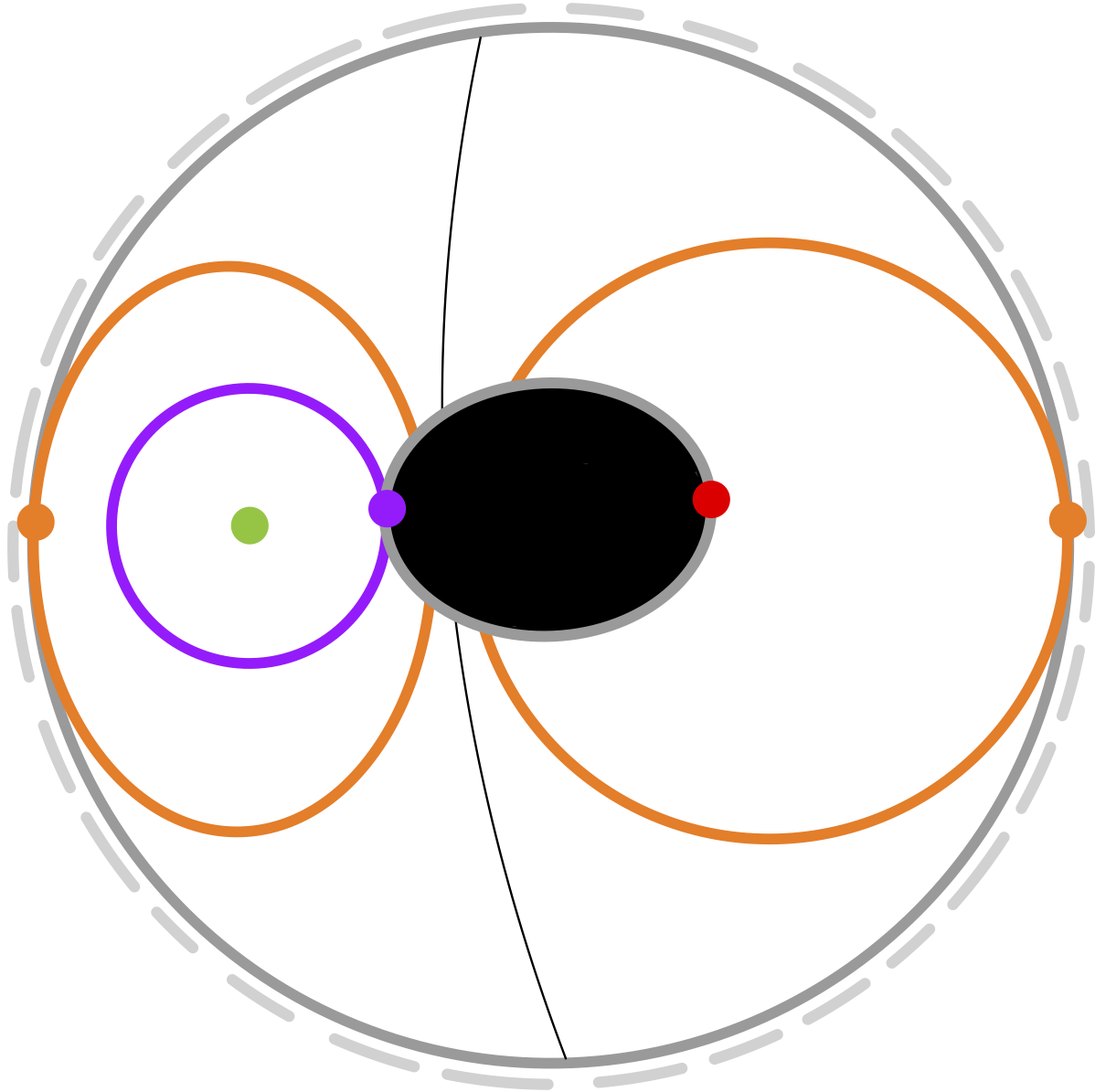}}\\[5ex]
    \includegraphics[width=\linewidth]{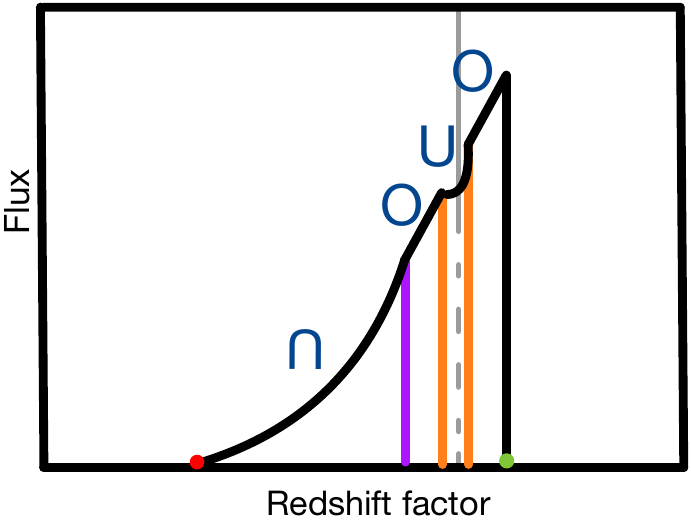}&
    \includegraphics[width=\linewidth]{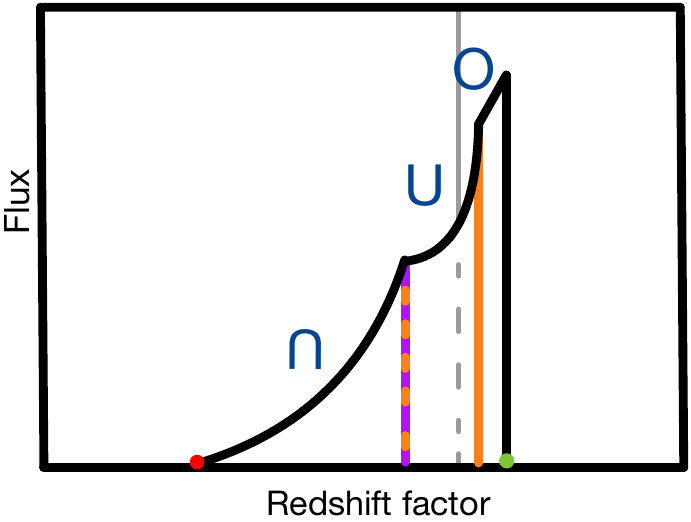}&
    \includegraphics[width=\linewidth]{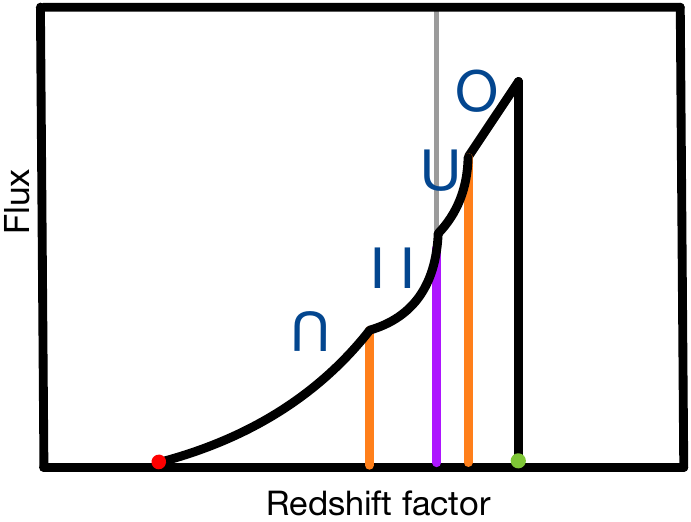}&
    \includegraphics[width=\linewidth]{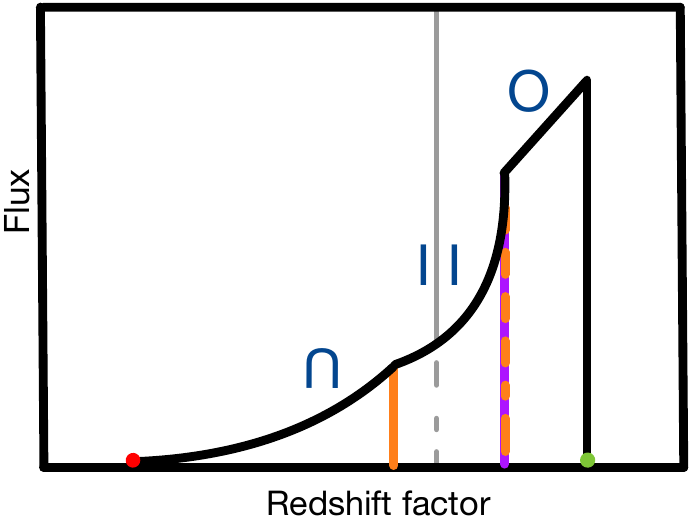}&
    \includegraphics[width=\linewidth]{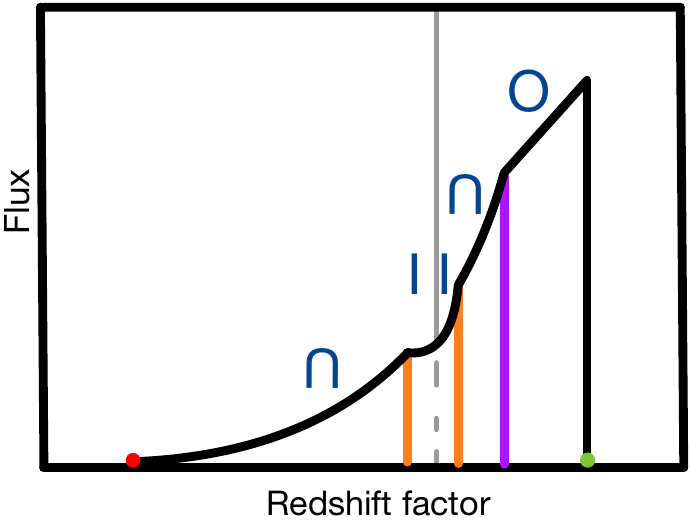}\\
    &&&&\\[5ex]
    \hline
    &&&&\\
    \Huge\underline{$a/M=0.7, \ \theta_{\rm o}=40^\circ, \ r_{\rm out}/M=200$}&
    \Huge\underline{$a/M=0.7, \ \theta_{\rm o}=45^\circ,\ r_{\rm out}/M=150$}&
    \Huge\underline{$a/M=0.7, \ \theta_{\rm o}=50^\circ,\ r_{\rm out}/M=200$}&
    \Huge\underline{$a/M=0.7, \ \theta_{\rm o}=55^\circ,\ r_{\rm out}/M=100$}&
    \Huge\underline{$a/M=0.7, \ \theta_{\rm o}=60^\circ,\ r_{\rm out}/M=200$}\\[2.5ex]
    \includegraphics[width=\linewidth]{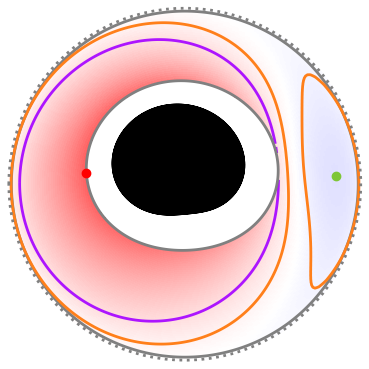}&
    \includegraphics[width=\linewidth]{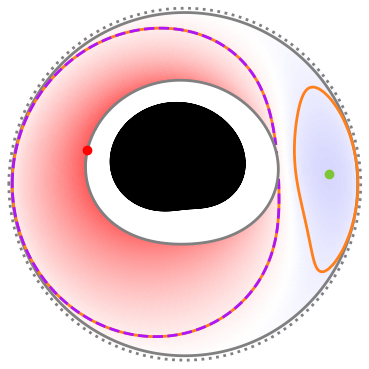}&
    \includegraphics[width=\linewidth]{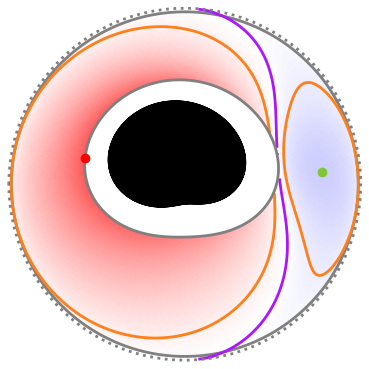}&
    \includegraphics[width=\linewidth]{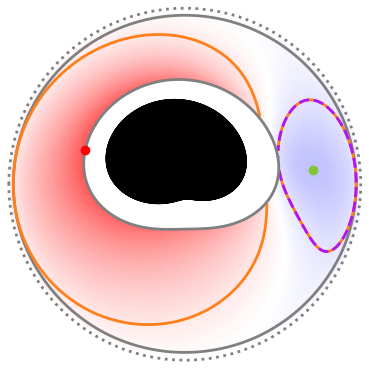}&
    \includegraphics[width=\linewidth]{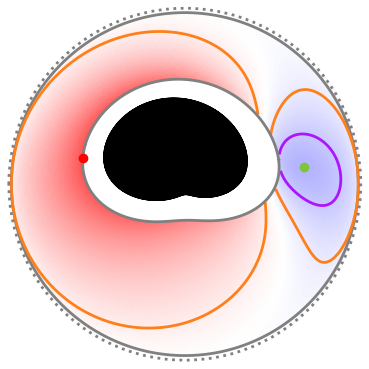}\\
    \includegraphics[width=\linewidth]{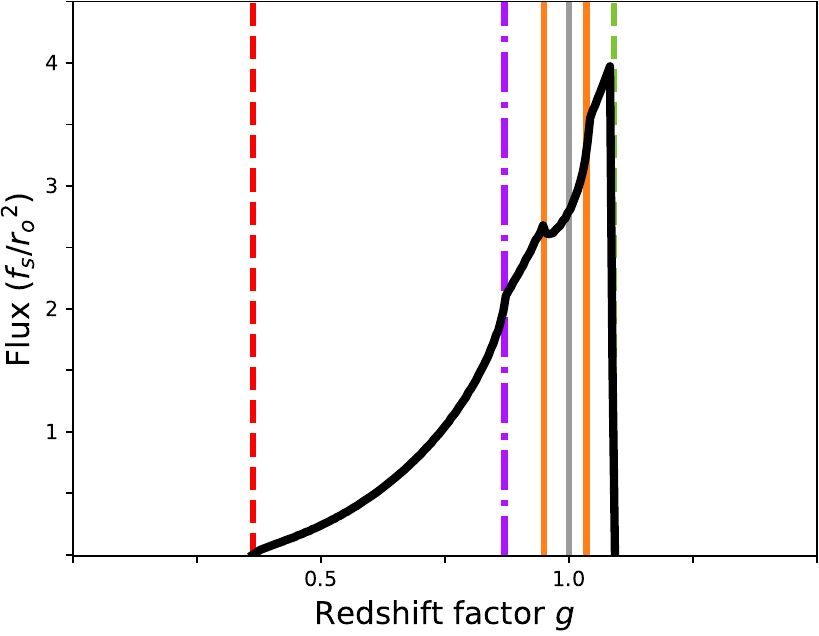}&
    \includegraphics[width=\linewidth]{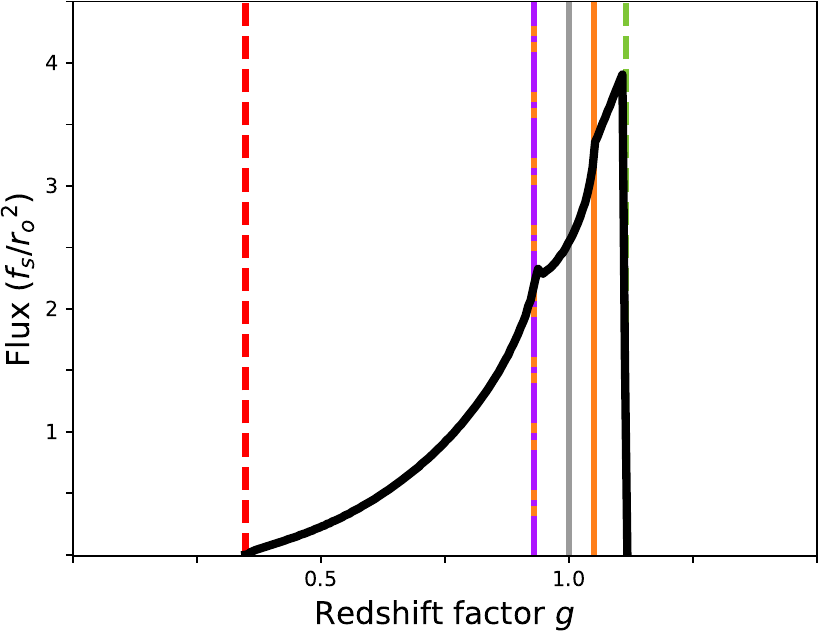}&
    \includegraphics[width=\linewidth]{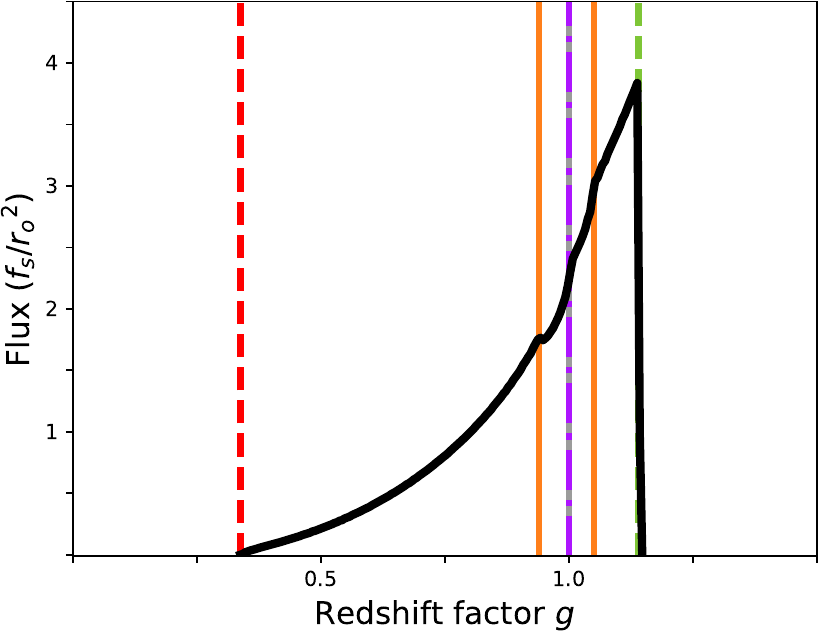}&
    \includegraphics[width=\linewidth]{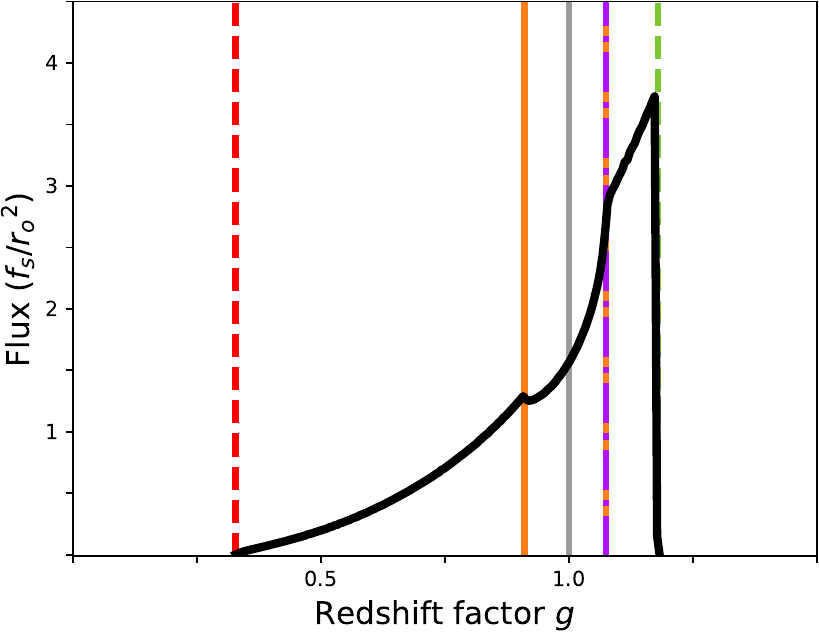}&
    \includegraphics[width=\linewidth]{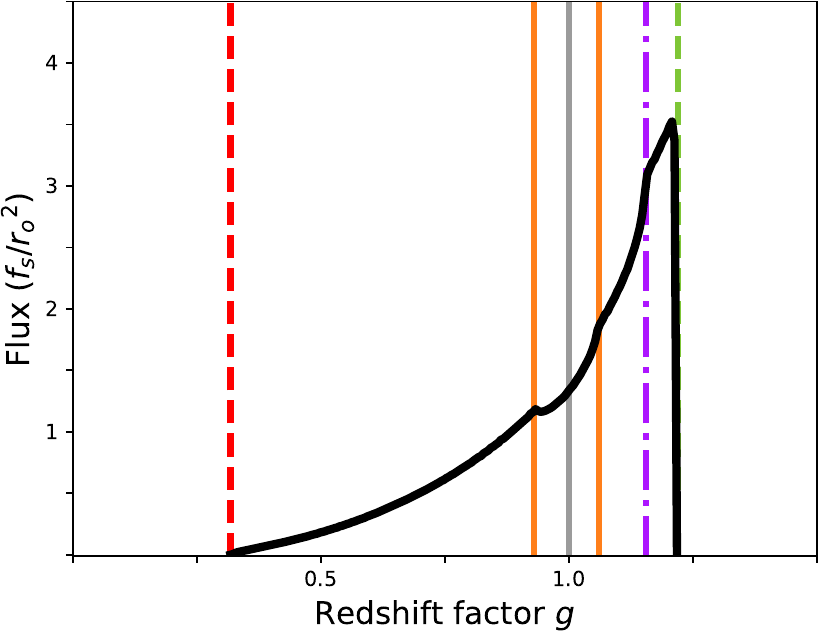}\\[1ex]
    \hline
    \end{tabular}}\\[3ex]}
    \begin{tabular}{ccccc}
    {Type I Disk}&  &{Type II FFP-exclusive Disks}& &{Type III Disk}\\
    {\resizebox{.14\linewidth}{!}{
    \begin{tabular}{|c|}
    \hline\\ 
    {\Huge\underline{$\bigcirc$}}\\[5ex]
    \includegraphics[width=\linewidth]{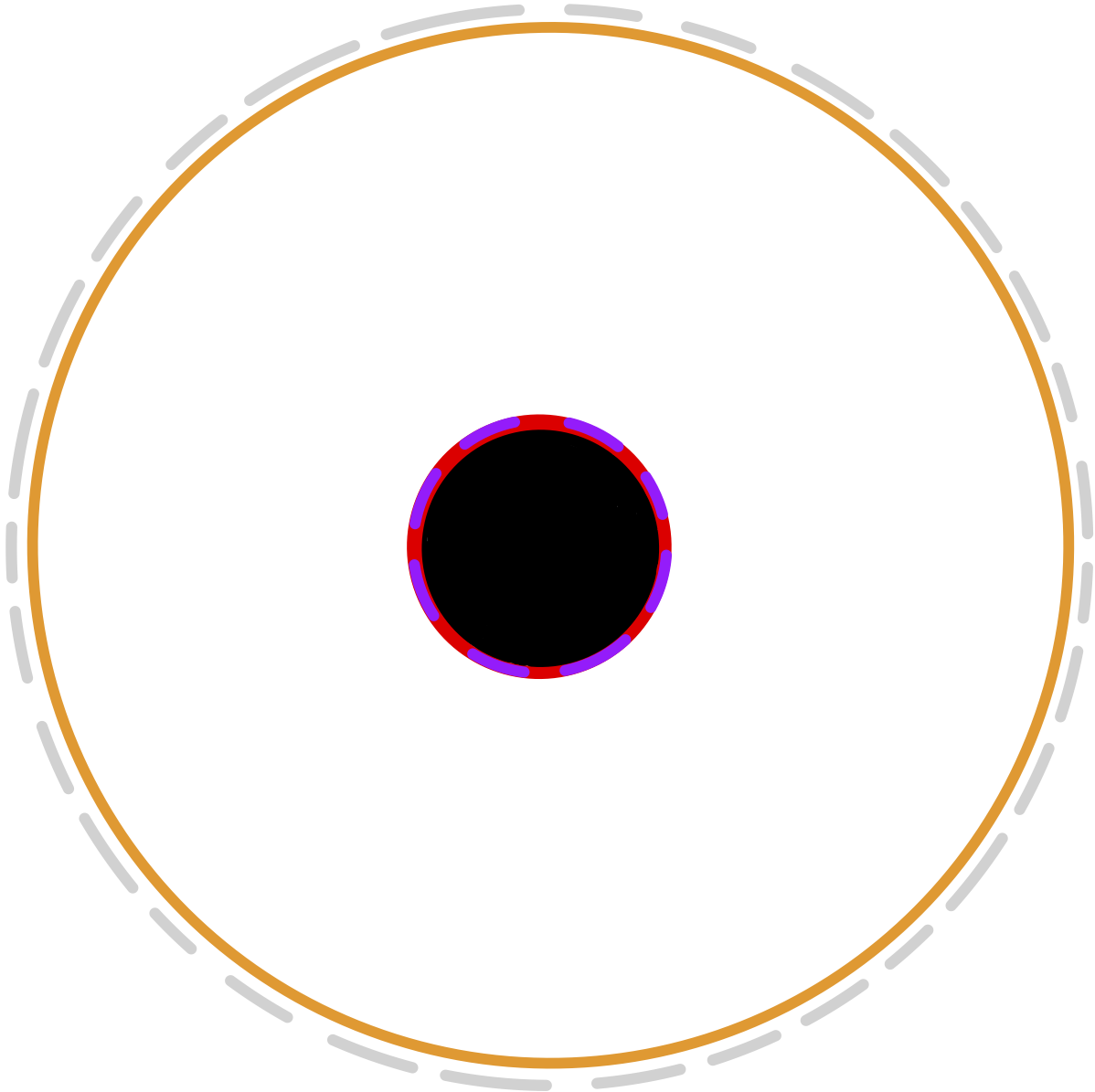}\\[5ex]
    \includegraphics[width=\linewidth]{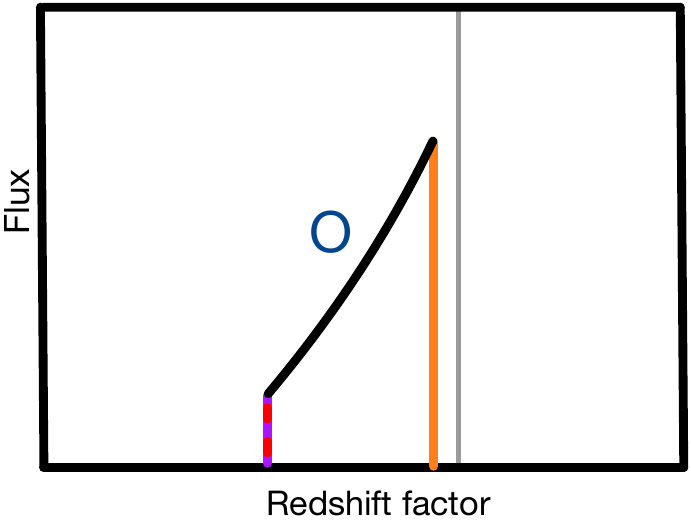}\\[5ex]
    \hline\\
    \Huge\underline{$a/M=0.7, \ \theta_{\rm o}=0^\circ,\ r_{\rm out}/M=50$}\\[2.5ex]
    \includegraphics[width=\linewidth]{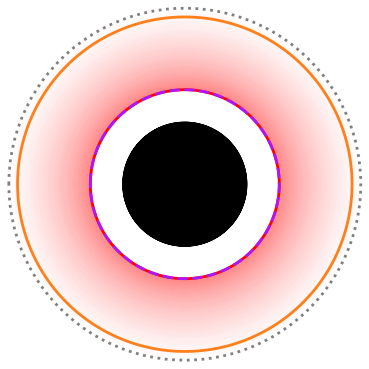}\\
    \includegraphics[width=\linewidth]{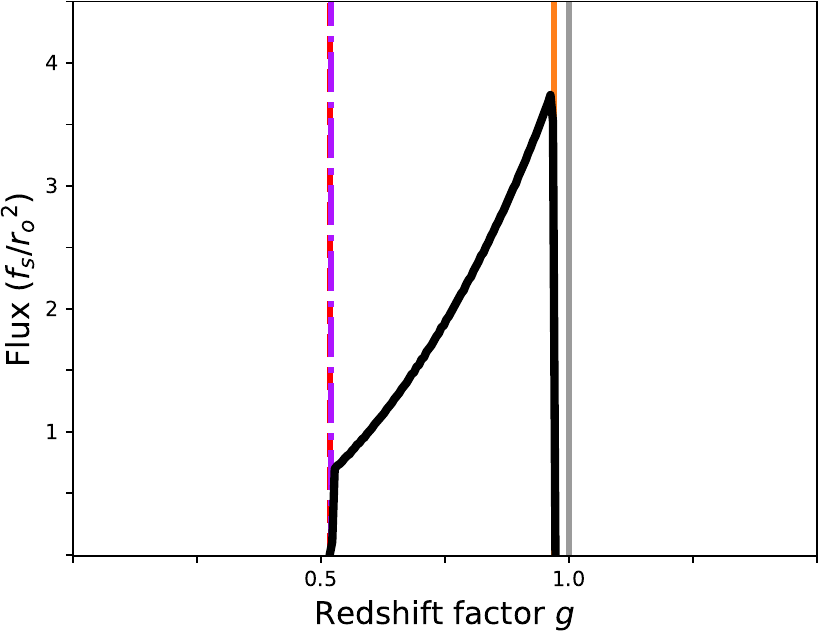}\\[5ex]
    \hline
    \end{tabular}}}
    &\quad\quad\quad\quad &
    {\resizebox{.42\linewidth}{!}{
    \begin{tabular}{|c|c|c|}
    \hline
    &&\\
    \Huge\underline{$\CutD~\bigcirc~\CutU$ (FFP~${}^\bullet$exc~A)}&
    \Huge\underline{$\CutD~\CutU$ (FFP~${}^\bullet$exc~B)}&
    \Huge\underline{$\CutD~\vertvert~\CutU$ (FFP~${}^\bullet$exc~C)}\\[2.5ex]
    \VF{\includegraphics[width=\linewidth]{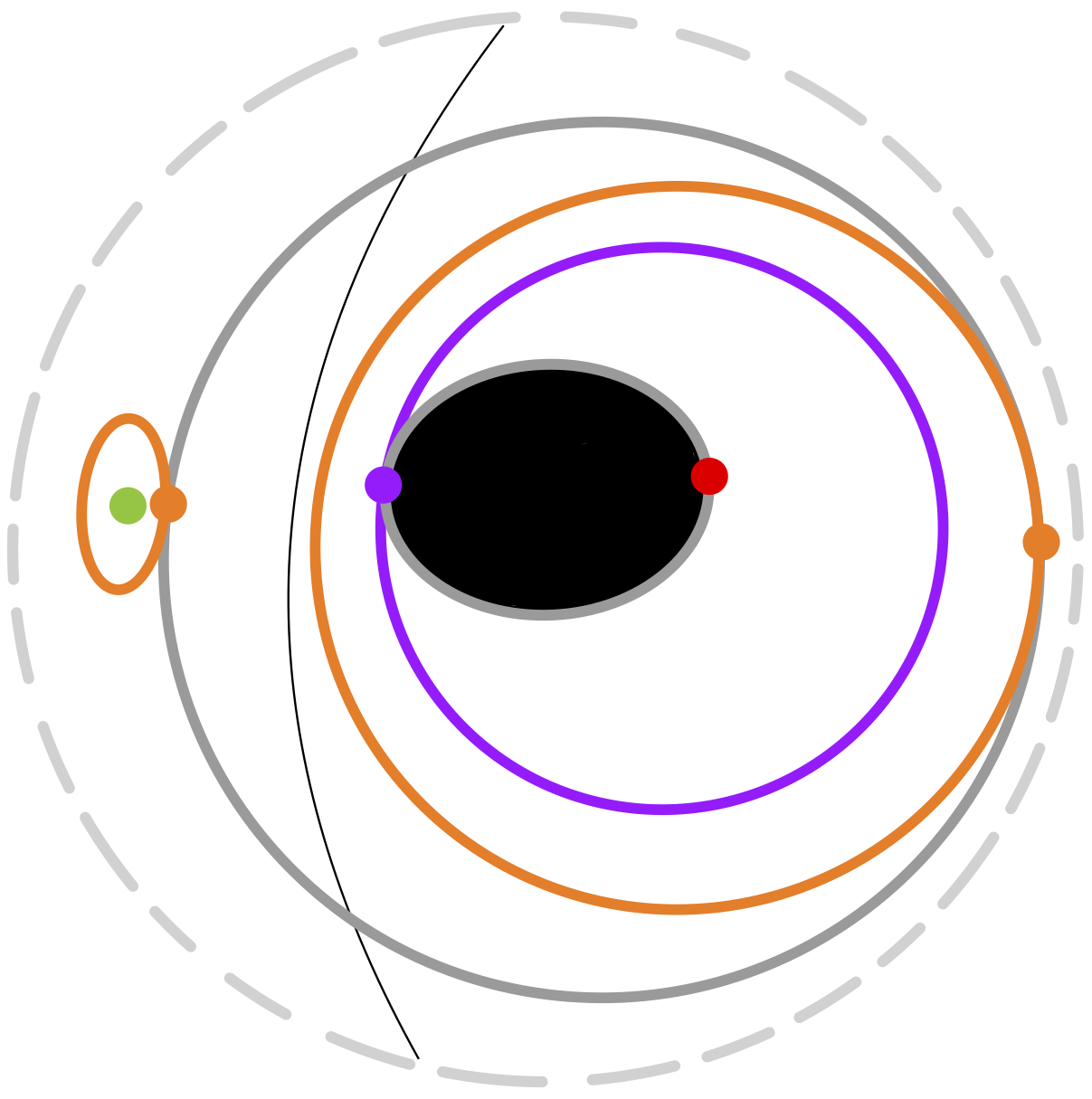}}&
    \VF{\includegraphics[width=\linewidth]{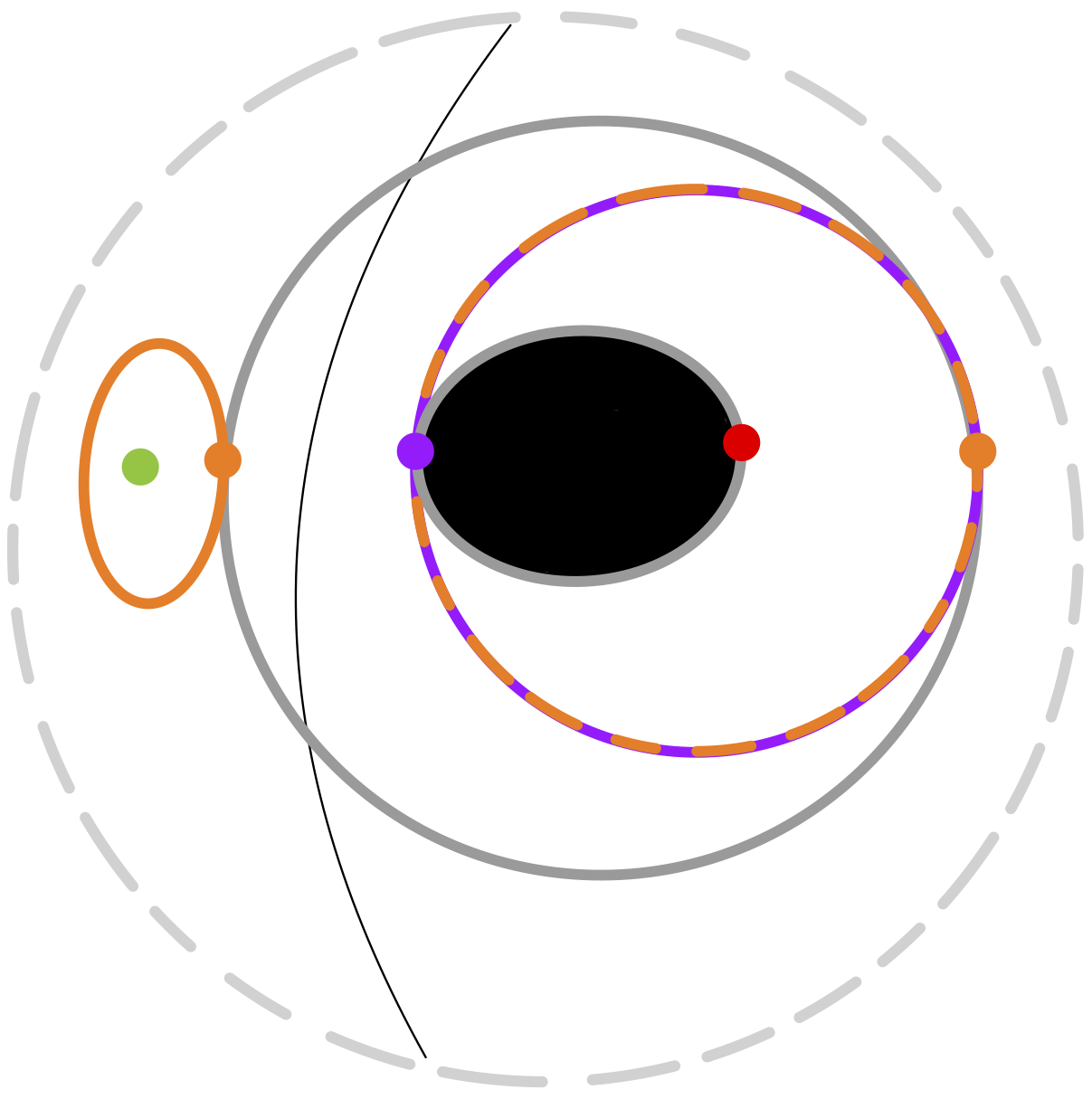}}&
    \VF{\includegraphics[width=\linewidth]{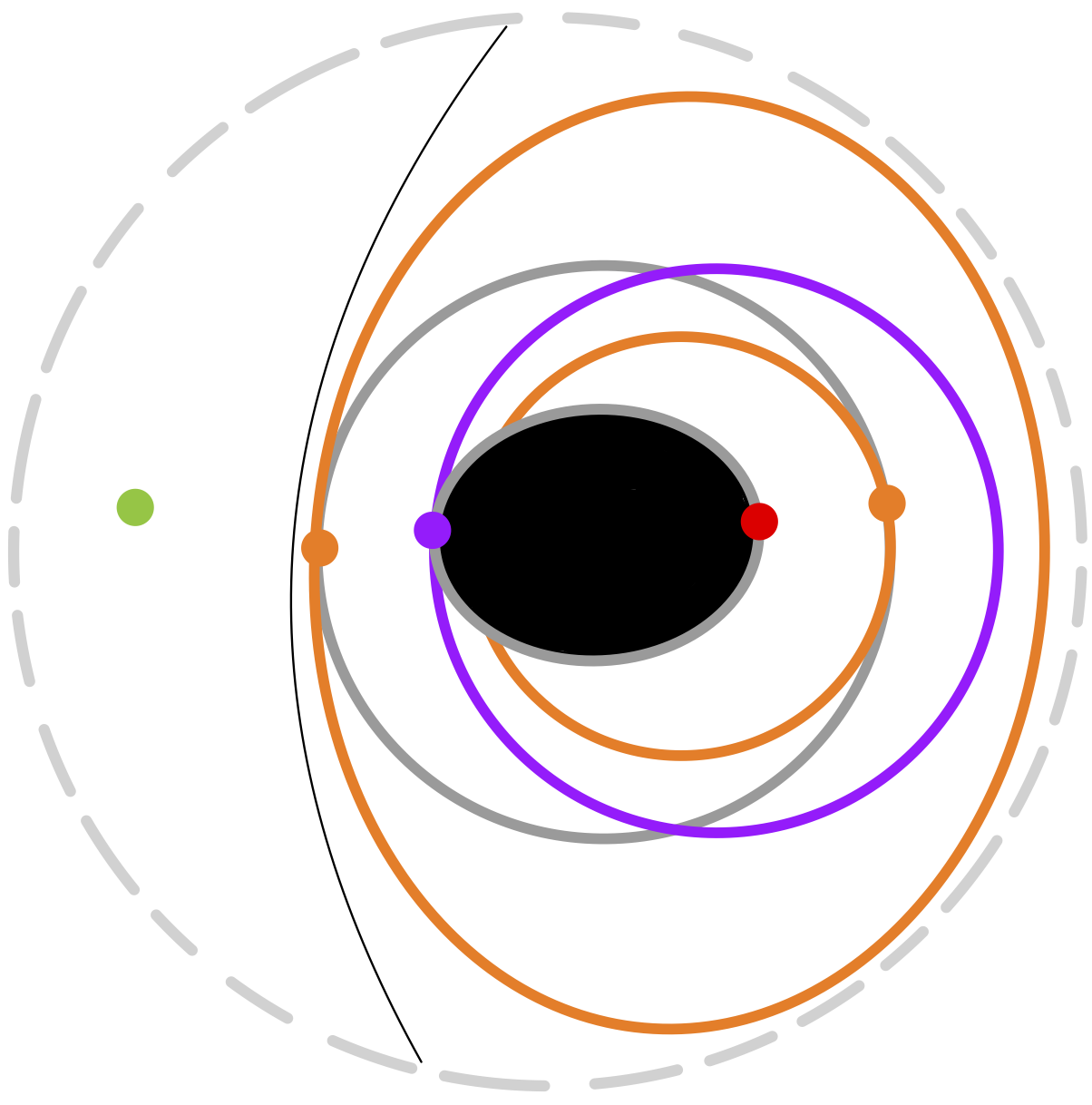}}\\[5ex]
    \includegraphics[width=\linewidth]{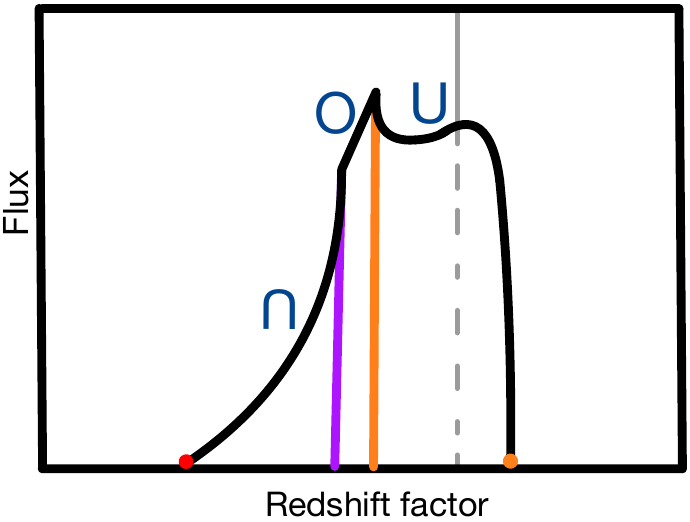}&
    \includegraphics[width=\linewidth]{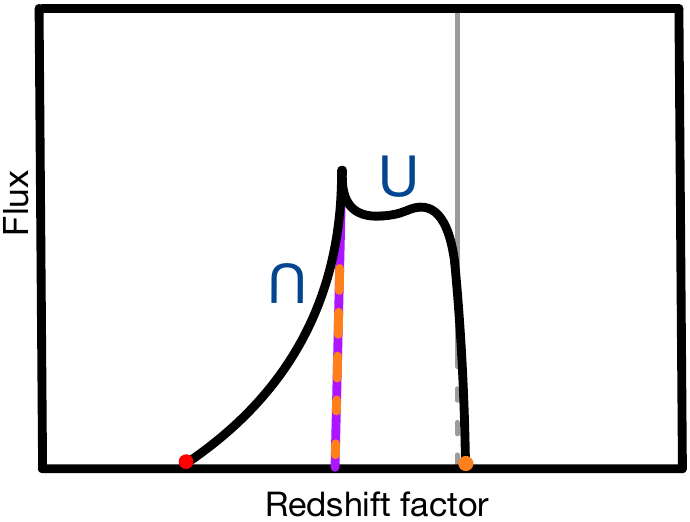}&
    \includegraphics[width=\linewidth]{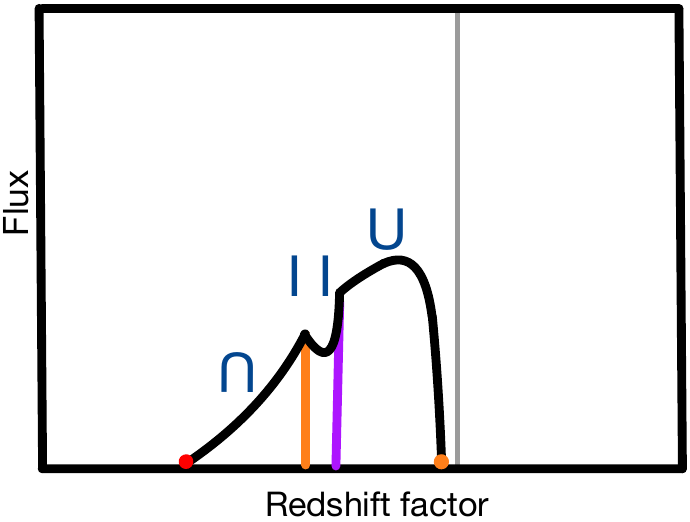}\\
    &&\\[5ex]
    \hline
    &&\\
    \Huge\underline{$a/M=0.7, \ \theta_{\rm o}=30^\circ, \ r_{\rm out}/M=20$}&
    \Huge\underline{$a/M=0.7, \ \theta_{\rm o}=30^\circ,\ r_{\rm out}/M=12.5$}&
    \Huge\underline{$a/M=0.7, \ \theta_{\rm o}=30^\circ,\ r_{\rm out}/M=7$}\\[2.5ex]
    \includegraphics[width=\linewidth]{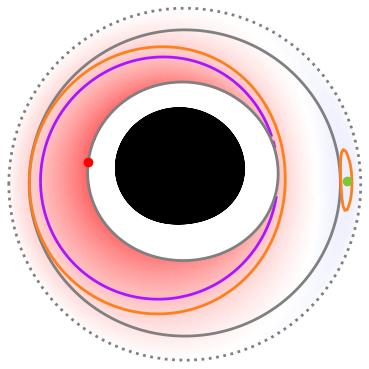}&
    \includegraphics[width=\linewidth]{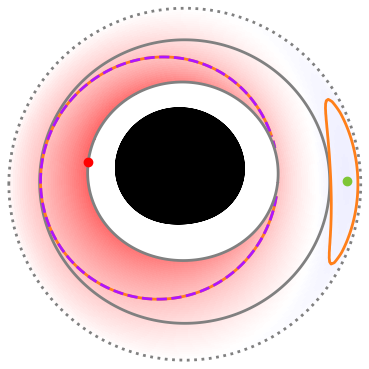}&
    \includegraphics[width=\linewidth]{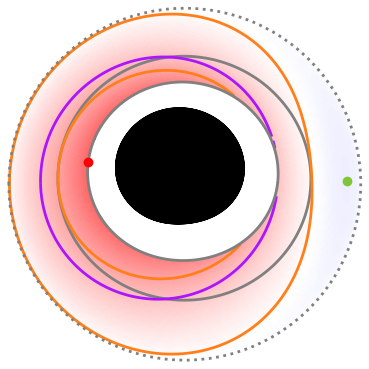}\\
    \includegraphics[width=\linewidth]{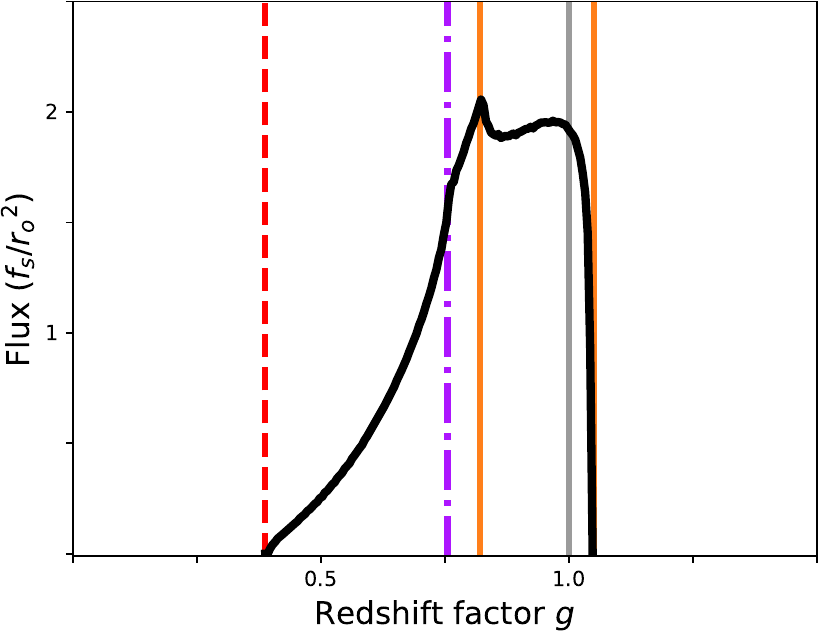}&
    \includegraphics[width=\linewidth]{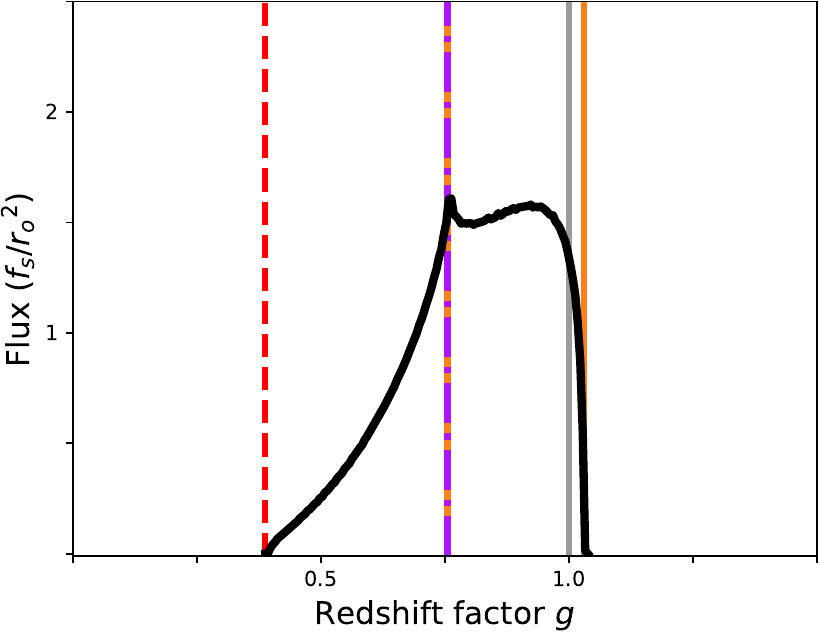}&
    \includegraphics[width=\linewidth]{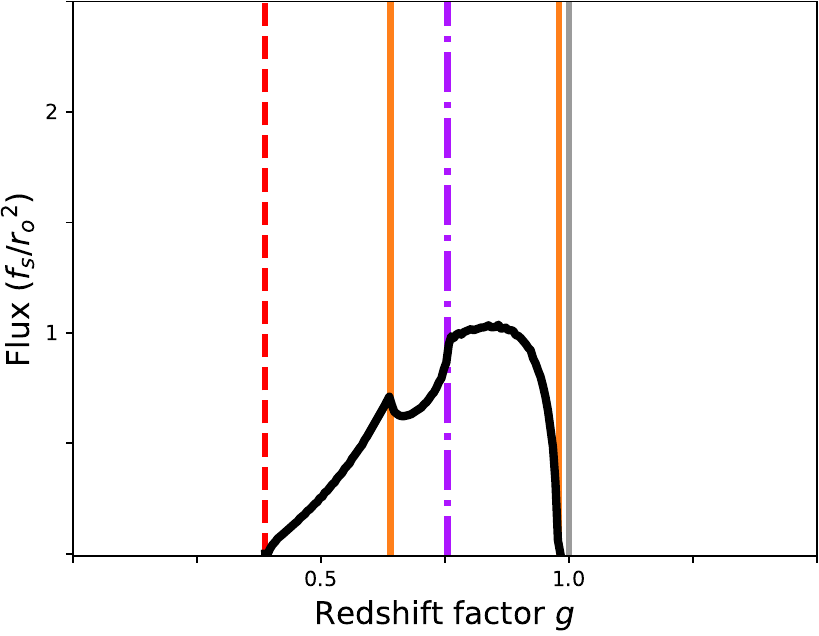}\\[1ex]
    \hline
    \end{tabular}}} 
    &\quad\quad\quad\quad &
    {\resizebox{.14\linewidth}{!}{
    \begin{tabular}{|c|}
    \hline\\ 
    {\Huge\underline{$\CutD~\vertvert~\CutD$}}\\[5ex]
    \VF{\includegraphics[width=\linewidth]{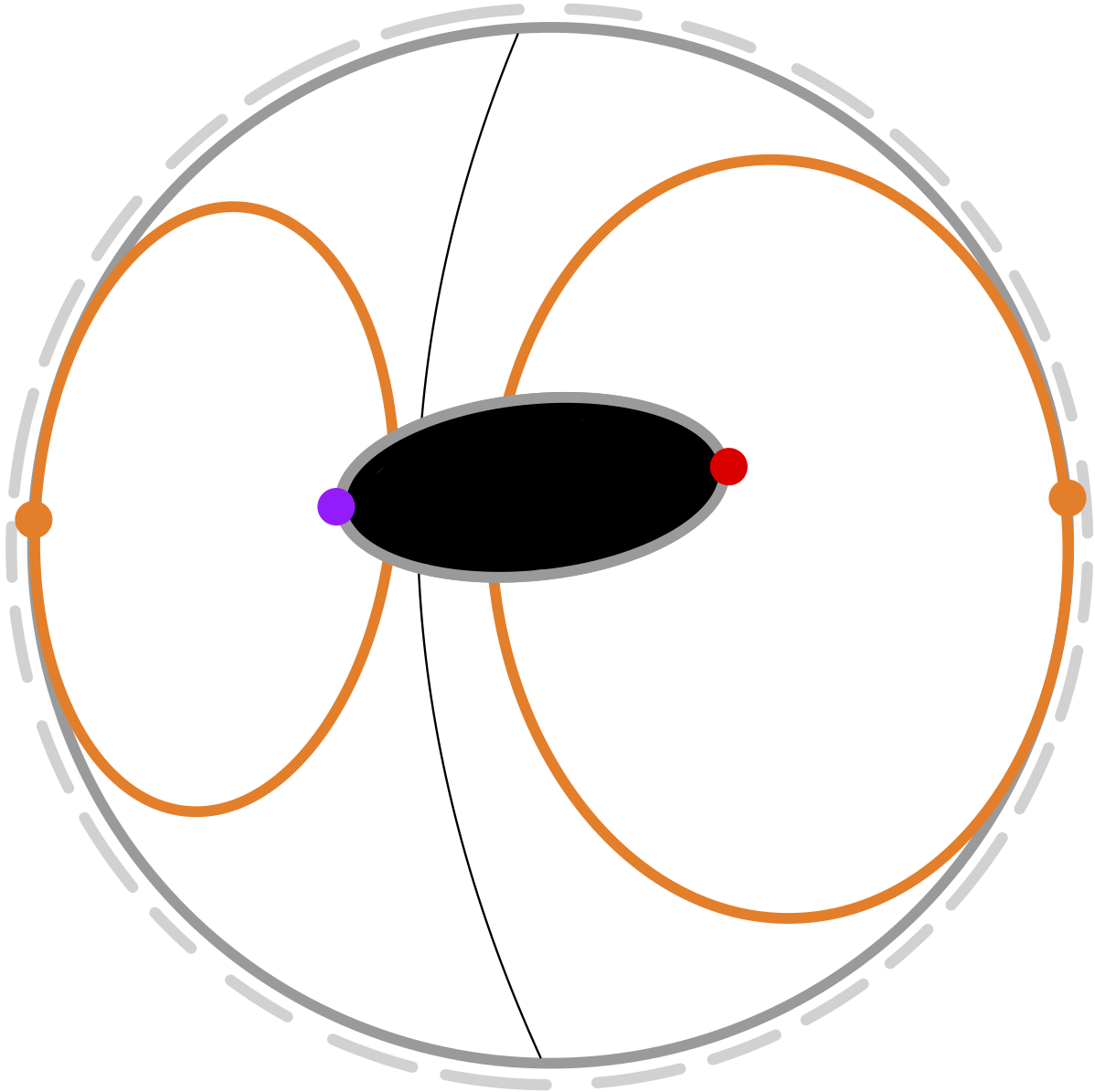}}\\[5ex]
    \includegraphics[width=\linewidth]{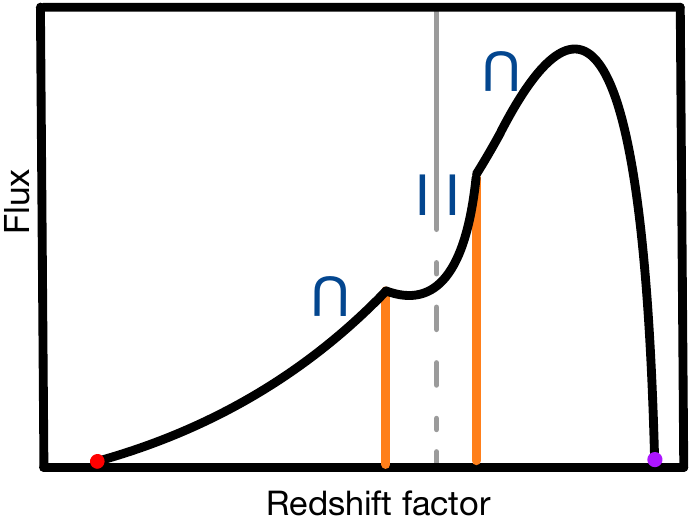}\\[5ex]
    \hline\\
    \Huge\underline{$a/M=0.7, \ \theta_{\rm o}=75^\circ, r_{\rm out}/M=50$}\\[2.5ex]
    {\includegraphics[width=\linewidth]{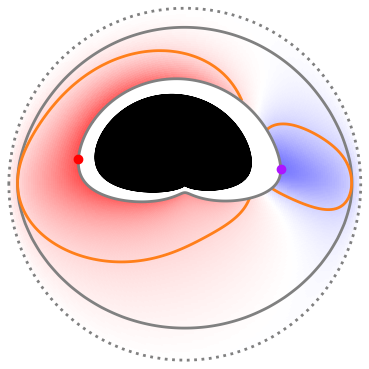}}\\
    \includegraphics[width=\linewidth]{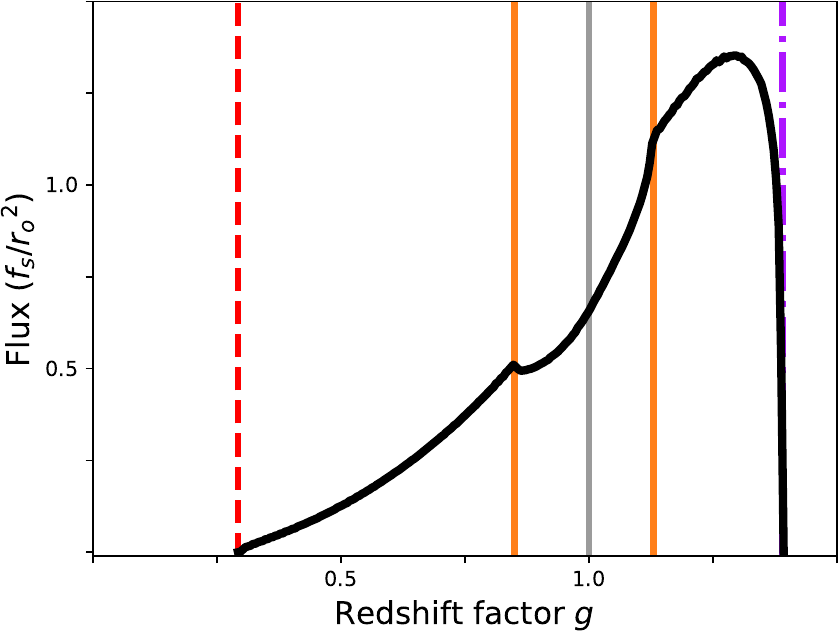}\\[5ex]
    \hline
    \end{tabular}}}
    \end{tabular}
    \caption{Line profile morphologies associated with the redshift factor configurations for finite disks. As in Fig.~\ref{fig:InfiniteDisks}, each column features four images: a schematic of the disk on the observer's screen in compactified coordinates (first row), a schematic of the associated line profile (second row), an example of a finite disk on the observer's screen in compactified coordinates (Eq.~\ref{eq:CompactBardeenRadius}) colored by redshift factor with inner disk radius $r_{\rm in}=\risco$ (third row), and the associated line profile of the above example disk for emissivity $I_{\rm s}=f_{\rm s} r_{\rm s}^{-3}$ (fourth row). We show the inner and outer disk radii $r_{\rm in,out}$ (solid gray curves) and $r=\infty$ as a dotted gray curve on the observer's screen. We highlight the extremal redshift factors of the inner and outer radii $g_{\rm in -}$ (red), $g_{\rm in +}$ (purple), $g_{{\rm out}\pm}$ (orange), and the FFP redshift factor $\gffp$ (green). To guide the reader, in the line profiles we have highlighted $g=1$ with a vertical gray line.}
    \label{fig:FinteDisks}
\end{figure*}

Let us now examine the changes a finite $r_{\rm out}$ imposes upon the redshift contour configuration and determine the corresponding line profile morphology. We find the Type I and Type III~finite disks, each only having one redshift contour configuration. In contrast, the Type II finite disks have eight redshift contour configurations, which can be further divided into cases when the FFP is included or excluded from the FCR. We summarize the resulting configurations in Tab.~\ref{tab:FiniteDisk}, exemplifying associated line profiles in Fig.~\ref{fig:FinteDisks}, and provide their detailed derivations below.

\begin{table}
{\centering
\begin{tabular}{c| c| c} 
\hline
 Disk Type  & Condition & Configuration \\ 
\hline
\hline
I & $x_{\rm o}=0$, No FFP & $\bigcirc$ \\ 
\hline
 & $\rffp\in\pa{r_{\rm in},r_{\rm out}}$ & $\CutD~\cdots~\bigcirc$  \\[1ex] 
II & or &  or \\
  & $\rffp\geq r_{\rm out}$ &  $\CutD~\cdots~\CutU$ \\ 
\hline
III  & $x_{\rm o}\in(0,1)$, No FFP or $\rffp\leq r_{\rm in}$ & $\CutD~\vertvert~\CutD$  \\
\hline
\end{tabular}}\\[2.5ex]
{\centering FFP-inclusive Configurations}\\ 
{\centering
\begin{tabular}{c| c| c| c} 
\hline
Condition & $g_{\rm in +}<g_{\rm out -}$ & $g_{\rm in +}=g_{\rm out -}$ & $g_{\rm in +}>g_{\rm out -}$ \\
\hline
\hline
$g_{\rm in +}<g_{\rm out +}$  & $\CutD~\bigcirc~\CutU~\bigcirc$ & $\CutD~\CutU~\bigcirc$ & $\CutD~\vertvert~\CutU~\bigcirc$\\
\hline
$g_{\rm in +}=g_{\rm out +}$  & --- & --- & $\CutD~\vertvert~\bigcirc$\\[1ex] 
\hline
$g_{\rm in +}>g_{\rm out +}$ & --- & --- & $\CutD~\vertvert~\CutD~\bigcirc$\\
\hline
\end{tabular}}\\[2.5ex]
{\centering FFP-exclusive Configurations}\\
{\centering
\begin{tabular}{c| c| c| c} 
\hline
Condition & $g_{\rm in +}<g_{\rm out -}$ & $g_{\rm in +}=g_{\rm out -}$ & $g_{\rm in +}>g_{\rm out -}$ \\
\hline
\hline
$g_{\rm in +}<g_{\rm out +}$  & $\CutD~\bigcirc~\CutU$ & $\CutD~\CutU$ & $\CutD~\vertvert~\CutU$ \\ 
\hline
\end{tabular}}
\caption{Disk types and possible configurations of constant redshift factor contours on the observer's screen for disks with a finite extension.}
\label{tab:FiniteDisk}
\end{table}

\underline{\emph{Type I:}}
For on-axis viewing inclination, the infinitely large disk has redshift contour configuration $\bigcirc$. In this case, constant redshift factor are concentric circles which are one-to-one with source radius. Shrinking the FCR removes on-screen redshift contours of values $g>g(r_{\rm out})$, and the remaining redshift contours of $g<g(r_{\rm out})$ are still topology $\bigcirc$, so the redshift contour configuration remains unchanged. Imposing a finite outer radius for the disk simply cuts off the right-hand side of the infinite disk line profile at $g(r_{\rm out})$, giving the finite disk line profile a finite flux MOB $g_\mathrm{mob}=g(r_{\rm out})$. We will shorthand the Type I configuration $\bigcirc$ as ``I.''

\underline{\emph{Type III:}}
When the spin and inclination are such that there is no FFP, $g_{\rm s-}$ and $g_{\rm s+}$ are strictly decreasing and increasing in $r_{\rm s}$, respectively; so, $\gmor=g_{\rm in -}< g_{\rm out -}<1<g_{\rm out +}< g_{\rm in +}=\gmob$. 
The infinite disk has redshift contour configuration $\CutD$, implying the on-screen redshift contours in the interval $g\in(g_{\rm out -}, g_{\rm out +})$ are amputated by $r_{\rm out}$ and are now topology $\vertvert$. Thus, introducing a disk cutoff takes a ``bite'' out of the infinite disk line profile, resulting in configuration $\CutD~\vertvert~\CutD$. As the MOB is not effected by the outer edge of the disk, the line profile maintains a finite flux MOB which is given by $g_\mathrm{MOB}=g_{{\rm {in}} +}$. We will shorthand the Type III configurations $\CutD$ (infinite disk) and $\CutD~\vertvert~\CutD$ (finite disk) as ``III~$\infty$'' and ``III~finite'', respectively.

\underline{\emph{Type II:}}
When the inclination and spin are such that there exists an FFP, $g_{\rm s-}$  is strictly decreasing in $r_{\rm s}$, and consequently $\gmor=g_{\rm in -}\leq g_{\rm out -}<g_{\rm out +}< \gffp$. The location of $g_{\rm{in} -}$ relative to $g_{{\rm out}\pm}$ depends on the location of $r_{\rm out}$. Starting with the on-screen redshift factor configuration of the infinite disk {$\CutD~\bigcirc$} (which we will shorthand ``I~$\infty$''), we can deduce the topologies of the redshift contours near the MOR and MOB. Changes in the redshift contour topologies can only occur at $g_{\rm in +}$ and $g_{{\rm out}\pm}$. Therefore, all the redshift contours are topology $\CutD$ for $g\in(g_{\rm in +},\min\cu{g_{\rm in +},g_{\rm out -}})$.

As the maximum redshift factor of emission from a given disk radius $g_{\rm s+}(r_{\rm s})$ is increasing on the interval $r_{\rm s}\in[r_{\rm in},\rffp]$, and decreasing on $r_{\rm s}\in[\rffp,\infty)$, the first delineation in configurations we make depends on whether $r_{\rm out} \gtrless \rffp$. When $r_{\rm out}> \rffp$, the FFP is still included in the FCR, $\gmob=\gffp$, and the redshift contours are topology $\bigcirc$ for $g\in(\max\cu{g_{\rm in +},g_{\rm out +}},\gffp)$; when $r_{\rm out}\leq \rffp$, the FFP is not included in the FCR, $\gmob=g_{\rm out +}$, and the redshift contours are topology $\CutU$ for $g\in(\max\cu{g_{\rm in +},g_{\rm out -}},g_{\rm out +})$. Thus, the FFP-inclusive configurations (``II~${}^\bullet$inc'' for short) takes the form $\CutD\cdots\bigcirc$, where the ellipsis represents the various internal topologies that transition at redshift factors $\cu{g_{\rm in +},g_{{\rm out}\pm}}$. Lastly, the FFP-exclusive configurations (``II~${}^\bullet$exc'' for short) take the form $\CutD\cdots\CutU$ with internal topologies that transition at redshift factors $\cu{g_{\rm in +},g_{\rm out -}}$.

{{\emph{Finite flux point-inclusive disks---}}}For $r_{\rm out}>\rffp$, $g_{\rm out -}<1<g_{\rm out +}$ and the redshift contour configuration depends on the value of  $g_{\rm in +}$ relative to $g_{{\rm out}\pm}$. The on-screen contours in the interval $g\in(g_{\rm out -},g_{\rm out +})$ are amputated where they intersect with $r_{\rm out}$. The different configurations are as follows:

\begin{itemize}
    \item $g_{\rm in +}\leq g_{\rm out -}< g_{\rm out +}$: The on-screen redshift contours in the interval $g\in(g_{\rm out -}, g_{\rm out +})$ are now topology $\CutU$. Thus, introducing a disk cutoff takes a ``bite'' out of the topology $\bigcirc$ region of the infinite disk's line profile, resulting in configurations $\CutD~\bigcirc~\CutU~\bigcirc$ (``II~${}^\bullet$inc~A'') when $g_{\rm in +}<g_{\rm out -}$, and $\CutD~\CutU~\bigcirc$ (``II~${}^\bullet$inc~B'') when $g_{\rm in +}=g_{\rm out -}$ .
    
    \item $g_{\rm out -}<g_{\rm in +}< g_{\rm out +}$: The on-screen redshift contours in the intervals $g\in(g_{\rm out -}, g_{\rm in +})$ and $g\in(g_{\rm in +}, g_{\rm out +})$ are now topology $\vertvert$ and $\CutU$, respectively. Thus,  introducing a disk cutoff takes a ``bite'' out of the infinite disk line profile, which straddles the inner radius wing and the topology $\bigcirc$ region, resulting in configuration $\CutD~\vertvert~\CutU~\bigcirc$ (``II~${}^\bullet$inc~C'').
    
    \item $g_{\rm out -}< g_{\rm out +}\leq g_{\rm in +}$:  The on-screen redshift contours in the interval $g\in(g_{\rm out -}, g_{\rm out +})$ are now topology $\vertvert$. Thus, introducing a disk cutoff  takes a ``bite'' out of the inner radius wing of the infinite disk's line profile, resulting in configurations $\CutD~\vertvert~\bigcirc$ (``II~${}^\bullet$inc~D'') when $g_{\rm out +}=g_{\rm in +}$, and $\CutD~\vertvert~\CutD~\bigcirc$ (``II~${}^\bullet$inc~E'') when $g_{\rm out +}<g_{\rm in +}$.
\end{itemize} 

{{\emph{Finite flux point-exclusive disks---}}}For $r_{\rm out}<\rffp$, $g_{\rm in +}< g_{\rm out +}$ and the redshift factor configuration is determined according to $g_{\rm in +}\gtrless g_{\rm out -}$. The on-screen contours in the interval $g\in(g_{\rm out -},g_{\rm out +})$ are amputated where they intersect with $r_{\rm out}$. The different configurations are as follows:
\begin{itemize}
    \item $g_{\rm in +}\leq g_{\rm out -}$: Only the on-screen redshift contours in $g\in\pa{g_{\rm out -},g_{\rm out +}}$ are amputated by $r_{\rm out}$ and are now topology $\CutU$, resulting in configurations $\CutD~\bigcirc~\CutU$ (``II~${}^\bullet$exc~A'') when $g_{\rm in +}<g_{\rm out -}$, and $\CutD~\CutU$ (``II~${}^\bullet$exc~B'') when $g_{\rm in +}=g_{\rm out -}$.
    
    \item $g_{\rm out -}<g_{\rm in +}$: The on-screen redshift contours in $g\in\pa{g_{\rm out -},g_{\rm out +}}$ are amputated by $r_{\rm out}$. Contours in $g\in\pa{g_{\rm out -},g_{\rm in +}}$ and $g\in\pa{g_{\rm in +},g_{\rm out +}}$ are now topology $\vertvert$ and $\CutU$, respectively, resulting in a configuration $\CutD~\vertvert~\CutU$ (``II~${}^\bullet$exc~C'').
\end{itemize}

For all the FFP-inclusive configurations, the associated line profiles extend from $g_{\rm in -}$ to $\gmob=\gffp$ and have a finite MOB flux value. However, the line profiles associated with the FFP-exclusive configurations extend from $g_{\rm in -}$ to $\gmob=g_{\rm out +}$ and have a vanishing MOB flux value~\footnote{
Note that when $r_{\rm out}$ is smaller than but near to $\rffp$, the rightmost topology $\bigcirc$ region may be hard to (numerically or observationally) resolve as $g_{\rm out +}$ is close to $\gffp$; nonetheless, the MOB flux is finite as the FFP is still within the FCR. In such cases, the FFP-inclusive configurations have line profiles that appear similar to FFP-exclusive configurations, but with a finite MOB flux.}.

\subsection{The Redshift Contour Configuration Phase Space}

\begin{figure*}
    {\resizebox{\linewidth}{!}{\includegraphics{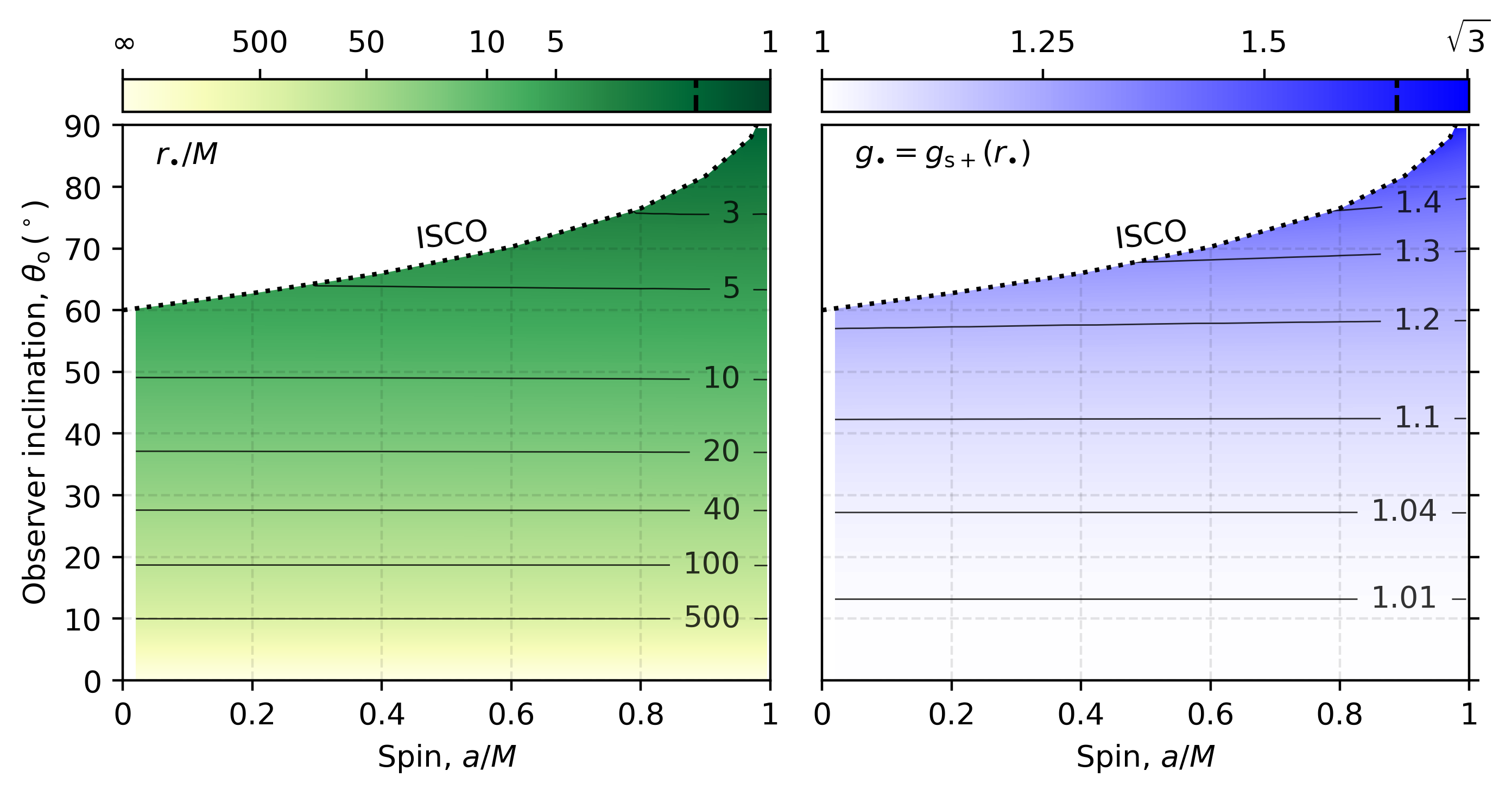}}}\\
    \caption{Finite flux point (FFP) source radius $\rffp$ (left) and redshift factor $\gffp$ (right) as functions of the BH's spin and the observer's inclination. The line of solutions to $\risco(a)=\rffp(a,x_{\rm o})$ (black dotted line) delineates the parameters for which the observer's screen does or does not have an FFP (shaded and white regions, respectively). For spins  $a<\tilde a \simeq 0.9788M$, there is an FFP at all inclinations. At $(a, x_{\rm o})=(\tilde a,1)$, the FFP redshift factor is maximized $\gffp\simeq 1.67$ corresponding to the maximal spin for which the FFP is sourced by the ISCO $\rffp=\risco(\tilde a)\simeq 1.63M$ (dashed lines on  color bars).}
    \label{fig:MOBSourceRad}
\end{figure*}

\begin{figure*}
    {\resizebox{\linewidth}{!}{
    \begin{tabular}{c c}
    \bf A: Configuration Types I-III & \quad\quad\quad \bf \quad B: Cross Section at Fixed $(a,r_{\rm in})$ with $\tilde{x}(a,r_{\rm in})<1$ \quad\quad\quad \\
    {\resizebox{.4\linewidth}{!}{
    \begin{tabular}{c}
         \resizebox{\columnwidth}{!}{\begin{tabular}{|c| c| c| c|} 
         \hline
         $a$ range & $r_{\rm in}$ range & $x_{\rm o}$ range & Type \\
         \hline
         \hline 
         \cellcolor{Red} $[0,1]$ & \cellcolor{Red} $[\risco,\infty]$ & \cellcolor{Red} $0$ & \cellcolor{Red} I\\
         \hline
         \cellcolor{Green} $[0,\tilde a]$ & \cellcolor{Green} $[\risco,\infty]$ & \cellcolor{Green} $(0,\tilde{x}(a,r_{\rm in}))$& \cellcolor{Green} II \\
         \hline
         \cellcolor{Blue} $[0,\tilde a]$ & \cellcolor{Blue} $[\risco,\infty]$  & \cellcolor{Blue} $[\tilde{x}(a,r_{\rm in}),1)$ & \cellcolor{Blue} III\\ 
         \hline
         \cellcolor{Green} $(\tilde a,1]$ & \cellcolor{Green} $(\tilde r(a),\infty]$ & \cellcolor{Green} $(0,\tilde{x}(a,r_{\rm in}))$ & \cellcolor{Green} II\\
         \hline
         \cellcolor{Blue} $(\tilde a,1]$ & \cellcolor{Blue} $(\tilde r(a),\infty]$ & \cellcolor{Blue} $[\tilde{x}(a,r_{\rm in}),1)$ & \cellcolor{Blue} III\\ 
         \hline
         \cellcolor{Green} $(\tilde a,1]$  & \cellcolor{Green} $[\risco,\tilde r(a)]$ &  \cellcolor{Green} $(0,1)$ & \cellcolor{Green} II\\ 
         \hline\end{tabular}}\\
         \\[16ex]
         \bf{\large \underline{Key for Plot B}}\\
         \includegraphics[width=.75\columnwidth]{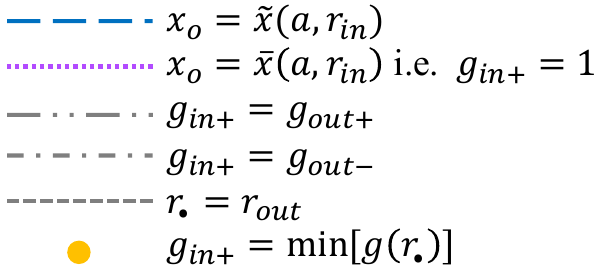}
    \end{tabular}}}  & \quad 
    {\resizebox{.47\linewidth}{!}{
    \begin{tabular}{c}
         
         \includegraphics{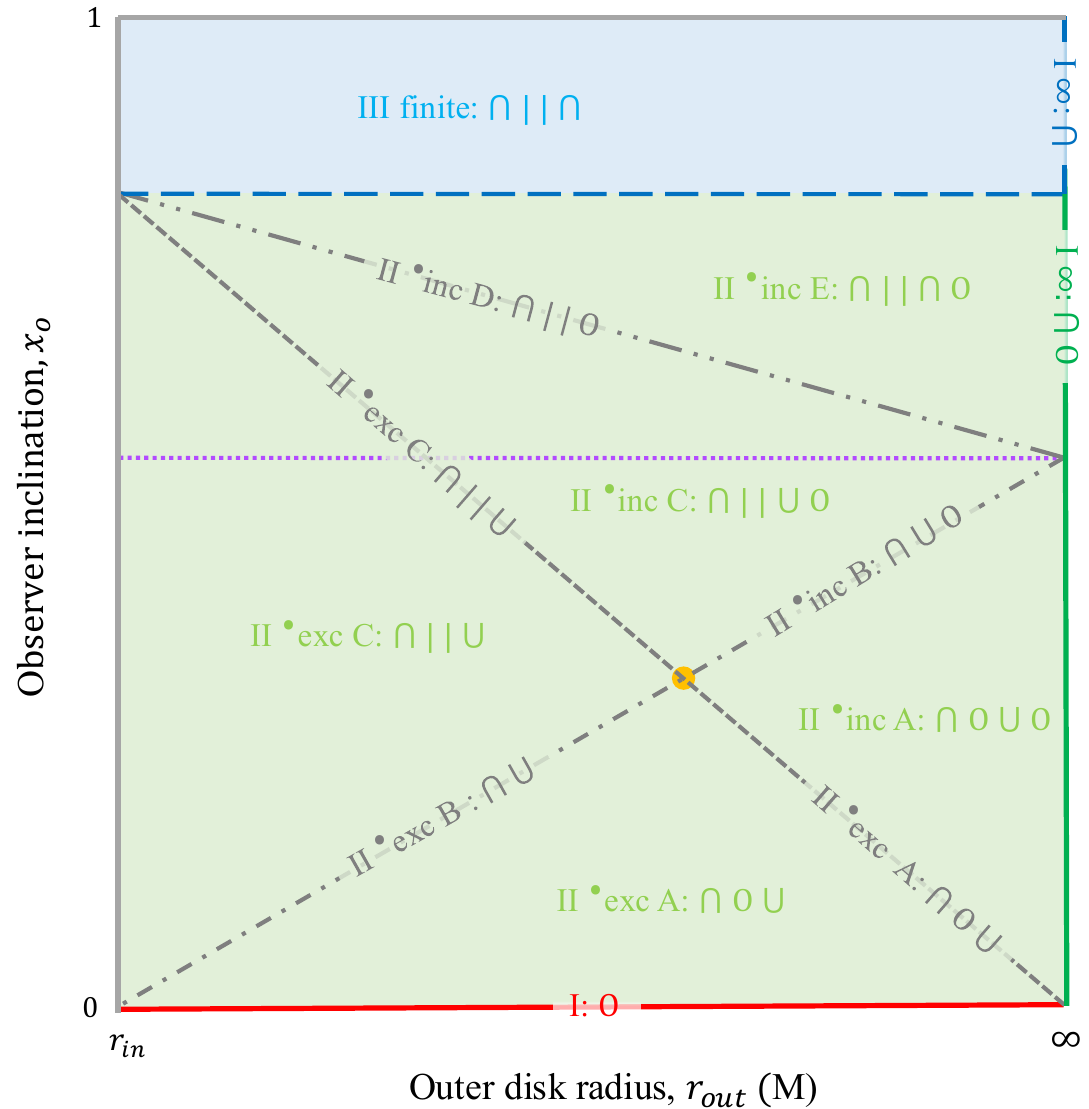}\\ 
    \end{tabular}}} 
    \end{tabular}}}
    \caption{The types of redshift contour configurations (Tabs.~\ref{tab:InfiniteDisk} and \ref{tab:FiniteDisk}) discussed in this work for the Standard disk model can be visualized in the phase space of the BH-disk parameters: BH's spin $a$, the observer's inclination $x_{\rm o}$, and the inner and outer disk radius, $r_{\rm in}$ and $r_{\rm out}$, respectively. \textbf{A}: The landscape of configuration Types. \textbf{B}: Sketches for the phase space cross section at fixed spin and inner disk radius $(a,r_{\rm in})$, such that $\tilde{x}(a,r_{\rm in})<1$. We color code the regions of Type I (red line), II (green lines and green shaded area), and III (blue lines and blue shaded area) configurations. Within the Type II region, we indicate co-dimension one surfaces at which the configurations change (gray lines). Note that this sketch is not to scale and that the gray lines are depicted as lines, for simplicity, to illustrate that they are monotonic. When $\tilde{x}(a,r_{\rm in})=1$ or $\nexists \, \tilde{x}_{\rm o}(a,r_{\rm in})$, the  phase space cross section is isomorphic to the depicted cross section in the range $x_{\rm o}\in[0,\tilde{x}(a,r_{\rm in})]$ or $x_{\rm o}\in\cu{[0,x_{\rm cut}]}$ where $\bar{x}<x_{\rm cut}<\tilde{x}$, respectively. The legend for the sketch is shown on the bottom left. The colors (red, green, and blue for disk Types I, II, and III, respectively) are matched to those of panel A.}
    \label{fig:ConfigLandscape}
\end{figure*}

Let us now present a generalized picture of the redshift contour configurations as a function of the BH's spin $a$, the observer's inclination $x_{\rm o}$, inner disk radius $r_{\rm in}$, and outer disk radius $r_{\rm out}$.
First, we will only consider spin, inclination, and the inner disk radius, since we developed our classifications (Types I-III) via resolved images of disks in the limit of an infinite outer disk radius.

The Type I configuration occurs when $x_{\rm o}=0$, while the Type II and III configurations occur when $x_{\rm o}>0$. The Type II configurations occur when the observer's screen has an FFP with source radius larger than the inner disk radius $\rffp>r_{\rm in}$. Type III configurations occur when the observer's screen does not have an FFP or has an FFP with source radius at or smaller than the inner disk radius $\rffp\leq r_{\rm in}$. 
To determine where in phase space the disk is Type II or III, 1) we must determine if there is an FFP for each spin and inclination pair; and 2) where there is an FFP, we must invert $r_{\rm in}=\rffp(a,x_{\rm o})$ for $\tilde{x}(a,r_{\rm in})$.

In Fig.~\ref{fig:MOBSourceRad}, we present the FFP source radius $\rffp$ and the FFP redshift $\gffp$ as a function of spin and inclination. For spin values below the critical value $\tilde a \simeq 0.9788M$, there exists an FFP for inclinations $x_{\rm o}<\tilde{x}(a,\risco)$; while for sufficiently rapid spins $a>\tilde a$, there is always an FFP from a source radius $\rffp>\risco$~\cite{Gates2020}. The FFP source radius at fixed spin is monotonically decreasing in the inclination $x_{\rm o}$. Hence, for $a>\tilde a$, we can define a critical radius $\tilde r(a)\equiv \rffp(a,x_{\rm o}=1)$ such that any disk with an inner radius $r_{\rm in}<\tilde r(a)$ will be of Type II for all inclinations $x_{\rm o}>0$. In this super-critical spin regime, the inner edge of the disk can source very blue photons, as emitting orbiters in this region will have the highest relativistic speeds. These photons can, in principle, reach observers far from the BH. However, they must travel on trajectories that cross the equatorial plane at least once, consequently contributing to what is called the ``photon ring,''  the higher lensed image of the disk (see e.g., Ref.~\cite{Lupsasca:2024xhq}, for an overview of this concept and efforts to measure it). The theoretical limit for the value of the MOB when taking into account all higher lensed images is $g=\sqrt{3}$ ~\cite{Gates2020}. This value corresponds to photons leaving ISCO emitters in the direction of their motion and received by observers at an inclination $x_{\rm o} = 1$. In this work, we do not take into account these higher-order contributions, as the flux they add to the line profile is very small~\cite{Aldi:2016ntn}.  

The phase space regions for the configurations are summarized in Fig.~\ref{fig:ConfigLandscape}. The Type I region of the phase space is a co-dimension one surface at $x_{\rm o}=0$ (red solid line in Fig.~\ref{fig:ConfigLandscape}) for both the finite and infinite disk. The Type II and Type III regions of phase space can be separated into their finite configurations (shaded green and blue regions, respectively) and infinite configurations (vertical green and blue lines, respectively), i.e., as co-dimension zero finite disk regions extending from $r_{\rm out}\in(r_{\rm in},\infty)$, and as co-dimension one regions at $r_{\rm out}=\infty$, respectively.

The Type II finite disk region of the phase space can be divided into regions corresponding to each configuration (five FFP-inclusive cases and three FFP-exclusive cases). The FFP-inclusive configurations are separated by the co-dimension one surfaces of $\rffp=r_{\rm out}$ (gray dashed line in Fig.~\ref{fig:ConfigLandscape}), with the boundary surface, $\rffp=r_{\rm out}$ being a part of the FFP-exclusive region. Further, the co-dimension one surfaces $g_{\rm in +}=g_{\rm out -}$ (gray dash-dot line) and $g_{\rm in +}=g_{\rm out +}$ (gray dash-dot-dot line) also subdivide the Type II subset of the phase space. In particular, the Type II B FFP-exclusive and inclusive configurations are co-dimension one regions along $g_{\rm in +}=g_{\rm out -}$ surface, while the Type II D FFP-inclusive configurations are the co-dimension one regions along $g_{\rm in +}=g_{\rm out +}$. Note that $g_{\rm out +}=g_{\rm out -}$ only occurs for $r_{\rm out}=\infty$, where $g_{{\rm out}\pm}=1$. At a fixed point $(a,r_{\rm in})$, the co-dimension one surface $g_{\rm in +}=1$ defines a critical angle $\bar x(a,r_{\rm in})$ (purple dotted line). Lastly, the $g_{\rm in +}=g_{\rm out -}$ and $r_{\rm out}=\rffp$ surfaces intersect at a co-dimension two surface $g_{\rm in +}=\min(g(\rffp))$ (yellow dot).

\section{Drawing Inferences from the Morphology of the Line Profile}
\label{sec:MeasureSpinIncRout}

In previous sections, we have described how morphological features of the line profile arise from features in resolved images of the observed accretion disk around a BH. Now, we will shift our focus to how these line profile morphological features constrain the BH's spin, the observer's inclination, and the disk's inner and outer radii. The first step is determining the configuration associated with a given line profile.

\subsection{Distinguishing Configurations Based on the Line Profile Morphology}

\begin{table*}
{\centering
\begin{tabular}{c c| c| c| c| c| c| c| c| c} 
\multicolumn{2}{c|}{Configuration} & \# Transitions & MOB Flux & $D_1$  & $D_2$ & $f_3(g_1^c)=F_g(g_1^c)$& $f_1(g_2^c)=F_g(g_2^c)$ & $f_4(g_2^c)=F_g(g_2^c)$ & $f_2(g_3^c)=F_g(g_3^c)$\\
\hline
\hline
I & $\bigcirc$ & 0 &  finite&$\cdots$&$\cdots$&$\cdots$&$\cdots$&$\cdots$&$\cdots$\\
III~$\infty$ & $\CutD$  & 0 & vanishing &$\cdots$&$\cdots$&$\cdots$&$\cdots$&$\cdots$&$\cdots$\\
\hline
II~$\infty$ & $\CutD~\bigcirc$  & 1 &  finite &$+$&$\cdots$&$\cdots$&$\cdots$&$\cdots$&$\cdots$\\
II~${}^\bullet$exc~B & $\CutD~\CutU$ & 1 & vanishing &$\pm$&$\cdots$&$\cdots$&$\cdots$&$\cdots$&$\cdots$\\
\hline
II~${}^\bullet$inc~B & $\CutD~\CutU~\bigcirc$  & 2 &  finite &$\pm$&$+$&true&false&false&$\cdots$\\
II~${}^\bullet$inc~D & $\CutD~\vertvert~\bigcirc$  & 2 &  finite &$-$&$+$&false&false&false&$\cdots$\\
II~${}^\bullet$exc~A & $\CutD~\bigcirc~\CutU$  & 2 & vanishing &$+$&$-$&false&false&false&$\cdots$\\
II~${}^\bullet$exc~C & $\CutD~\vertvert~\CutU$ & 2 & vanishing &$-$&$+$&false&false&false&$\cdots$\\
III~finite & $\CutD~\vertvert~\CutD$ & 2 & vanishing &$-$&$+$&true&true&false&$\cdots$\\
\hline
II~${}^\bullet$inc~A & $\CutD~\bigcirc~\CutU~\bigcirc$ & 3 &  finite &$+$&$-$&false&false&true&true\\
II~${}^\bullet$inc~C & $\CutD~\vertvert~\CutU~\bigcirc$  & 3 &  finite &$-$&$+$&false&false&false&false\\
II~${}^\bullet$inc~E & $\CutD~\vertvert~\CutD~\bigcirc$  & 3  &  finite &$-$&$+$&true&true&false&false\\
\end{tabular}}
\caption{Morphological features associated with the line profiles for each redshift contour configuration. We list the number of allowed transitions (kinks), the MOB's flux, the relative steepness across the transitions, Eq.~\ref{Eq:Steepness} (a positive value means the right-hand side is steeper than the left-hand side, and vice versa), and the comparison of the functional form of the line profile in the $i$-th region evaluated at the $j$-th kink, $f_i(g^{c}_j)$, to the actual value of the line profile at the $j$-th transition, $F_g(g^{c}_j)$.}
\label{tab:ConfigFeatures}
\end{table*}

As each configuration has a different line profile morphology, it is possible to distinguish between configurations using the line profiles. In this section, we review the morphological features (number of line profile kinks, the flux of the MOB, and behavior between critical redshift factors) of the various configurations that allow for configuration discrimination. We summarize these features in Tab.~\ref{tab:ConfigFeatures} and describe them below.

Although astrophysical disks cannot be infinitely large, a line profile with insufficient spectral resolution or low count rates per energy bin (high uncertainties due to Poisson noise) may make it difficult to distinguish between a large disk and an infinitely large one. Therefore, we will include the infinite disk configurations in this discussion and consider a discretized line profile. 

In particular, recall that as $r_{\rm out}\to\infty$, $g_{\rm out}\to 1$ for $x_{\rm o}=0$ and $g_{{\rm out}+}-g_{{\rm out}-}\to 0$ for $x_{\rm o}>0$. Thus, for the on-axis viewing ($x_{\rm o}=0$), when $g_{\rm out}$ falls into the same energy bin as $g=1$, the line profile will be indistinguishable from the infinitely large Type I disk. Likewise, for off-axis viewing ($x_{\rm o}>0$), when $g_{{\rm out}\pm}$ are in the same energy bin as $g=1$, the line profile will be indistinguishable from the infinitely large Type II or III disks.  

The initial discrimination for the line profile can be achieved by analyzing the number of kinks alongside the MOB flux (refer to the second and third columns of Tab.~\ref{tab:ConfigFeatures}). The line profiles without kinks or only one, can be distinguished by the MOB value. The configurations I and III~$\infty$ both have zero line profile kinks; but, the configuration I has a MOB with finite flux at a value $\gmob\leq1$, while the configuration III~$\infty$ has a MOB with vanishing flux at a value $\gmob>1$. (In principle, as discussed earlier, the infinite Type I disk can have a MOB with vanishing or infinite flux. However, this only happens if the disk is infinite. Such a feature cannot arise as an apparent feature in the finite disk line profile due to the limitations of spectral resolution.) The configurations II~${}^\bullet$exc~B and II~$\infty$ both have one kink, with the vanishing and finite MOB flux, respectively.

For the line profile with two and three kinks, we can perform another level of discrimination using the fall-off of the line profile near these transitions. Recall, at a kink in the line profile, the relative steepness from the left and right is determined by whether the critical value is a local maximum or minimum, Eq.~\eqref{eq:RelativeSteepness}. 
For a transition including the closed topology $\bigcirc$ and any other curve topology ($\CutU$, $\CutD$, or $\vertvert$), the line profile will be less steep on the $\bigcirc$ side of the critical value.
Similarly, for a transition including the double open curve topology $\vertvert$ and any other topology ($\CutU$ or $\CutD$, or $\bigcirc$), the line profile will be steeper on the $\vertvert$ side of the critical value. 
Lastly, in the case of a transition from $\CutD$ to $\CutU$, the line profile need not be steeper on either side of the critical value. 
We can diagnose the left-right comparative steepness of the line profile using the function
\begin{align}
\label{Eq:Steepness}   D_i\equiv&\sign\pa{\abs{F_g'(g_i^{c-})}-\abs{F_g'(g_i^{c+})}},
\end{align}
where, recall, $\cu{g_i^c| i\in\cu{1,2,...,n}}$ are the critical redshift values at which kinks occur. The left-right comparative steepness at the first two kinks, $D_{1,2}$, are listed in the fourth and fifth columns of Tab.~\ref{tab:ConfigFeatures}. For three-kink line profiles, $D_3$ is always positive as the topologies of the rightmost portion of the configuration is $\bigcirc$, and is not useful for discrimination.

The last diagnostic for configuration discrimination is the behavior of the line profile between critical redshift values.
Each line profile with $n$ kinks is divided into intervals $\cu{G_i=(g_{i-1}^c,g_{i}^c)|i\in[1,2,...n+1]}$, where $g_{0}^c=\gmor$ and $g_{n+1}^c=\gmor$. In each of these intervals, the line profile can be written as $F_g(g)=f_i(g)$. Recall, intervals of the line profile arising from redshift contour curves with the same topology have the same functional form. Thus, we can discriminate line profiles by identifying disjoint regions of $F_g(g)$ with the same topology. For the configurations II~${}^\bullet$inc~E and III~finite, the topology is $\CutD$ on $G_1$ and $G_3$; for the other configurations $f_1(g)\neq f_3(g)$. For the configuration II~${}^\bullet$inc~B, the critical value $g_1^c$ marks a transition between the topologies $\CutD$ and $\CutU$, so the topology of the $g_1^c$ is $\bigcirc$; as the topology for $(g_{2}^c,g_{3}^c)$ is also $\bigcirc$, $f_3(g_1^c)=F_g(g_1^c)$. Only for the configuration II~${}^\bullet$inc~A, for which the topology is $\bigcirc$ on $G_2$ and $G_4$, $f_2(g)=f_4(g)$.

Note that as the line profile cannot be explicitly written analytically, as discussed in Sec.~\ref{sec:LineProfileMorpology}, one has to approximate $f_i(g)$ by, e.g., interpolating $F_g(g)$ on $(g_{i-1}^c,g_{i}^c)$. However, to calculate $f_i(g)$ where $g$ is not in the range $(g_{i-1}^c,g_{i}^c)$, one will inevitably have to extrapolate. To minimize the length over which extrapolation has to be carried out, we can discriminate between the different configurations that have two or three kinks, by numerically evaluating the following statements: 
$f_3(g_1^c)=F_g(g_1^c)$, $f_1(g_2^c)=F_g(g_2^c)$, $f_4(g_2^c)=F_g(g_2^c)$, and $f_2(g_3^c)=F_g(g_3^c)$ (listed in the last four columns of Tab.~\ref{tab:ConfigFeatures}).

Determining a function $f_i(g)$ is difficult when the length of the interval $(g_{i-1}^c,g_{i}^c)$ is small. In the finite MOB configurations, $G_{n+1}$ becomes small when the FFP source radius is close the disk's inner or outer radius, and we lose the ability to use $D_{n}$ and $f_{n+1}$ for the discrimination. In these cases, we lose access to $f_4(g)$ for the three-kink line profiles but can continue using $D_1$, $f_1(g)$, and $f_3(g)$ for the discrimination. On the other hand, the two-kink cases become harder to distinguish when we lose access to $f_3(g)$. If $D_1$ is non-negative, the configuration is necessarily II~${}^\bullet$inc~B; however when $D_1$ is negative, the configuration may be either II~${}^\bullet$inc~B or II~${}^\bullet$inc~D. It is not apparent how significant a problem this situation causes, as these configurations, II~${}^\bullet$inc~B or II~${}^\bullet$inc~D, are co-dimension one sets of the whole phase space of configurations. Additionally, for the vanishing MOB cases, the intervals $G_1$ and $G_{3}$ become small, as the outer disk radius is close to the inner disk radius. Here, losing access to $f_1(g)$ and $f_3(g)$, respectively, can present an obstruction to distinguishing these configurations. Only the configuration II~${}^\bullet$exc~A has $D_1$ negative and $D_2$ positive; meanwhile, both the configurations II~${}^\bullet$exc~B and III~finite have $D_1$ positive and $D_2$ negative. Fortunately, $\gmob\leq1$ necessarily indicates the configuration II~${}^\bullet$exc~B, as the configuration III~finite requires $\gmob>1$; however, these configurations remain indistinguishable when $\gmob>1$.

Once the configuration of the line profile has been identified, one can readily pinpoint which critical values correspond to $g_{{\rm in} \pm}$ and $g_{{\rm out}\pm}$. The values of $g_{{\rm in} \pm}$ and $g_{{\rm out}\pm}$ can then be used to constrain the BH's spin, the observer's inclination, and the disk's inner and outer radii, as we will describe in the next section. 

\subsection{Measuring the BH's Spin and the Observer's Inclination}
\label{sec:MeasureSpin&Inclination}

\begin{figure*}
    {\resizebox{\linewidth}{!}{\includegraphics{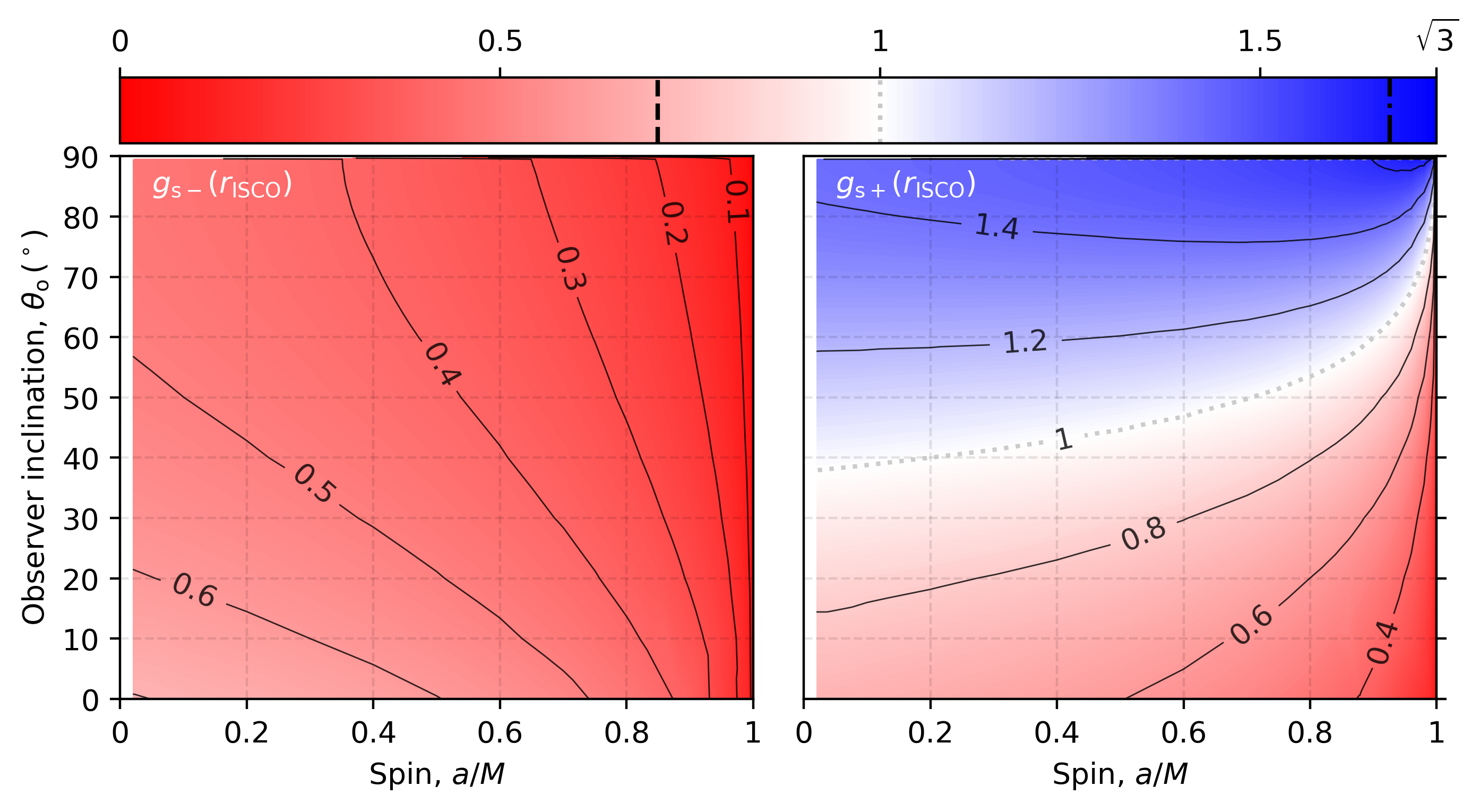}}}
    \caption{Minimum (left) and maximum (right) redshift factor for the direct emission from an ISCO orbiter, as functions of the BH's spin and observer's inclination. The redshift factor goes up to $g=\sqrt{3}$, the theoretical maximum limit of an orbiter emitting from the ISCO radius. The minimum and maximum ISCO redshifts values vanish when $a=M$, as the ISCO reaches the horizon and $g_{\rm s-}(\risco)=g_{\rm s+}(\risco)=0$. The maximum values for the ISCO redshifts are at $(a,\sin\theta_{\rm o})=(0,0)$ (left panel) and $(a,\sin\theta_{\rm o})=(0.9788,1)$ (right panel), where $g_{\rm s-}(\risco)=1/\sqrt{2}$ and $g_{\rm s-}(\risco)\approx 1.65$ (dashed lines on color bar), respectively.}
    \label{fig:gMinMaxISCO}
\end{figure*}

Let us now discuss how to use the critical redshift values, $\cu{g_{{\rm in}\pm},\gffp}$, to constrain the BH's spin $a$ and the observer's inclination $x_{\rm o}$. In the Standard disk model, the inner disk radius is at the ISCO ($r_{\rm in}=\risco$) and the critical redshift values take the form $g_{{\rm in}\pm}=g_{{\rm s}\pm}(\risco)$, as shown in Fig.~\ref{fig:gMinMaxISCO} for the entire parameter space. A constraint of the spin and inclination can be obtained because $(g_{\rm s-}(\risco),g_{\rm s+}(\risco))$ is one-to-one with $(a,x_{\rm o})$. We note that $g_{\rm s-}(\risco)$ is the MOR and has vanishing flux, and for Type III disks, $g_{\rm s+}(\risco)$ is the MOB and also has vanishing flux. Consequently, the portion of a line profile with flux above the noise floor (i.e., instrumental noise and weaker near-by emission lines) can be used to place an upper bound on $g_{\rm s-}(\risco)$ and a lower bound on $g_{\rm s+}(\risco)$. In the case of Type III FFP-inclusive configurations, where we can identify $\gmob=\gffp$ (right panel Fig.~\ref{fig:MOBSourceRad}), one can also uniquely map $(g_{\rm s-}(\risco),\gffp)$ or $(g_{\rm s+}(\risco),\gffp)$ to $(a,\theta_{\rm o})$, with the latter being more useful as both critical values have finite flux. 

\begin{figure}
    \resizebox{\linewidth}{!}{
    \includegraphics{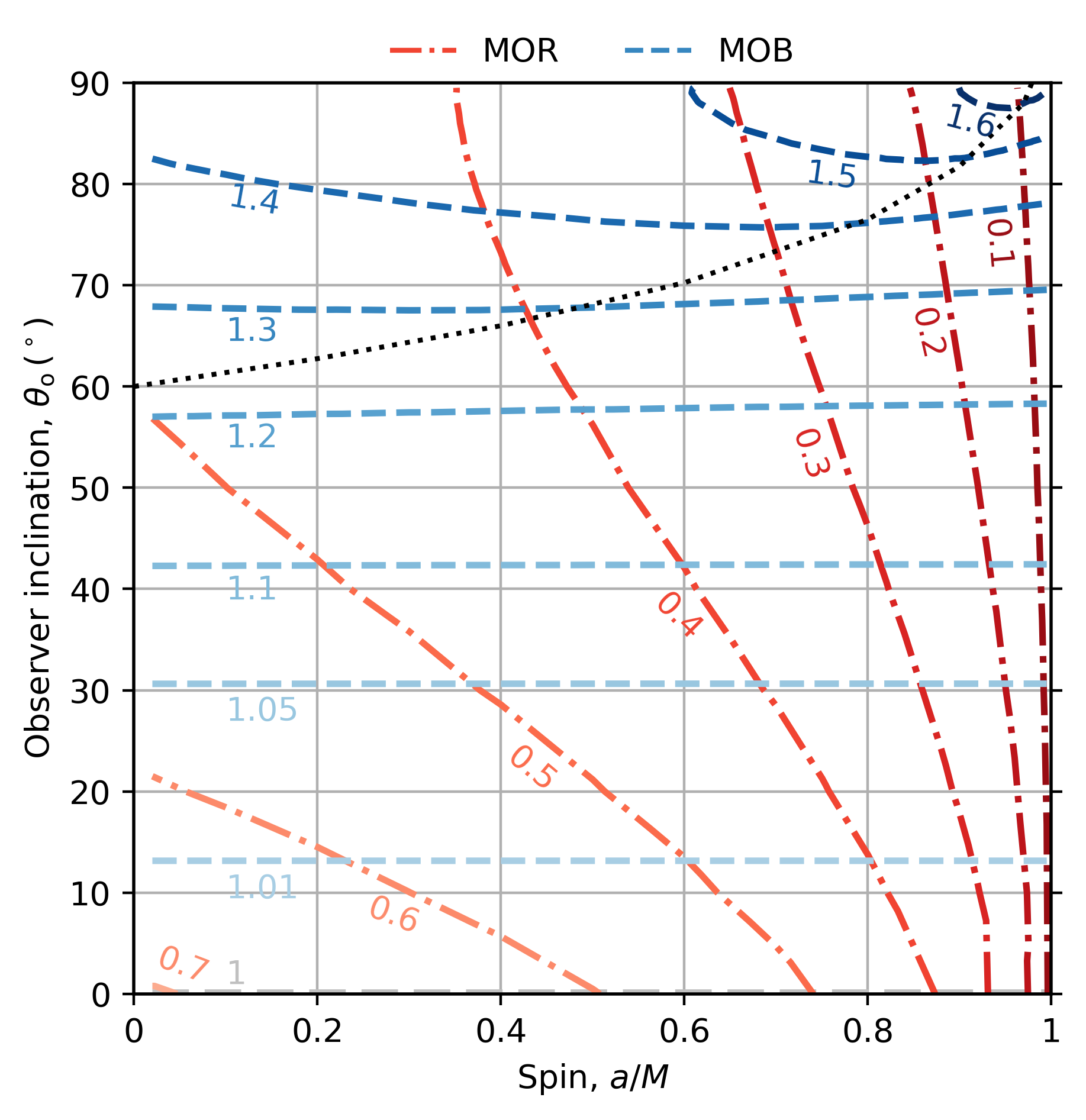}}
    \caption{The way the values of the maximum frequency shifts, the maximum observable redshift (MOR) and the maximum observable blueshift (MOB), which foliate the observer's inclination and BH's spin space, can be used to infer these parameters. The red-dot-dashed contours show constant values of the MOR, while the blue-dashed contours show the MOB, computed assuming the Standard disk model maximally extended. The contours get darker as the frequency shift gets more extreme. For this disk model, the MOR is always sourced from the innermost stable circular orbit (ISCO) radius. Below the black-dotted line, the MOB is sourced from a point in the geometry with a radius larger than the ISCO radius. At and above the dotted line, the MOB is sourced by the ISCO radius. Under this astrophysical model, the Standard disk model, measuring these two quantities provides the values of the spin and inclination.}
    \label{fig:StandardMOB&MOR}
\end{figure}

For the maximally extended disk, where $r_{\rm in}=\risco$ and $r_{\rm out}=\infty$, one obtains a MOR ($\gmorhat$) and a MOB ($\gmobhat$) given by
\begin{align}
    \gmorhat=&g_{\rm s-}(\risco),\\
    \gmobhat=&\begin{cases}
        1, \quad & \text{Type I}\\
        \gffp, & \text{Type II}\\
        g_{\rm s+}(\risco), & \text{Type III}
    \end{cases},
\end{align}
as shown in Fig.~\ref{fig:StandardMOB&MOR}. 
The spin-inclination plane can be foliated by contours of constant MOR, which we denote $\Cmorhat(g)$, in the redshift factor range $1/\sqrt{2}\geq g>0$; and, by contours of constant MOB, denoted $\Cmorhat(g)$, in the range $1 \leq g\lesssim 1.67$.
The inference of the parameters of the system can be performed using the MOR and the MOB values. The MOR is monotonically decreasing in both spin and inclination and takes values between $\gmorhat= 1/\sqrt{2}$ (at $(a,x_{\rm o})=(0,0)$) and $\gmorhat=0$ (for $a=M$ for all the values of the inclination). Mild values, i.e., values closer to $g=1$, of the MOR ($1/\sqrt{2} \geq  \gmorhat \gtrsim 0.47$) provide upper bounds to the spin and inclination values, while more extreme values ($0.47 \gtrsim \gmorhat >0$) can only provide lower bounds to the spin. The MOB value, which increases with inclination but not monotonically with the spin, takes values from $\gmobhat=1$ (when the $x_{\rm o}=0$ for all spin values) to $\gmobhat\simeq 1.67$ (for $x_{\rm o}=1$ and $a\simeq 0.9788 M$)~\cite{Gates2020}. Mild values of the MOB ($1\leq\gmobhat\lesssim 1.41$) can only constrain the inclination, while extreme values ($1.41 \lesssim \gmobhat \lesssim 1.67$) can provide lower bounds for both the spin and inclination. (Additionally, there is a small range of extreme values, $1.61 \lesssim \gmobhat \lesssim 1.67$, for which the MOB also imposes a spin upper bound.) Furthermore, the point $(\gmorhat,\gmobhat)$ is one-to-one with $(a,x_{\rm o})$, and thus the values of the MOR and MOB uniquely constrain the BH's spin and observer's inclination (i.e., when the contours $\Cmorhat$ and $\Cmobhat$ intersect in the spin-inclination space shown in Fig.~\ref{fig:StandardMOB&MOR}).

Constraining the spin and inclination with the MOR and MOB is particularly attractive for two reasons. Firstly, it does not require one to resolve the kinks of the line profile. One just needs the extent of the line. Secondly, it allows for constraints even when the disk is not maximally extended. This is because the maximally extended disk MOR ($\gmorhat$) and MOB ($\gmorhat$) are the extrema values for the model, i.e., 
\begin{subequations}
\begin{align}
    \gmorhat\leq&\gmor(a,x_{\rm o},r_{\rm in},r_{\rm out}),\\
    \gmobhat\geq&\gmob(a,x_{\rm o},r_{\rm in},r_{\rm out}).
\end{align}   
\end{subequations}
Thus, when allowing $r_{\rm in}\geq \risco$ and $r_{\rm out}<\infty$, the MOR and MOB allow us to place lower bounds on the BH's spin and observer's inclinations.

Let us detail the constraints that the MOR ($\gmor$) and MOB ($\gmob$) impose when the disk is \emph{not} maximally extended (i.e., the inner edge of the disk may not go all the way down to the ISCO, or its outer edge may not go to infinity). Measuring these quantities provides bounds on the allowed combinations of spin and inclination $(a,x_o)$ in the spin-inclination plane shown in Fig.~\ref{fig:StandardMOB&MOR}, where contours of these quantities are depicted. The constraints arise from the intersection of the measurements. An ideal measurement results in a single point in the spin-inclination plane, but the uncertainty associated with the measurement leads to bands of possible values. Depending on the assumptions of the model, the resulting inference can vary.

When $\gmorhat\leq\gmor$, the MOR restricts the BH spin and observer inclination to the region bounded from above by $\Cmorhat (\gmor)$, i.e., the  $\gmorhat=\gmor$ contour; we denote this region as $\Rmorhat (\gmor)$. For MOR values $1/\sqrt{2} \gtrsim \gmor \gtrsim 0.47$, spin constraints cannot be derived. From extreme values of the MOR $0.47 \gtrsim \gmor > 0$, we can derive a general spin lower bound. The MOR comes from the disk's inner edge and can be used to put a lower bound on the spin by finding the minimum value of $a$ for which $\gmor=\gmorhat(a,x_o)$ has a solution. In contrast, with the maximally extended disk, no MOR values allow us to put an upper bound on the spin or inclination. 

On the other hand, when $\gmobhat\geq\gmob$ the MOB restricts $(a,x_{\rm o})$ in the spin-inclination plane to the region bounded below by $\Cmobhat (\gmob)$, i.e., the $\gmobhat=\gmob$ contour; we denote this region as $\Rmobhat (\gmob)$. When $\gmob > 1$, we can derive a general inclination lower bound from the MOB by finding the minimum value of $x_{\rm o}$ for which $\gmob=\gmobhat(a,x_{\rm o})$ has a solution. Furthermore, when $\gmob\gtrsim1.41$ we can also derive a general spin lower bound from the MOB by finding the minimum value of $a$ for which $\gmob=\gmobhat(a,x_{\rm o})$ has a solution. Additionally, when $\gmob \gtrsim  1.61$, the MOB also imposes a spin upper bound. In contrast with the maximally extended disk, no MOB values allow us to put an upper bound on the inclination.

\begin{figure*}
    \resizebox{\linewidth}{!}{
    \includegraphics{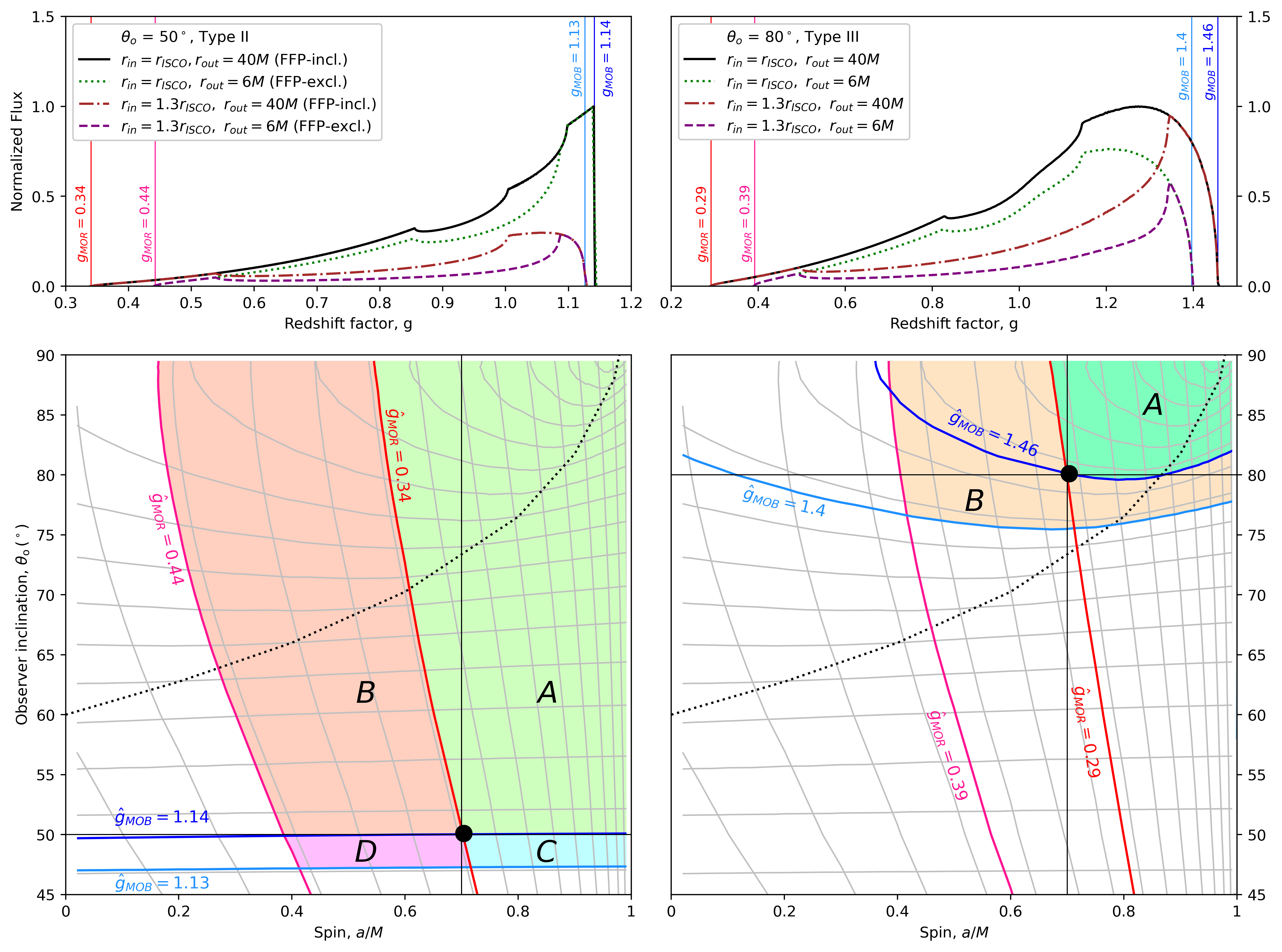}}
    \caption{
    Spin and inclination constraints on the BH-disk system made using the maximal observable redshift (MOR) and maximal observable for specific line profiles. 
    \textbf{Top:} Line profiles for finite disks for a BH of spin $a/M=0.7$ viewed at inclinations $\theta_{\rm o}=50^\circ$ (left) and $\theta_o=80^\circ$ (right), with $r_{\rm in}\in\cu{\risco,\ 1.3\risco}$ and $r_{\rm out}\in\cu{6M,\ 40M}$. 
    For all the line profiles, the MOR is set by the inner edge of the disk.
    In both panels, we mark the MOB of the line profiles with $r_{\rm in}=\risco$ and $1.3\risco$ with vertical red and pink lines, respectively.
    We mark the MOB of the Type II configuration line profiles with $r_{\rm out}=40M$ (which are FFP-inclusive) and $r_{\rm out}=6M$ (which are FFP-inclusive) with a vertical dark blue line and vertical light blue line, respectively. 
    In the right column, we mark the MOB of Type III configuration line profiles with $r_{\rm in}=\risco$ and $1.3\risco$ with a vertical dark blue line and a vertical light blue line, respectively.
    \textbf{Bottom:} Constraints of the BH-disk system in the spin-inclination plane for the line profiles shown in the corresponding panels above. 
    Both panels show the true spin-inclination values of BH-disk system as black points.
    Without knowledge of the line profile configuration, the MOR constrains $(a,\theta_{o})$ to the region above $\Cmorhat(\gmor)$; and the MOB constrains $(a,\theta_{o})$ to the region above $\Cmorhat(\gmor)$. As a result, the combined constraints of the MOR and MOB values create a smaller, more limited area of allowed values $(a,\theta_{\rm o})$.
    We label the $\Cmorhat(\gmor)$ contours in red and pink, and the $\Cmobhat(\gmob)$ contours in dark blue and light blue, to match the MOR and MOB values of the above panels.
    (Additionally, we foliate the spin-inclination plane with $\Cmorhat(g)$ and $\Cmorhat(g)$ for redshift factors $g$ spaced by $\Delta g= 1/32$ as gray-dot-dashed and gray-dashed lines, respectively.)
    The extent of the line profile constrains the BH-disk system to $\Rmorhat(\gmor)\cap\Rmobhat(\gmob)$. In the left column, the line profiles with $(r_{\rm in},\ r_{\rm out})\in\cu{(\risco,40M),\ (\risco,6M),\ (1.3\risco,40M),(1.3\risco,6M)}$, are restricted to shaded regions $\cu{A,\ A\cap B,\ A\cap C,\ A\cap B\cap C\cap D}$, respectively. In the right column, the line profiles with $r_{\rm in}\in\cu{40M,\ 6M}$ are restricted to shaded region $\cu{A,\ A \cap B}$, respectively.
    }
\label{fig:MORMOBConstraints}
\end{figure*}

Lastly, combining the MOR and MOB values, the extent of the line profile restricts $(a,x_{\rm o})$ to the intersection of the regions allowed by each extremal redshift factor $\Rmorhat (\gmor)\cap \Rmobhat (\gmob)$. In Fig.~\ref{fig:MORMOBConstraints}, we present examples of the constraints that can be derived for specific line profiles. Therein, we present the allowed spins and inclinations for the exemplar line profiles (top row), with the contours $\Cmorhat (\gmor)$ (pink and red lines in bottom row), the contours $\Cmobhat (\gmob)$ (light blue and dark blue lines in bottom row), and regions $\Rmorhat (\gmor) \cap \Rmobhat (\gmob)$ (shaded regions in bottom row). Note, if $\gmor>1/\sqrt{2}$ and $\gmob< 1$, we cannot derive any spin and inclination constraints from the MOR and MOB.

When the line profile is resolved enough to determine the configuration, one can further restrict the BH spin and observer inclination of the BH-disk system. For all finite disks, if the inner edge is at the ISCO ($r_{\rm in}=\risco$), the pain $(a,\theta_o)$ is restricted to lie along the contour $\Cmorhat(\gmor)$ (the black dots lie along the red lines in the bottom row of Fig.~\ref{fig:MORMOBConstraints}). If the inner edge of the disk is not at the ISCO ($r_{\rm in}>\risco$), the pair $(a,x_o)$ lies strictly in the region $\Rmorhat(\gmor)$ (the black dots lie above the pink lines in the bottom row of Fig.~\ref{fig:MORMOBConstraints}).

For the finite Type I configuration, the MOB comes from the disk's outer edge and has $\gmob<1$, so it cannot be used to constrain spin and inclination.  For the FFP-inclusive Type II configurations, where the MOB comes from the FFP on the disk ($r_{\rm in} <\rffp < r_{\rm out}$), we can restrict the pair $(a,x_o)$ to the contour $\Cmobhat(\gmob)$ (the black dot lies along the dark blue line in the left bottom panel of Fig.~\ref{fig:MORMOBConstraints}). For the FFP-exclusive Type II configuration where the MOB comes from the outer edge of the disk ($r_{\rm in}  < r_{\rm out} < \rffp$), we can restrict the pair $(a,\theta_o)$ to the region $\Rmobhat(\gmob)$ (the black dot lie above the light blue line in the left bottom panel of Fig.~\ref{fig:MORMOBConstraints}). Lastly, for the Type III configuration, the MOB comes from the disk's inner edge. If we assume the disk terminates at the ISCO radius, we can constrain $(a,x_o)$ to the contour $\Cmobhat(\gmob)$ (the black dot lies along the dark blue line in the right bottom panel of Fig.~\ref{fig:MORMOBConstraints}). If we assume the disk does not extend down to the ISCO radius, we constrain $(a,x_o)$ to the region $\Rmobhat(\gmob)$ (the black dot lies above the light blue line in the right bottom panel of Fig.~\ref{fig:MORMOBConstraints}). 

It is important to emphasize that if the disk is mistakenly assumed to end at the ISCO radius when the truncation radius is greater than the ISCO radius, one may infer incorrect values for the spin and inclination. For Type II FFP-inclusive and Type III disks, the true value of $(a,x_{\rm o})$ lies at the intersection of the $\Cmorhat(\gmor)$ and $\Cmobhat(\gmob)$ contours when the disk terminates at the ISCO radius. Considering the Type II FFP-inclusive and Type III line profiles for which $r_{\rm in}=1.3 \risco$, as illustrated in Fig.~\ref{fig:MORMOBConstraints}, if one assumes that the disks terminate at the ISCO, one erroneously underestimates both the spin and inclination.
As shown in the left column of Fig.~\ref{fig:MORMOBConstraints}, if one assumes that the green-dotted line profile (top panel) comes from a disk that terminates at the ISCO radius, one would conclude that $(a,x_o)$ lies at the intersection of the pink and dark blue lines (bottom panel); though, the error in inclination estimate is extremely small. Indeed, for the Type II FFP-inclusive disk (where the MOB tightly constrains the inclination), one may underestimate the BH spin, particularly in the region where the contours $\Cmobhat$ are approximately horizontal lines. In the right column of Fig.~\ref{fig:MORMOBConstraints}, assuming that the green-dotted and purple-dashed line profiles (top panel) come from disks that terminate at the ISCO radius, one would conclude that $(a,x_o)$ lies at the intersection of the pink and light blue lines (bottom panel). For Type II FFP-exclusive disks, the true value of $(a,x_{\rm o})$ lies along the portion of the $\Cmorhat(\gmor)$ curve which falls above the $\Cmobhat(\gmob)$ contour when the disk terminates at the ISCO radius. Considering the brown-dot-dashed line profile for which $r_{\rm in}=1.3 \risco$ in the top left panel of Fig.~\ref{fig:MORMOBConstraints}, if one assumes that the disks terminate at the ISCO, one erroneously would underestimate the BH spin concluding $(a,x_{\rm o})$ is restricted to the pink curve above light blue curve.

Likewise, assuming an incorrect truncation radius, the potential to overestimate spin may also exist. For example, for the BH-systems of spin and inclination values shown in Fig.~\ref{fig:MORMOBConstraints}, if there exists a choice of $\cu{r_{\rm in}>\risco,r_{\rm out}}$, such that the MOR values are just over the values show in red and the MOB values are near the values shown in light blue, the erroneous $(a,x_o)$ constraint would appear near the intersection of the red and light blue lines in the bottom panels. The potential to overestimate inclination also exists. For example, for a BH-system of spin and inclination $(a=0.7M,\theta_{\rm o}=80^\circ)$, as shown in the right column of Fig.~\ref{fig:MORMOBConstraints}, if there exists a choice of $\cu{r_{\rm in}>\risco,r_{\rm out}}$ such that the MOR value is close to $0.39$ and the MOB value is just under $1.46$, the erroneous spin-inclination constraint would appear near the intersection of the pink add dark blue lines. 

To definitively determine if overestimation of spin or inclination can occur when wrongly assuming an ISCO truncation radius for a BH-system of fixed parameters $\cu{a,x_{\rm o},r_{\rm out}}$, one needs to calculate the intersection point of the curves $\Cmorhat(\gmor(r_{\rm in}))$ and $\Cmobhat(\gmob(r_{\rm in}))$ for all choices of $r_{\rm in}\in\pa{\risco,r_{\rm out}}$, and check whether any intersection point correspond to spin greater than $a$ or inclination greater than $x_{\rm o}$.

Lastly, all constraints on the spin and inclination made using the MOR and MOB are subject to uncertainty set by the spectra resolution of the observed line profile. For an observation of spectral resolution $\Delta E_{\rm o} (E_{\rm o})$, the corresponding line profile $F_g(g)$ of line emission with rest energy $E_s$ has resolution $\Delta g (g=E_{\rm o}/E_{\rm s})=\Delta E_{\rm o} (E_{\rm o})/E_{\rm s}$. For instance, a redshift factor resolution $\Delta g=1/32$ translates to a $\Delta E_{\rm o}=200$eV resolution of Fe K$\alpha$ line emission (i.e., $E_{\rm s}=6.4$~KeV). The spacing of the contours $\Cmorhat$ and $\Cmobhat$ are nonuniform in spin-inclination space, with the contours of each becoming more tightly paced at higher spin and higher inclination where the MOR and MOB values are more extreme. This concentration of values in a small portion of the spin and inclination space leads to smaller uncertainties on spin and inclination when more extreme MOB and MOR values are observed. To illustrate this point, we have foliated the spin-inclination plane in Fig.~\ref{fig:MORMOBConstraints} with MOR contours $\Cmorhat(g)$ and MOB contours $\Cmobhat(g)$ with uniform spacing $\Delta g=1/32$, shown as gray-dot-dashed lines and gray-dashed lines, respectively. Lastly, we note that while such analysis allows us to model the uncertainty on spin and inclination measurements that the MOR and MOB produce  (i.e., the extent of the box the black point falls into for the $\Cmorhat(g)$-$\Cmobhat(g)$ foliation of resolution $\Delta g$), one must first be able to reliably locate which of the line profile's redshift factor bins the critical redshift factors (kinks/edges) fall into. The minimal spectra resolution needed to identify individual critical redshift factor values is determined by the ``sharpness'' the corresponding line profile kink/edge, i.e., the size of the difference in line profile's derivative as we approach the critical redshift value $g_i^c$ from the left and the right $\abs{F_g'(g_i^{c-})-F_g'(g_i^{c+})}$. The sharpness of line profile kinks/edges is highly dependent on all parameters of the BH-disk system $\cu{a,x_o,r_{\rm in},r_{\rm out}}$ and the emissivity profile of the emission line. The kink/edge sharpness and the resolution needed to detect them warrant further study.

\subsection{Measuring the Extent of the Accretion Disk}
\label{sec:DiskExtent}
The critical values of the redshift, $\cu{g_{{\rm in}\pm},g_{{\rm out}\pm},\gffp}$, can also be used to constrain the inner and outer radii of the disk. This is because one can invert the maximal/minimal redshifts at fixed source radius $g_{{\rm }\pm}(a,x_{\rm o},r_{\rm s})$, for the radii of maximal/minimal redshifts factors $r_{\rm s}^{\pm}(a,x_{\rm o},g)$. Even if the spin and inclination values are unknown, one may still be able to provide constraints on the disk edges using these inversion formulae. On the other hand, if the spin and inclination values are known, $r_{\rm s}$ can be directly measured with either of $r_{\rm s}^{\pm}(a,x_{\rm o},g)$. If the inner disk radius is the ISCO, and BH's spin and disk inclination are measured as outlined in the Sec.~\ref{sec:MeasureSpin&Inclination}, $g_{\rm out \pm}$ can be used to measure the outer disk radius.

For all the line profiles, bounds for the radius of each of the disk's edges, $r_{\rm e}\in\cu{r_{\rm in}, r_{\rm out}}$, can be obtained using either of its critical values $g_{{\rm e}\pm}$. Since a given radius can only produce a finite range of redshift values, for all spin and inclination values, these critical redshifts map to a set of radii in the interval $r_{\rm e}\in\br{r_{\rm e}^{L\pm},r_{\rm e}^{R\pm}}$, where
\begin{subequations}
\begin{align}
    r_{\rm e}^{L\pm}&=\min_{(a,x_{\rm o})}\pa{r_{\rm s}^\pm(a,x_{\rm o},g_{{\rm e}\pm})},\\ 
    r_{\rm e}^{R\pm}&=\max_{(a,x_{\rm o})}\pa{r_{\rm s}^\pm(a,x_{\rm o},g_{{\rm e}\pm})}. 
\end{align}    
\end{subequations} 
If we have both critical values for the disk's edge, $g_{{\rm e}\pm}$, we can further constrain the radius to $r_{\rm e}\in\br{r_{\rm e}^{L},r_{\rm e}^{R}}$ where
\begin{align}
    r_{\rm e}^{L}=\max\pa{\cu{r_{\rm e}^{L-},r_{\rm e}^{L+}}},\ \ r_{\rm e}^{R}=\min\pa{\cu{r_{\rm e}^{R-},r_{\rm e}^{R+}}}.
\end{align}   

For the Type I configuration, where $x_{\rm o}=0$, $g(a,r_{\rm s})\equiv g_{\rm s-}(a,0,r_{\rm s})=g_{\rm s+}(a,0,r_{\rm s})$, this implies $r_{\rm s}(a,g)\equiv r_{\rm s}^-(a,0,g_{\rm s-})=r_{\rm s}^+(a,0,g_{\rm s+})$. Since $g(a,r_{\rm s})$ monotonically decreases in both spin and source radius, the disk edge constraints simplify to
\begin{align}
    r_{\rm e}^{L}={r_{\rm s}(1,g_{\rm e})},\quad r_{\rm e}^{R}={r_{\rm s}(0,g_{\rm e})}.
\end{align} 
On the other hand, for Type II and III configurations where $x_{\rm o}>0$, the $g_{\rm s-}(a,x_{\rm o},r_{\rm s})$ is monotonically decreasing in all its arguments, and the edge radius constraints become
\begin{align}
    r_{\rm s}^{L-}={r_{\rm s}^-(1,1,g_{{\rm e}-})},\quad r_{\rm s}^{R-}={r_{\rm s}^-(0,0,g_{{\rm e}-})}.
\end{align}    
When $x_{\rm o}>0$, $g_{\rm s+}(a,x_{\rm o},r_{\rm s})$ is not monotonic in all its arguments, so there is no apparent simplification for the constraints $\cu{r_{\rm s}^{L+},r_{\rm s}^{R+}}$.

For Type II configurations, the additional FFP redshift factor value $\gffp$ can be used to constrain the disk edges because the FFP is sourced by a radius $\rffp(a,x_{\rm o})\equiv r_{\rm s}^+(a,x_{\rm o},\gffp)$. For a given value of $\gffp$, the FFP can only be sourced from a fixed range of radii $\rffp\in[\rffp^{L},\rffp^{R}]$, where
\begin{align}
    \rffp^{L}=\min[\rffp(a,x_{\rm o})], \quad \rffp^{R}=\max[\rffp(a,x_{\rm o})],
\end{align}
as shown in Fig.~\ref{fig:MOBSourceRad}.
For the FFP-inclusive disk, the inner and outer radii obey $r_{\rm in} <\rffp^{L}\leq \rffp\leq \rffp^{R}<r_{\rm out}$. While this bound is generally very weak, it is useful when the rightmost topology $\bigcirc$ region is narrow, as it indicates that the FFP is near the edge of the disk. In particular, this applies to the configurations III~${}^\bullet$inc~A, B and, C when the FFP is near the outer disk radius; and, the configurations III~${}^\bullet$inc~D and E when the FFP is near the inner disk radius. For the FFP-exclusive disk, $r_{\rm in}<r_{\rm out}\leq \rffp<\rffp^{R}$.

Thus far, we have assumed that a measurement can resolve the values $g_{{\rm e}\pm}$. However, the resolution of the energy bins can impact the inferences of the location of the disk edges. In particular, if the line profile is under-resolved, i.e., when the critical values do not appear as sharp kinks but can still be roughly localized, the previous analyses must be considered over the whole range of redshifts where the critical value $g_{{\rm e}\pm}$ may occur. 
Additionally, when the disk is large and the resolution is such that the configuration cannot be distinguished from the infinitely large disk, the resolution of the bin including $g=1$ should still allow one to place a lower bound on $g_{\rm out}$ for the Type I configuration, and lower/upper bounds on $g_{{\rm out}\pm}$, respectively, for the Type II and III configurations.

The previous analyses rely on the inversions of $g_{\rm s-}(a,x_{\rm o},r_{\rm s})$ and $g_{\rm s+}(a,x_{\rm o},r_{\rm s})$, which, unfortunately, cannot in general be performed analytically. However, the functions $r_{\rm s}^-(a,x_{\rm o},g_{\rm s-})$ and $r_{\rm s}^+(a,x_{\rm o},g_{\rm s+})$ can be approximated analytically if one assumes that $g_{\rm s-}$ and $g_{\rm s+}$ arrive to the observer from a large source radius. We provide the analytic expressions under this approximation in App.~\ref{app:LargeRApprox}.

\subsection{Impact of the Emissivity Profile on Parameter Constraints}
\label{sec:Emissitivty}

So far, we have not explicitly addressed the role of emissivity on the line profile morphology. In the previous sections, we have only assumed it is possible to identify $\cu{g_{{\rm in}\pm},g_{{\rm out}\pm}}$ in the line profile, as kinks and edges for constraining the inner and outer disk radii. As shown in the line profiles presented in this work and most prior works (see, e.g., Refs.~\cite{Reynolds:2007rx,Reynolds:2013qqa, Bambi:2020jpe} and references therein for several examples), the critical values associated with the inner and outer disk radii are kinks (exhibiting discontinuous derivatives) in well-resolved line profiles. These kinks are produced because the line emission emissivity is assumed to be discontinuous at the inner and outer disk radii (e.g., for models assuming power-law fall off, broken power-law fall off, or reflection from a lamppost-like corona).

\begin{figure}
    \resizebox{\linewidth}{!}{
    \includegraphics{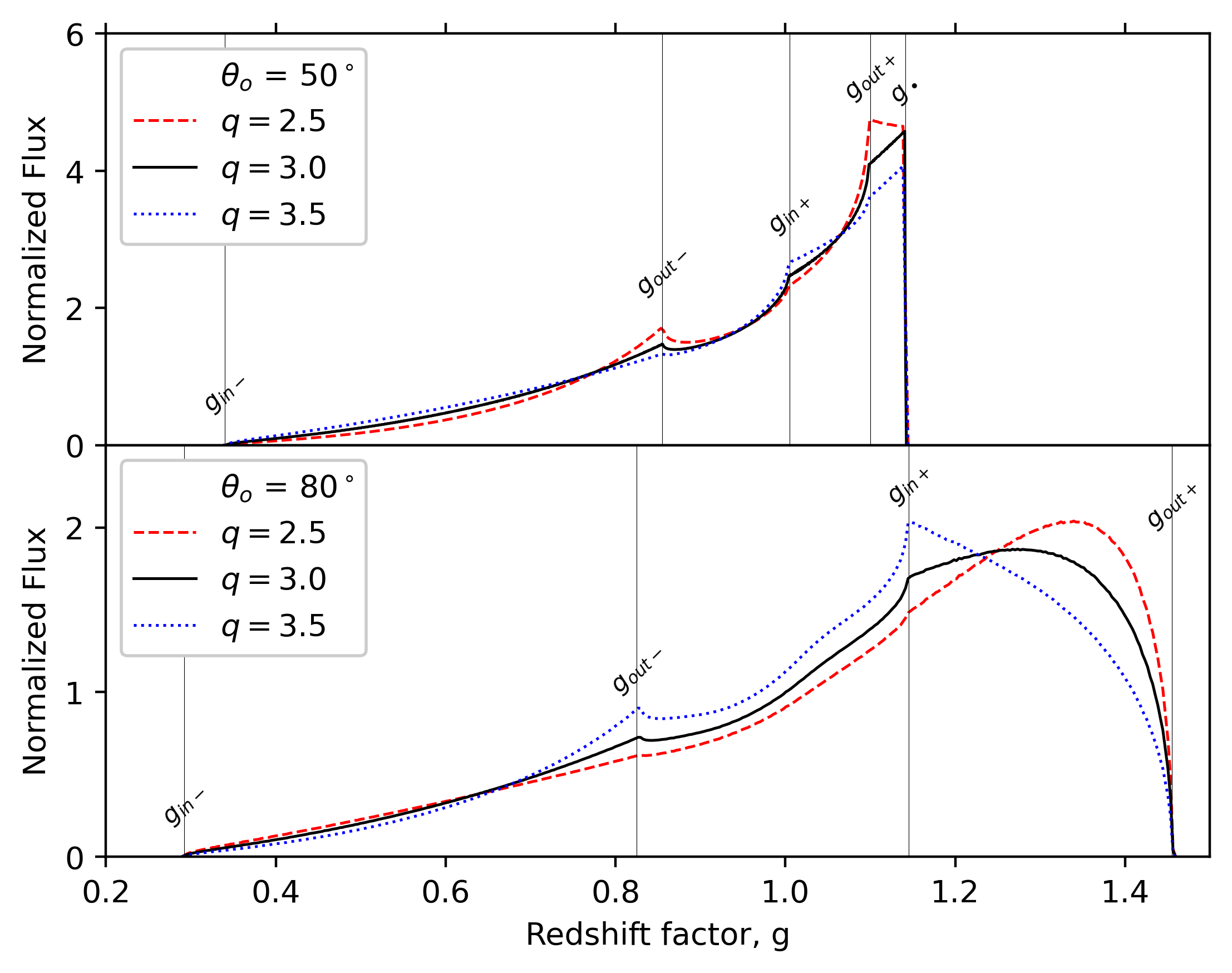}}
    \caption{Changing the emissivity prescription of the line emission on the disk does not impact the location of the sharp features in the line profile; however, it does change the behavior of the line profile on the intervals between adjacent kinks/edges. In these examples, we impose a power-law fall-off for the emissivity $I_s\propto r^{-q}$ and vary the power law exponent. The red-dashed, black-solid, and blue-dotted line corresponds to $q=2.5, 3.0, 3.5$, respectively. We have set $a=0.7M$, $r_{\rm in}=\risco$, and $r_{\rm out}=40M$ for both panels and $\theta_{\rm o}=50^\circ$ and $85^\circ$ for the top and bottom panels, respectively. In each panel, we indicate the critical redshift values with vertical lines. Each line profile is normalized so the area under the curve equals unity.}
\label{fig:ChangingEmissivity}
\end{figure}

In models where the emissivity is discontinuous at the inner and outer disk radii, the positions of the kinks and edges in the line profile remain fixed, regardless of how the emissivity behaves within the range $r_s\in\pa{r_{\rm in},r_{\rm out}}$. Instead, the emissivity dictates the specific functional form of the line profile $f_i(g)$ within the intervals between adjacent critical values $g\in\pa{g^c_i,g^c_{i+1}}$. In Fig.~\ref{fig:ChangingEmissivity}, we present several examples where the sharp features remain unchanged even though the emissivity varies.

Deriving constraints on BH spin and observer inclination based on the location of critical redshift values, without relying on a specific emissivity model, allows us to use the shape of the line profile between these critical values to fit the shape of the emissivity profile independently. Constraints on the shape of the emissivity profile, in turn, can be used to study viable coronal geometries (as studied in, e.g., Refs.~\cite{Wilkins:2012zm,Dauser:2013xv,Gonzalez:2017gzu,Zhang:2024ahe}).

\begin{figure*}
    \resizebox{\linewidth}{!}{
    \includegraphics{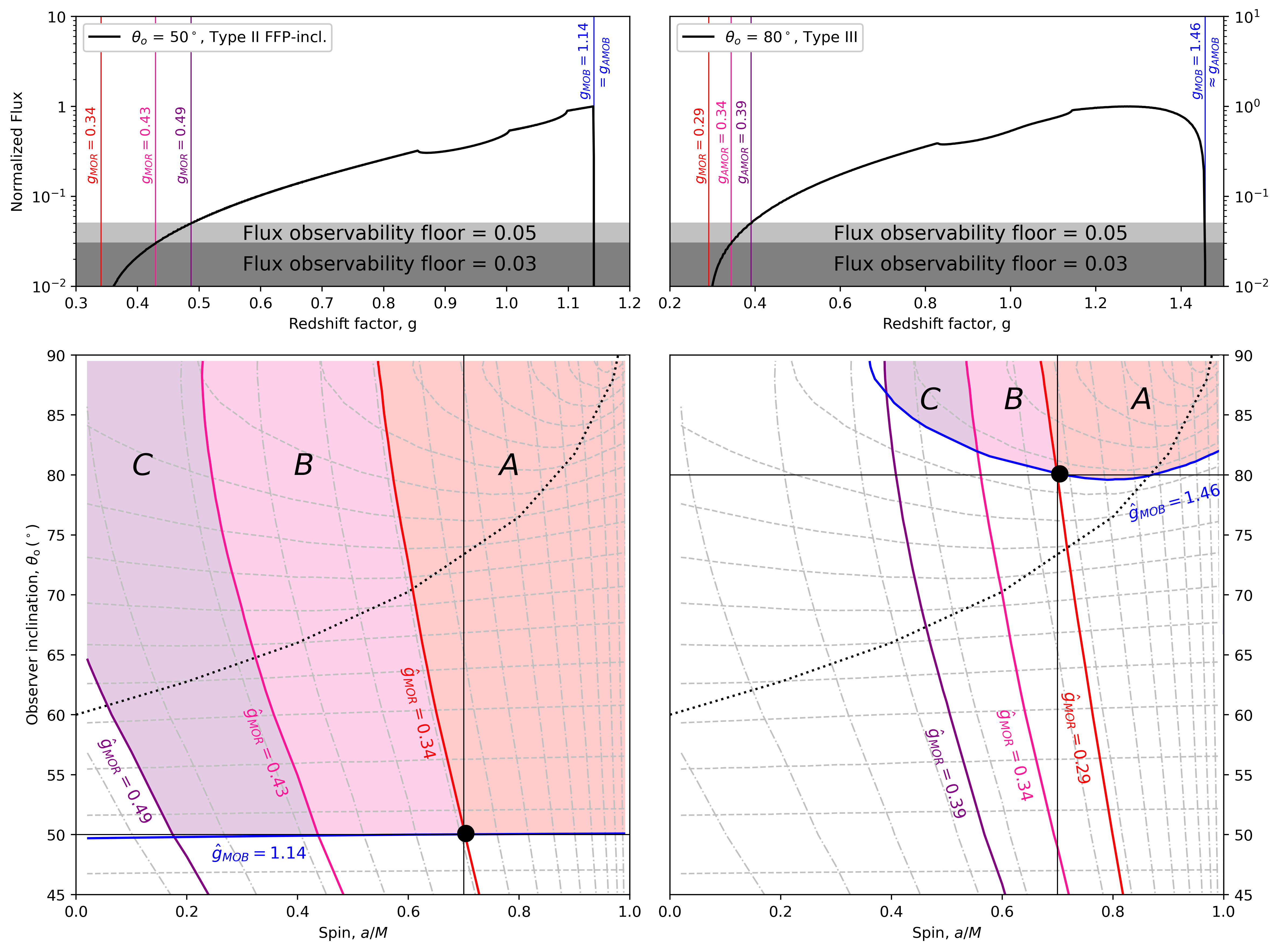}}
    \caption{Spin and inclination constraints on the BH-disk system are made using the apparent maximal observable redshift (AMOR) and an apparent maximal observable blueshift (AMOB) for specific line profiles.
    \textbf{Top:} Line profiles for finite disks for a BH of spin $a/M=0.7$ viewed at inclinations $\theta_{\rm o}=50^\circ$ (left) and $\theta_o=80^\circ$ (right) surrounded by a disk with inner and outer radii $(r_{\rm in},r_{\rm out})=(\risco,40M)$ and emissivity $I_{\rm s}\propto r^{-3}$ (the same line profiles shown in black in Fig.~\ref{fig:ChangingEmissivity}). These lines are normalized such the peak line profile flux is unity ($\max(F_g(g))=1$). We mark the MOR and MOB with vertical red and blue lines, respectively. We illustrate two values for flux observability floor, $0.03 \max(F_g(g))$ and $0.03 \max(F_g(g))$ as shaded dark gray and light gray areas, and mark the resulting AMOR values with vertical pink and purple lines, respectively. In the left panel, the Type II FFP-inclusive line profile has the AMOB equal to the MOB. In the right panel, the Type III line profile has the AMOB approximately equal to the MOB for both values of the nominal flux observability floor ($\gamob-\gmob\sim10^{-3}$), so we do not plot them. 
    \textbf{Bottom:} Constraints of the BH-disk system in the spin-inclination plane for the line profiles shown in the corresponding panels above. We label the $\Cmorhat(\gmor)$ contours in purple and pink, the $\Cmorhat(\gmor)$ contour in red, and the $\Cmobhat(\gmob)$ contour in blue to match the values of the above panels. We have also foliated the spin-inclination plane with $\Cmorhat(g)$ and $\Cmorhat(g)$ for redshift factors $g$ spaced by $\Delta g= 1/32$ as gray-dot-dashed and gray-dashed lines, respectively. Together, the MOR and MOB can restrict the BH-disk system to the region $\Rmorhat(\gamor)\cap\Rmobhat(\gamob)$.
    The MOR and MOB restrict the pair $(a,\theta_{\rm o})$ to the shaded region $A$. For flux observability floor values of $0.03\max(F_g(g))$ and $0.05\max(F_g(g))$, the AMOR and AMOB restrict $(a,\theta_{\rm o})$ to the shaded regions $A \cap B$ and $A \cap B\cap C$, respectively.}
    \label{fig:AMOB&AMOR}
\end{figure*}

Line emission used to constrain BH-disk parameters must be disentangled from other components of an observed spectrum (e.g., thermal continuum, power-law component, or the reflected Compton hump). As such, the shape of a line profile depends sensitively on correctly modeling the shape of the other spectrum components, which must be subtracted from the spectrum to isolate the line profile. If there is a systematic error in the model of non-fluorescent line components, the overall shape of the line profiles will be affected. However, the positions of the critical redshift factors remain unchanged. As a result, while these systematic errors can influence the calculated emissivity profile, the BH-disk constraints derived from the critical redshift values are more reliable. 

Additionally, the emissivity plays an important role in the observability of the MOR and MOB. The MOR and MOB, marking the redshift factor values at which the line profile flux drops to zero, may not be easy to determine in data given the noise and other nearby fluorescent lines. Since the red end of the line profile gradually drops to zero, given a specific flux floor of line profile observability, we can only directly observe an \emph{apparent MOR} (AMOR) and \emph{apparent MOB} (AMOB), representing the lowest and highest redshift factors, respectively, for which the line profile has flux at or above the flux observability floor. For all configurations, the AMOB is larger than the MOR ($\gamor>\gmor$), and the AMOB is less than or equal to the MOB ($\gamob \le \gmob$). For Types I and III configurations, the AMOB is always less than the MOB ($\gamob < \gmob$). For the Type II configuration, the AMOB usually equals the MOB ($\gamob < \gmob$); however, in extreme cases, one may find $\gamob < \gmob$. 

Since the broadening of line emission is used for spin constraints by fitting observed spectra to simulated line profiles under a particular emissivity model (e.g., power law fall out, broken power-law fall off, reflection from a lamppost corona reflect), a direct measurement of the MOR and MOB is not required. Without assuming a particular emissivity model, the AMOR and AMOB may be used for general bounds of the BH-disk system following the understanding derived in Sec.~\ref{sec:MeasureSpin&Inclination}, wherein the MOR and MOB were used for this exact purpose.

\begin{figure}
    \resizebox{\linewidth}{!}{
    \includegraphics{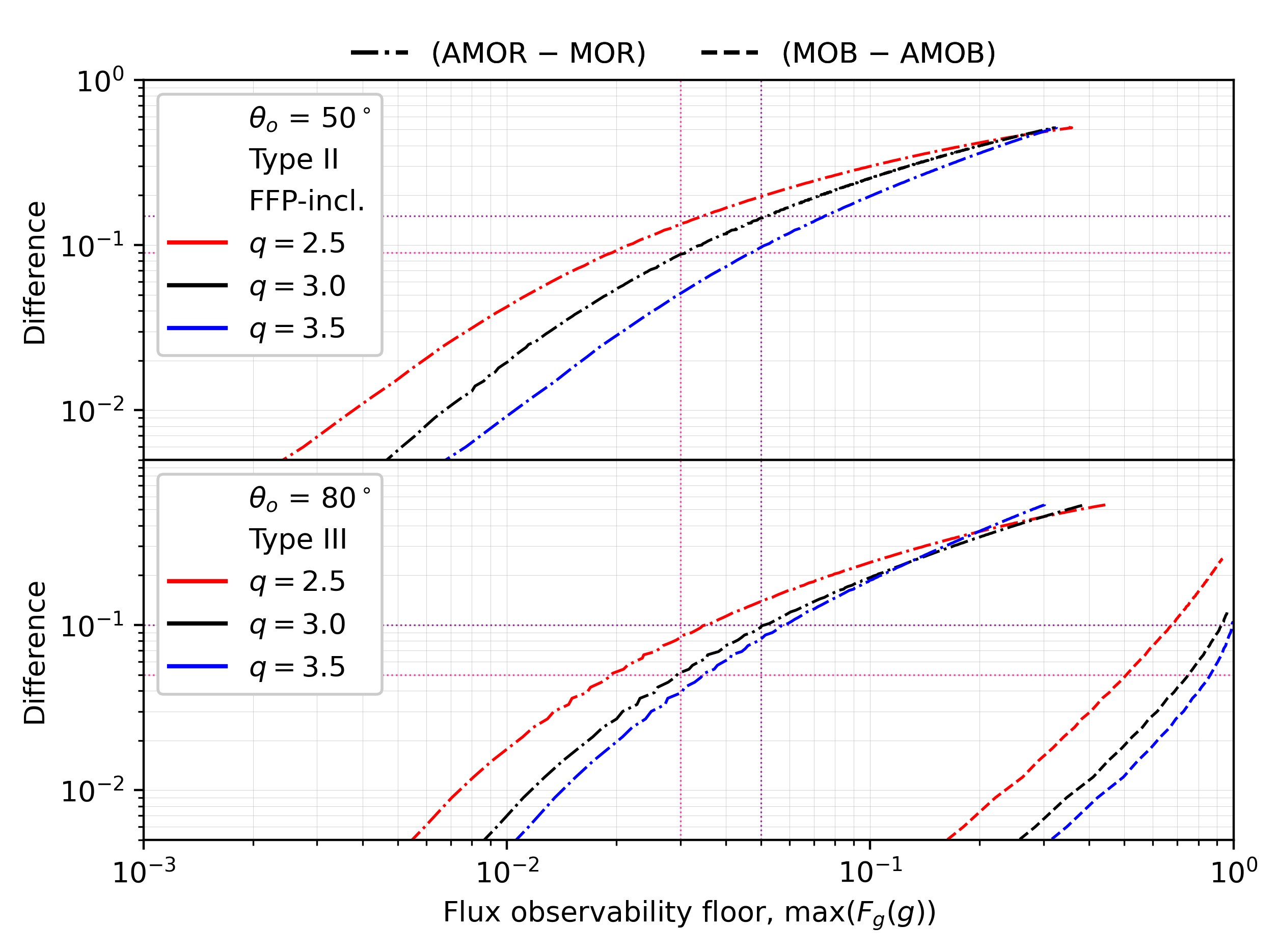}}
    \caption{The difference between the AMOR and AMOR ($\gamor-\gmor$), and the difference between the MOR and MOR ($\gmob-\gamob$) as a function of flux observability floor in units of maximum line profile flux, $\max(F_g(g))$, for the line profiles of Fig.~\ref{fig:ChangingEmissivity}. This difference corresponds to range of the line profile that can be observed above a baseline of flux. The line profiles come from systems of a BH spin $a=0.7M$, disk extent $(r_{\rm in},r_{\rm out})=(\risco,40M)$, observer inclination $\theta_{\rm o}=50^\circ$ (top) and $85^\circ$ (bottom), and emissivity $I_s\propto r^{-q}$. The red, black, and blue lines correspond to $q=2.5,\ 3.0,\ 3.5$, respectively. In both panels, we show $\gamor-\gmor$ and $\gmob-\gamob$ with dot-dashed and dashed lines, respectively. (In the top panel, $\gmob-\gamob=0$ and therefore not show.) Additionally, we mark the value of the MOR-AMOB difference of the $q=3.0$ line profile for flux observability floor values of $0.03$ (pink) and $0.05$ (purple) times the maximum line profile flux shown in Fig.~\ref{fig:AMOB&AMOR} (black curves).}
\label{fig:MOBMORDeviations}
\end{figure}

Comparing the AMOR and AMOB to the MOR and MOB of the maximally extended disk may allow for constraints on the BH spin and observer inclination. In the spin-inclination plane, the AMOB and MOR constrain the spin and inclination to the region $\Rmorhat(\gamor)\cap\Rmobhat(\gamob)$, as shown in Fig.~\ref{fig:AMOB&AMOR}. For all the line profiles, the AMOR provides a weaker constraint than the MOR because $\gmor>\gamor$. For Type II FFP-inclusive configurations where $\gmob=\gamob$, the AMOB constraint matches the MOB constraint, while for Type III configurations where $\gmob<\gamob$, the AMOB provides a weaker constraint than the MOB. Fortunately, for Type III disks that terminate at the ISCO, the difference between the MOB and AMOB ($\gmob-\gamob$) typically grows slowly as we increase the flux observability floor. 
On the other hand, the AMOR-MOR difference ($\gamor-\gmor$) grows faster than the difference between the MOB-AMOB ($\gmob-\gamob$),
because the line profile flux approaches zero faster on the red end of the line profile than on the blue side. 
In Fig.~\ref{fig:MOBMORDeviations}, we show  $\gmob-\gmob$ and $\gamor-\gmor$ for the example line profiles shown in Fig.~\ref{fig:ChangingEmissivity}. For these examples, a flux observability floor of $0.1$ times the peak line profile flux yields $\gmob-\gmob\gg 10^{-2}$ but $\gamor-\gmor\sim\O{10^{-1}}$. When the flux observability floor is sufficiently large that the AMOR provides a weak bound or does not provide a bound, the kink $g_{\rm in+}$ (as shown for $r_{\rm in}=\risco$ in the right-hand panel of Fig.~\ref{fig:gMinMaxISCO}) may be helpful, as an additional input for spin-inclination constraint, for Type II configurations where $\gmob\neq g_{\rm in+}$.

Lastly, if the emissivity is continuous at one of the disk edges, $r_{\rm e}$, the line profiles will no longer exhibit these sharp kinks at $g_{{\rm e}\pm}$, making it more challenging to constrain the location of the inner and outer radii on which an emission line has support. Such considerations of modified emissivities may be relevant if the density of the disk drops closer to the edges of the disk. In particular, in the case of the so-called ``slim disk,'' the emissivity may extend to radii slightly below the ISCO radius. However, the density, and thus the emissivity, rapidly drops to zero outside the horizon. In the next section, we will consider two variations from the Standard disk model and how these changes impact the line profile morphology and our ability to infer parameters.

\section{Variations from the Standard Disk Model}
\label{sec:NonStandard}

So far, we have only described circular Keplerian orbiters. However, as already described in Sec.~\ref{Sec:Introduction}, for certain astrophysical sources, one may need to include ``slim disk'' modifications~\cite{1988ApJ...332..646A,2009ApJS..183..171S,Wu:2007sk}, which, among other effects, allow for particles moving on (slightly) sub-Keplerian trajectories in most of the disk and even change the topology of contours of fixed redshift $g$. One can also, for example, include the emission in the plunging region, i.e., within the ISCO, as prescribed in Ref.~\cite{Cunningham1975}. Introducing these changes in the disk dynamics will modify the line profile. Thus, if one assumes a model for the orbiters that does not correctly capture the astrophysics, the BH parameters one infers can be biased (see, e.g., Refs.~\cite{Fabian:2014tda,Draghis:2022ngm,Draghis:2023vzj} for examples and further discussions). In this section, we will first consider adding the emission within the ISCO (Sec.~\ref{Sec:CunninghamModel}), and then parametrically deviate from Keplerianity (Sec.~\ref{Sec:NonGeodesic}).

\subsection{Cunningham's Disk Model: Including the Plunging Region}
\label{Sec:CunninghamModel}

\begin{figure*}
    {\resizebox{\linewidth}{!}{\includegraphics{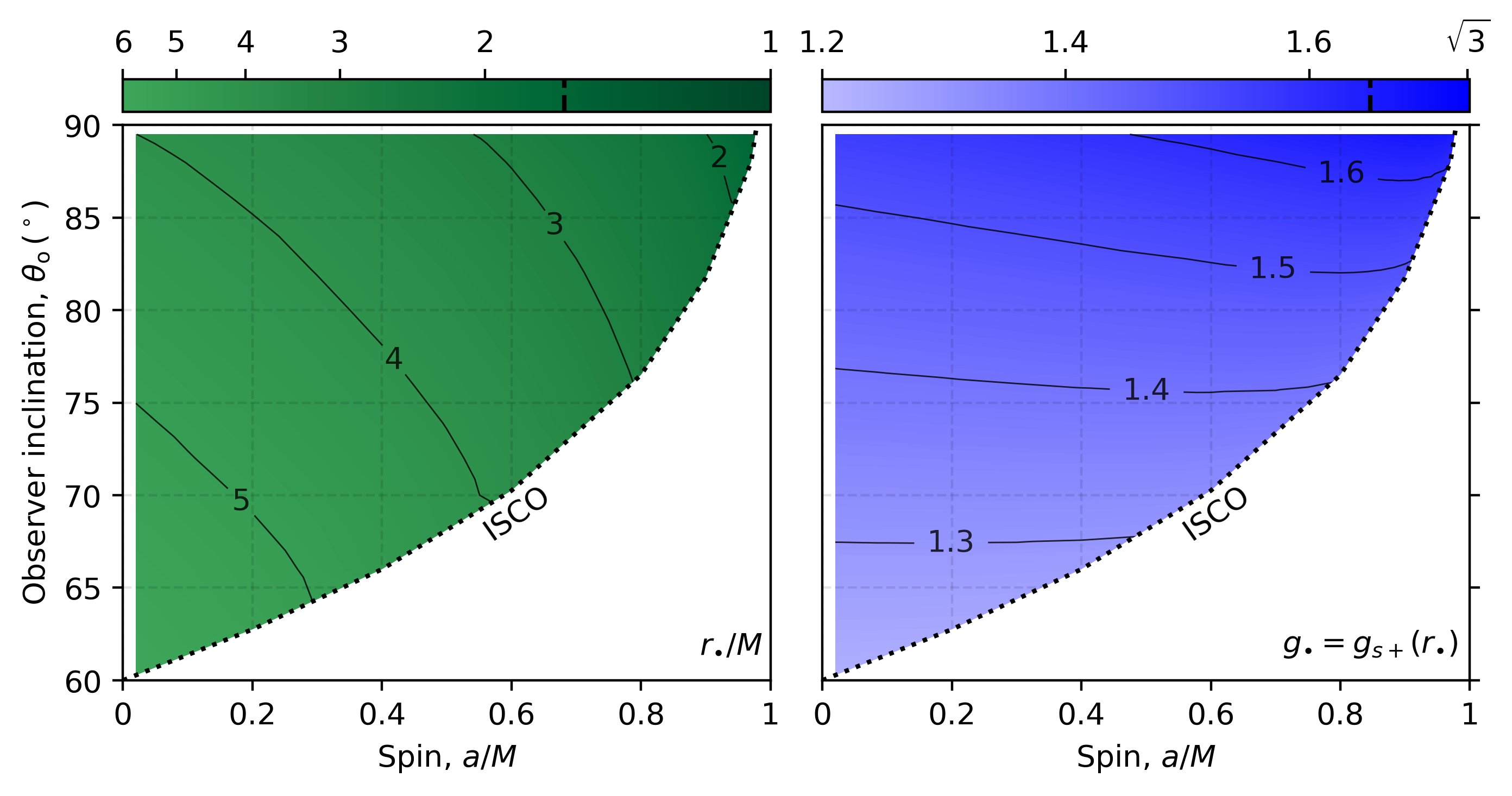}}}\\
    \caption{The source radius $\rffp$ (left) and redshift factor $\gffp$ (right) values for the finite flux point (FFP), as functions of the BH's spin and observer's inclination for the plunging region. The black dotted lines denote the value for which the FFP is sourced by the ISCO. For spin and inclinations occupying the white region below the black dotted lines, the FFP is sourced by the Keplerian orbiters outside the ISCO radius in the Cunningham model. The FFP source radii and redshift factor for the Keplerian region region are given in Fig.~\ref{fig:MOBSourceRad}.}
    \label{fig:MOBSourceRadPlunge}
\end{figure*}

Like the Standard disk model, Cunningham's model~\cite{Cunningham1975} assumes that accretion disk particles travel on circular orbits outside the ISCO, but it assumes that the particles within the ISCO are plunging towards the BH along geodesics while maintaining the conserved quantities of the ISCO orbiter.

In the plunging region, all on-screen redshift factor levels have topology $\bigcirc$ or $\CutU$. As in the Standard disk model case, when viewed from $x_{\rm o}=0$, all contours of constant radius have fixed redshift factor; and, when viewed from $x_{\rm o}>0$, all contours of constant radii $r_{\rm s}>r_{\rm h}$ larger than the BH horizon radius are a double cover of redshift factors $g_\in(g_{\rm s-},g_{\rm s+})$, while the $r_{\rm h}$ contour is uniformly $g(r_{\rm h})=0$. While the Keplerian portion of the disk $g_{\rm s-}(r_{\rm s})$ is monotonically decreasing with source radius at fixed inclination and spin, $g_{\rm s+}(r_{\rm s})$ may be maximized at any radius in $r_{\rm s}\in(r_{\rm h},\risco]$~\cite{Mummery:2024mrq}. When $g_{\rm s+}(r_{\rm s})$ is maximized at the ISCO, all redshift factor levels for $g>0$ have topology $\CutU$. When $g_{\rm s+}(r_{\rm s})$ is maximized at a radius smaller than the ISCO, it corresponds to photons arriving at an FFP with a source radius $\rffp$. In this case, the redshift factors have topology $\bigcirc$ for $g\in[0,g_{\rm s-}(\risco))$ and $g\in(g_{\rm s+}(\risco),\gffp)$, and topology $\bigcirc$ for $g\in[g_{\rm s-}(\risco),g_{\rm s+}(\risco)]$. We show the FFP source radius and value in Fig.~\ref{fig:MOBSourceRadPlunge} within the plunging region.

Further, when we include both the plunging and Keplerian portions of the disk, the redshift contours which touch the ISCO radius contour at two locations (topology $\CutU$ in plunging region and  $\CutD$ for the Keplerian region) join to form contours of topology $\bigcirc$~\footnote{This is not unique to the Cunningham's plunging region. It is true for any model which continuously extends the velocity of the accretion flow down to the horizon.}. Now, for \emph{all} viewing inclinations, the redshift factors contours have the topology $\bigcirc$. Thus, if the disk is maximally extended ($r_{\rm in}=r_{\rm h}$ and $r_{\rm out}=\infty$) the configuration will be $\bigcirc$. When the disk edges are between the horizon and infinity, the various configurations previously outlined for the Standard disk model are still allowed. However, leaving the outer edge of the disk at a finite radius, while extending the inner disk radius to the horizon, closes the contours which are open at the bottom, transforming the topology $\CutD$ to $\bigcirc$, and the topology $\vertvert$ to $\CutU$, leaving only three finite disk configurations: 1) $\bigcirc$ when $x_{\rm o}=0$; 2) $\bigcirc~\CutU~\bigcirc$ when $x_{\rm o}>0$ and $\rffp\in (r_{\rm h},r_{\rm out})$; and 3) $\bigcirc~\CutU$ when $x_{\rm o}>0$ and $\rffp\geq r_{\rm out}$.

When we include the plunging region for the maximally extended disk, the MOR is always zero for all spin and inclination values, since emission is allowed all the way down to the BH's horizon, where $g(r_\mathrm{h})=0$. As in the Standard disk case, the MOB at a fixed spin is strictly increasing with inclination, but varies non-monotonically with the spin. The MOB takes values from $\gmob=1$ when the $x_{\rm o}=0$, regardless of spin, to $\gmob\simeq 1.67$, which occurs when $x_{\rm o}=1$ and $a\simeq 0.94 M$. Mild values (i.e., $1<\gmob\lesssim 1.55$) can only bound the inclination, while extreme values (i.e., $1.55 \lesssim \gmob\lesssim 1.67$) can only provide lower bounds to the spin and inclination. 

\begin{figure}
    \resizebox{\linewidth}{!}{
    \includegraphics{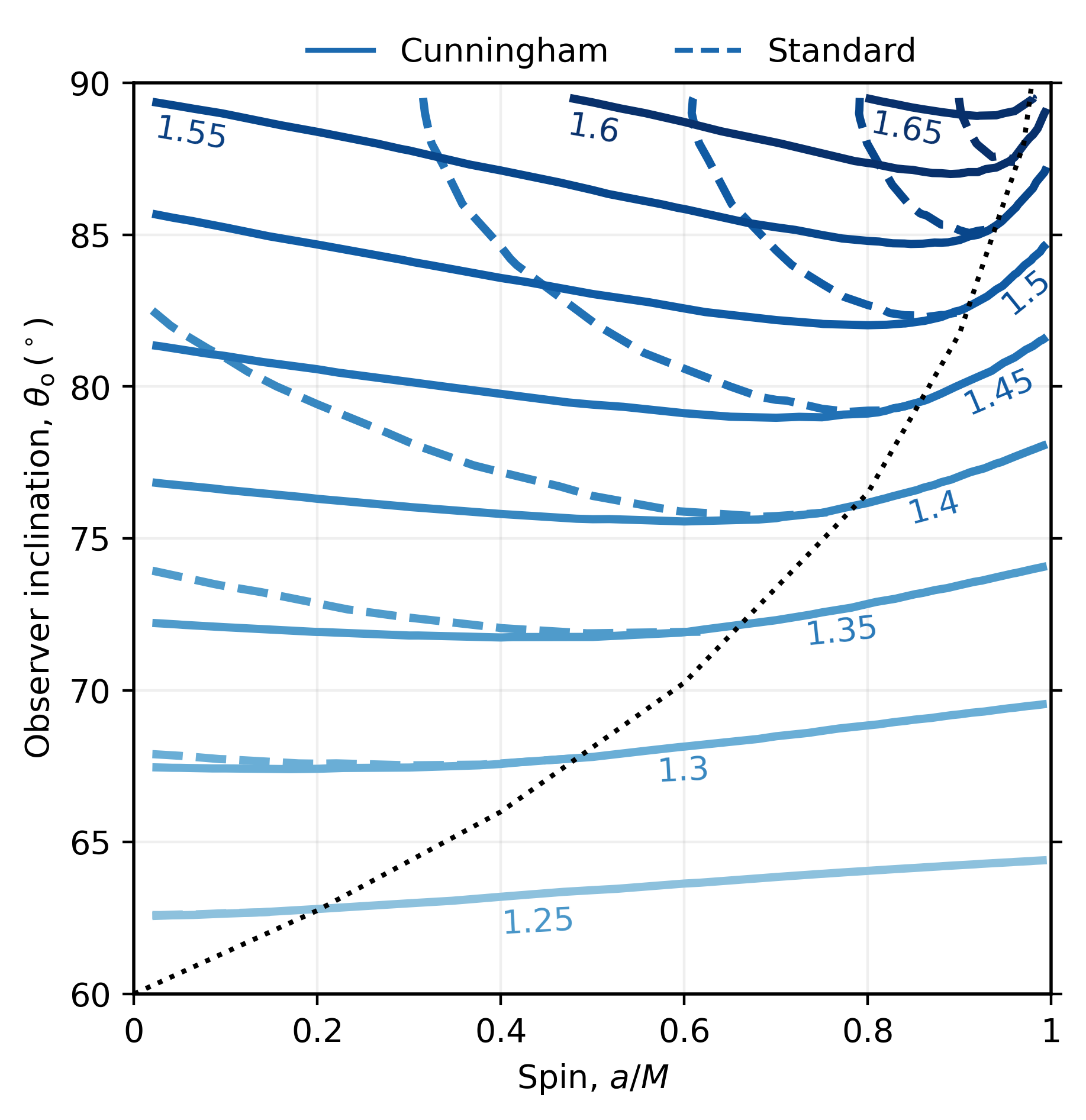}}
    \caption{Values of the  maximum observable blueshift (MOB) of maximally extended disks under the Standard and Cunningham disk models. The dashed and solid contours show constant MOB values under the Standard disk model and the Cunningham disk model, respectively. The black-dotted line marks where the Cunningham and Standard disk MOB values begin to diverge. Below the black-dotted line, the MOB is sourced from a particle on a Keplerian orbit with radius larger than the ISCO radius in both disk models; on the dotted line, the MOB is sourced at the ISCO in both disk models; above the dotted line the MOB is sourced by the ISCO in the Standard disk model and in the plunging region under the Cunningham disk model.}
    \label{fig:MOBcomparision}
\end{figure}

In Fig.~\ref{fig:MOBcomparision}, we compare the MOB of maximally extended disks under the Standard and Cunningham disk models~\footnote{In Fig.~\ref{fig:MOBcomparision}, we find the black-dotted line that bifurcates the regions where the two models' MOBs do and do not match as a function of the BH's spin, by performing a numerical search for the smallest inclination at which the Standard Model's MOB is sourced at the ISCO. Alternatively, this calculation could have been performed as a search for the inclination at which the Cunningham's MOB is sourced at the ISCO.}. As expected, the Cunningham disk MOB matches the Standard disk MOB when it is not sourced by a particle in the plunging region. For spin and inclination values where the MOBs of the two models differ, the Cunningham disk model's MOB value is, however, always greater. Figure~\ref{fig:MOBcomparision} generalizes the results presented in Ref.~\cite{2004ApJS..153..205D} for all the spin and inclination values of the Kerr geometry for this disk type.  

While the MOB value can substantially change when considering the plunging region, the morphology of the line profile will now also strongly depend on the radii at which the emissivity has support interior to the ISCO (see, e.g.,  Ref.~\cite{2010A&A...521A..15A} for several types of inner edges). If one, for example, considers that the same functional form of the emissitivity holds within the ISCO, there can be substantial changes in the morphology~\cite{Dovciak:2003jym,Wilkins:2020pgu,Cardenas-Avendano:2020xtw}. If this were the case for an astrophysical source using the Standard model, one could, for example, be misled to think that the source is rapidly rotating. This can be understood directly from Fig.~\ref{fig:MOBcomparision}. Consider a measurement of $\gmob=1.45$. Assuming the Standard Model, the spin has to be larger than $a\sim0.3M$ and the inclination larger than $x_{\rm o}\sim \sin (78^\circ)$. However, the inference using a model that includes the plunging region will not constrain the spin, and the inclination can only be around $\sin (78^\circ)\lesssim x_{\rm o}\lesssim \sin (81^\circ)$. 

Therefore, assuming the Standard disk model when emission from the plunging region has support within ISCO can, in principle, lead to a bias of the BH's spin and the observer's inclination. In particular, MOB values between $1.55\lesssim \gmob \lesssim 1.67$ can only bound the spin and inclination from below in both disk models, but, as shown in Fig.~\ref{fig:MOBcomparision}, the spin and inclination values inferred from the Standard disk model are systematically \emph{higher} than those inferred from the Cunningham disk model. The potential for parameter overestimation is intensified by the Cunningham disk MOR being zero, as one could incorrectly attribute the extreme redshifting to a rapidly rotating BH with an ISCO close to the horizon. On the other hand, $\gmob \lesssim 1.45$ can provide an upper bound of inclination in both disk models, but inclinations inferred from the Standard disk model are systematically \emph{higher} than those inferred from the Cunningham disk model.

However, one expects that the properties of the emitting plasma (such as the ionization) change once the dynamics of the gas change. In particular, a substantial density change past the ISCO~\cite{Reynolds:1997ek,Young:1998ha,Garcia:2013oma,Wilkins:2020pgu,Dong:2023bbd} is expected, making the material more, if not, fully ionized. Thus, how the line profile changes and our ability to infer parameters will inevitably strongly depend on the specific details of the astrophysics. 

In  Cunningham's disk model, as in the Standard disk model, decreasing the inner radius value increases the support of the line profile on the redshifted side (i.e., extending the line profile to lower redshift factors). If the disk is sufficiently inclined such that the FFP is interior to the ISCO radius, decreasing the inner radius will also increase the line profile support on the blueshifted side, extending the line profile to higher redshift factors. If the emissivity has support that quickly diminishes past the ISCO, there will not be the significant changes in the MOB and MOR values we discussed earlier. However, lacking a robust knowledge of the lower bound on the value of inner disk radius for the line emission means that the MOR, which is always sourced from the inner disk radius, cannot be used to help constrain BH-disk parameters, and the MOB sourced by the FFP in the plunging region is the only hard bound we can exploit from Cunningham's model when attempting to make spin and inclination inferences. This starkly contrasts with the inference in the case of the Standard disk model, where the lower bound on the inner disk radius sets a lower bound on the MOR, which is monotonic in both spin and inclination.

While the MOR and MOB are weaker constraint on the BH's spin and observer's inclination in the Cunningham disk model, we can still make constraints on the location of the inner and outer disk radii using the spin and inclination independent procedures outlined in Sec.~\ref{sec:DiskExtent}. This is because Cunningham model is an extension of the Standard disk model. Further, as the outer disk radius which is on the Keplerian portion of the disk, the location of $g_{{\rm out}\pm}$ is not affected by the addition of emission from the plunging region. In particular, if the disk is large, the spin and inclination agnostic constraints on the location of the disk's outer radius should be same in the two models.

\subsection{Non-geodesic Accretion Flows}
\label{Sec:NonGeodesic}

As another example of how changes in the dynamics of gas particles in the disk affect the morphology of the line profile and our ability to infer parameters from the system, in this section we will focus on parametrically modifying the specific angular momentum of the orbiting objects. Herein we will consider a disk of particles on non-geodesic circular orbits. Following Refs.~\cite{Penna:2013zga,Cardenas-Avendano:2022csp}, let us introduce a ``Keplerianity'' parameter, $\xi$, which represents the ratio of the fluid angular momentum to the Keplerian value. The details of this parametric modification of the four-velocity are presented in App.~\ref{App:NonKeplerian}. When $\xi=1$, one recovers the Keplerian motion described in the previous sections, while larger (smaller) values smoothly make the flow sub(super)-Keplerian. 

For each value of $\xi$, the analysis applied to the Standard disk model can be performed, and the same configurations are allowed. Changing the Keplerianity factor $\xi$, however, moves the location of critical redshift factors $\cu{g_{{\rm in}\pm},g_{{\rm out}\pm},\gffp}$. Considering a fixed BH's spin, observer's inclination, and emission radius $r_{\rm s}$, $g_{s-}$ ($g_{s+}$) decreases (increases) as $\xi$ increases. When there is an FFP $\gffp$, its value increases with $\xi$, and its source radius $\rffp$ may change. At fixed inclination slowing down (speeding up) the velocities of the particles in the disk pushes the FFP to larger (smaller) radii. Thus, disks with the same spin, inclination, and inner and outer radii values but differing Keplerianity factors may not even have the same configuration if $\xi$ causes the MOB source radius to reach one of the disk's edges.

\begin{figure}
    \resizebox{\linewidth}{!}{
    \includegraphics{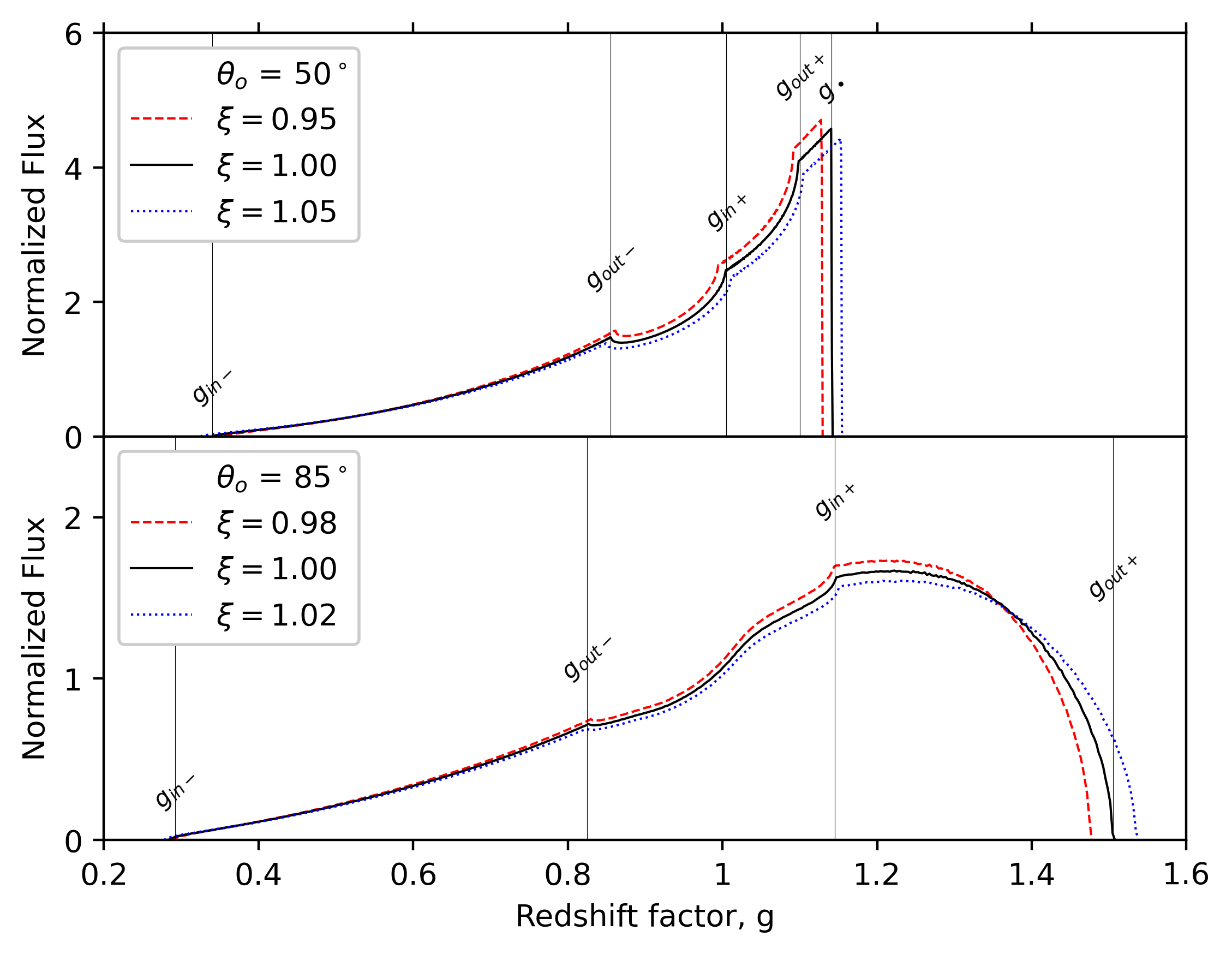}}
    \caption{Changing the velocity prescription of the particles moving on the disk impacts the location of the features in the line profile. In this example (where $a=0.7M$, $r_{\rm in}=\risco$, and $r_{\rm out}=40M$ for both panels, and $\theta_{\rm o}=50^\circ$ and $85^\circ$ for the top and bottom panels, respectively), we have parametrically changed the specific angular momentum of the particles to be sub(super)-Keplerian by multiplying the Keplerian specific angular momentum by a parameter $\xi$. The red-dashed (blue-dotted) line corresponds to a sub(super)-Keplerian with $\xi=0.95(1.05)$ for the top panel and $\xi=0.98(1.02)$ for the bottom panel. In each panel, we indicate the critical redshift values of the Keplerian profile with vertical lines. Each line profile is normalized so the area under the curve equals unity. The changes in the location of kinks as we shift $\xi$ are not merely an uniform shift---as $\xi$ decreases (increases), kinks corresponding to minimum/maximum shift right/left (left/right).}
\label{fig:SubSuperKeplerian}
\end{figure}

The changes in the morphology of the lines, influenced by the Keplerianity parameter, become more pronounced as the inclination increases. For example, as illustrated in Fig.~\ref{fig:SubSuperKeplerian}, a small change in the Keplerianity parameter—just a few percent—can lead to a substantial percentage change in the value of the MOB. These examples imply that fitting the morphology of the line profiles can lead to confusion in disk models due to inherent degeneracies and, therefore, systematic biases.

\section{Discussion}
\label{Sec:Discussion}

In this work, we have revisited the morphology of relativistically broadened line emission from geometrically thin, optically thick equatorial disks under the Standard disk model, Cunningham's disk model, and one that parametrically modifies the specific angular momentum of the gas particles in the disk. We have analytically decoded how parameters of the BH-disk system imprint characteristics features on the line profile, including their extent and number of sharp features. Focusing on the Standard model, we have mapped out how the BH-disk parameters can be inferred from line profile features and explicitly calculated the spin and inclination constraints the maximal frequency shifts can provide. In the Standard disk model, where relativistic broadening has been recognized as a method to constrain spin and inclination, we have demonstrated how these parameters are explicitly encoded in the MOR and MOB.

Further, we have shown that the location of the critical redshift factor values, marking edges and kinks, in the line profile are largely independent of the choice of emissivity profile. On the other hand, the shape of the line profile between critical values is very sensitive to the emissivity profile, which can can vary extensively with the geometry of the corona (as show in, e.g., Refs.~\cite{Wilkins:2012zm,Dauser:2013xv,Gonzalez:2017gzu,Zhang:2024ahe}). Therefore, using spin and inclination constraints derived from the location of line profile extents and kinks may help us more effectively leverage the line profile to infer the disk emissivity profile and the corona’s shape.

In addition to refining our understanding of relativistic lines, this work emphasizes the importance of model selection in interpreting observational data. We explored how different disk geometries and dynamics impact the observed spectra. For instance, in Cunningham's disk model, the maximal frequency shift provides vastly different constraints as the MOR is always zero, and the MOB primarily provides information on the observer's inclination, but little to no information on the BH's spin. Nonetheless, when one has access to all the line profile features, some constraints, e.g., the disk's outer edge location, may match across the Standard disk model and Cunningham's model. Furthermore, we found that the complexities introduced by non-Keplerian orbits have the potential to substantially degenerate all the sharp line profile features, which can alter the inferred properties of both the BH and its accretion disk. These results motivate further investigation into the physical processes governing these extreme environments, as the underlying assumptions can affect the robustness of our inferences. 

The presented mapping of line profile features to BH-disk parameters is valid within the confines of the model considered. The inferred relationships could differ in more realistic cases, such as slim disks where the finite thickness may render some parts of the disk invisible or transform closed redshift curves into open ones. However, our work provides a framework that can be generalized to other astrophysical models. In other words, the overall features will hold, e.g., kinks, but their number, ordering, and location will differ for other models.

The importance of relativistic lines extends beyond mere measurement; they also have the potential to be used as precision tools for testing theories of gravity (see, e.g., Ref.~\cite{Bambi:2016sac}). These inferences, however, are very challenging as the model parameters are intrinsically degenerate, and in general, the uncertainty on the accretion model parameters is larger than the uncertainty on the underlying theory of gravity~\cite{Cardenas-Avendano:2019pec,Bauer:2021atk}. In other words, the effects of modifying the spacetime geometry are expected to be subdominant with respect to changes in the astrophysics of the disk. 

While we have only studied the behavior of relativistic broadening on \emph{single line} emission, our theoretical results apply to the overall broadening of the spectrum. Spin constraint methods that rely heavily on emission close to the BH require a firm knowledge of both the disk's motion and emissivity. Moreover, the constraints derived from the sharp features in the line profile morphology discussed herein require a thorough understanding of the continuum emission, which must be subtracted from the observed spectra to isolate the relativistically broadened line emission. A deeper understanding of the changes the disk's motion and emissivity impose on relativistic broadening may allow disk model discrimination. Lastly, we acknowledge that further studies are needed to understand which features survive when considering spectra built from an ensemble of emission frequencies. We leave such investigations for future studies. 

\begin{acknowledgments}

It is our pleasure to thank Askar Abdikamalov, Cosimo Bambi, Laura Brenneman, Shahar Hadar, Aaron Held, Michael Johnson, Cole Miller, Jack Steiner, and George Wong for helpful discussions. We also thank the anonymous referee for their feedback, which enabled us to improve our results and manuscript presentation. D.G. acknowledges support from the Harvard Postdoctoral Fellowship for Future Faculty Leaders. D.G. acknowledges financial support from the National Science Foundation (AST-2307887). This publication is funded in part by the Gordon and Betty Moore Foundation (Grant \#8273.01). It was also made possible through the support of a grant from the John Templeton Foundation (Grant \#62286).  While at Princeton, D.G. was supported in this work by a Princeton Gravity Initiative postdoctoral fellowship and by a Princeton Future Faculty in the Physical Sciences fellowship. A.C.-A. acknowledges support from the DOE through Los Alamos National Laboratory (LANL) Directed Research and Development, grant 20240748PRD1, as well as by the Center for Nonlinear Studies. This work is authorized for unlimited release under LA-UR-24-32351. The simulations presented in this work were performed on computational resources managed and supported by Princeton Research Computing, a consortium of groups including the Princeton Institute for Computational Science and Engineering (PICSciE) and the Office of Information Technology's High Performance Computing Center and Visualization Laboratory at Princeton University.

The opinions expressed in this publication are those of the authors and do not necessarily reflect the views of these Foundations.

\subsection*{DATA AVAILABILITY} 
The data that support the findings of this article are openly available~\cite{GitHub}.

\end{acknowledgments}

\appendix

\section{Numerical Implementation of the Line Profile}
\label{sec:Discretize}

The line profile, which describes the relativistic broadening of a single spectral line of emission with rest energy $E_{\rm s}$, can be written as a function of the disk emissivity in the frame of the emitters, $I_{\rm s}$, and redshift factor of the observed photons, $g=E_{\rm o}/E_{\rm s}$, integrated over the observer's screen (see  Eq.~\eqref{eq:LineProfileEo}). After performing a coordinate transformation from the screen coordinates ${\bf x}$ to $\pa{g,z_{\rm s}}$, where $z_{\rm s}$ is taken to be the disk emission radius $r_{\rm s}$ (angle $\phi_{\rm s}$) when the disk is viewed off (on) the BH's spin axis $x_{\rm o}>0$ ($x_{\rm o}=0$), the line profile is given by $F_{E_{\rm o}}(E_{\rm o})=E_{\rm s} F_g(g=E_{\rm o}/E_{\rm s})$ (see  Eqs.~\eqref{eq:TotalFlux} and~\eqref{eq:LineProfileG}). 

For off-axis viewing, recall that each source radius corresponds to a finite set of redshift factors $g\in[g_{\rm s-}(r_s),g_{\rm s+}(r_s)]$. Similarly, each redshift factor corresponds to a finite set of source radii $r_{\rm s}\in[r_{\rm s-}(g),r_{\rm s+}(g)]$. When calculating the line profile, an additional coordinate transformation from $(g,r_{\rm s})$ to $(\hat g,r_{\rm s})$, where $\hat g=(g-g_{\rm s-})/(g_{\rm s+}-g_{\rm s-})\in[0,1]$, is typically considered. This transformation reduces the line profile to
\begin{align}
    F_g(g)&=\int_{r_{\rm in}}^{r_{\rm out}}  \ed r_{\rm s} \ f(g,r_{\rm s} ),\\ 
    f(g,r_{\rm s} )&=\frac{g^3  I_{\rm s}(g, r_{\rm s}) \psi(\hat g(g,r_{\rm s}),r_{\rm s})}{(g_{\rm s+}(r_{\rm s})-g_{\rm s-}(r_{\rm s}))\sqrt{\hat g(g,r_{\rm s})(1-\hat g(g,r_{\rm s}))}},
\end{align}
where
\begin{align}
    \psi(\hat g,r_{\rm s})=\sqrt{\hat g \pa{1-\hat{g}}} \abs{\frac{\partial\Omega(\alpha,\beta)}{\partial \hat g \partial r_{\rm s}}}
\end{align}
is Cunningham's ``transfer function,'' and $(\alpha,\beta)$ are the Cartesian coordinates on the observer's screen~\cite{Cunningham1975}. Since these functions can only be written in quadratures, $g_{{\rm s}\pm}(r_{\rm s})$ and $\psi(\hat g,r_{\rm s})$ are computed numerically to obtain the line profile. 

In this work, however, we take a different approach to calculating the line profiles. We work with a compactifed radial screen coordinate $\bar \rho \in[0, 1)$, which maps to the uncompactified screen radius in the range $\rho=\sqrt{\alpha^2+\beta^2} \in [0, \infty)$  (described in the next section).
A compactified screen coordinate is particularly useful for resolving the line emission from large disks, as the redshift factor and source radius change more rapidly at smaller screen radii. To ensure we completely resolve the outer disk radius $r_{\rm out}$, we set the outer limit of the screen to $\sim \bar \rho (\rho = r_{\rm out}+5M)$, as equatorial rings of radii $r_{\rm s}$ far from the BH appear on the observer screen as closed curves with largest screen radial extent $\rho\sim r_{\rm s} + M$~\cite{Gates2020}.

\subsection{Line Profile in Radially Compactified Polar Screen Coordinates}

In our line profile implementation, we analytically ray trace photons arriving at a distant observer with zero angular momentum, at screen locations ${\bf{x}}$, from an equatorial disk around a Kerr BH. We perform these integrations using the Adaptive Analytical Ray-Tracing (\texttt{AART}) code~\cite{Cardenas-Avendano:2022csp}. We discretize the observer's screen in polar coordinates with a compactified radial screen coordinate $\bar\rho=\rho/(\rho+1)$. Lastly, we build the line profile as a histogram of all emission received for a given set of energy bins.

For compactified radial screen coordinate $\bar\rho$ and angular screen coordinate $\varphi$, the line profile Eq.~\eqref{eq:LineProfileEo} becomes
\begin{align}
    F_{E_{\rm o}}(E_{\rm o})&= \frac{g^3}{r_{\rm o}^2} \int \ed  \bar\rho \ \abs{\frac{\partial\rho}{\partial\bar\rho}} \rho(\bar\rho) I_{\rm s}(E_{\rm s},\bar\rho,\varphi).
\end{align}
Discretizing the observer's screen coordinates and the observed energy,
we take
\begin{align}
    N_{\bar\rho}=\frac{\bar\rho_{\max}}{\Delta \bar\rho}, \quad
    N_\varphi=\frac{2\pi}{\Delta\varphi},\quad
    N_{E_{\rm o}}=\frac{E_{\max}}{\Delta E_{\rm o}},
\end{align}
to be integers and
\begin{subequations}
\begin{align}
    \bar\rho_i= i \Delta \bar\rho,& \quad i \in \br{1, N_{\bar\rho}},\\
    \varphi_j= j \Delta \varphi,& \quad j \in \br{1,N_\varphi},\\
    E_k= \pa{k-\frac{1}{2}} \Delta E_{\rm o},&  \quad k\in \br{1,N_{E_{\rm o}}},
\end{align}    
\end{subequations}
with the screen resolution $\Delta\bar\rho$ and $\Delta\varphi$ chosen such that the difference in the value of $E_k$ of all neighboring pixels on the observer's screen  is less the $\Delta E_{\rm o}$.

Evaluating the functions redshift factor $g$ and disk's emissivity $I_{\rm s}$ at each screen pixel, we define
\begin{subequations}
\begin{align}
    \abs{\frac{\partial\rho}{\partial\bar\rho}}_{\{i\}} &= \abs{\frac{\partial\rho}{\partial\bar\rho}(\bar\rho_i)},\\ 
    \rho_{\{i\}}&=\rho(\bar\rho_i),\\
    g_{\{i,j\}}&=g(\rho(\bar \rho_i),\varphi_j),\\
    I_{\{i,j\}}&=I_{\rm s}(\rho(\bar \rho_i),\varphi_j).
\end{align}    
\end{subequations}
Then, the line profile can be approximated as
\begin{align}
    F(E_{k}) =&H\pa{g_{\cu{i,j}}-{E_{\rm s}k \Delta E_{\rm o}}} \nonumber\\
    & \times H\pa{-g_{\cu{i,j}}+{E_{\rm s}(k+1)\Delta E_{\rm o}}} \nonumber\\
    &\times\frac{\Delta \bar\rho \Delta \varphi}{\Delta E_{\rm o}}\sum_{i,j} \abs{\frac{\partial\rho}{\partial\bar\rho}}_{\{i\}} \rho_{\{i\}} \ g_{\cu{i,j}}^3 I_{\{i,j\}} \rho_i,
\end{align}    
where $H$ is the Heaviside function.

In this work, we do not specify a rest energy $E_{\rm s}$, but instead we choose to discuss the line profile as a function of the redshift factor $F_g(g)$ \eqref{eq:LineProfileG}, as the same relativistic shifting applies equally to all rest energies. Thus, to generate the line profiles in this work, we set $E_{\rm s}=1$, which renders $E_{\rm o}=g$, $E_{\max}=g_{\max}$, and $\Delta E_{\rm o}=\Delta g$. Given the analytical understanding of the line profile morphologies outlined in this work, the resolution of the underlying numerical scheme should be selected such that the known number of sharp features for a given set of BH-disk parameters is resolved.

\section{Line Profile in Flat Space}
\label{app:NewtonianDisk}

\begin{figure}
    \includegraphics[width=\linewidth]{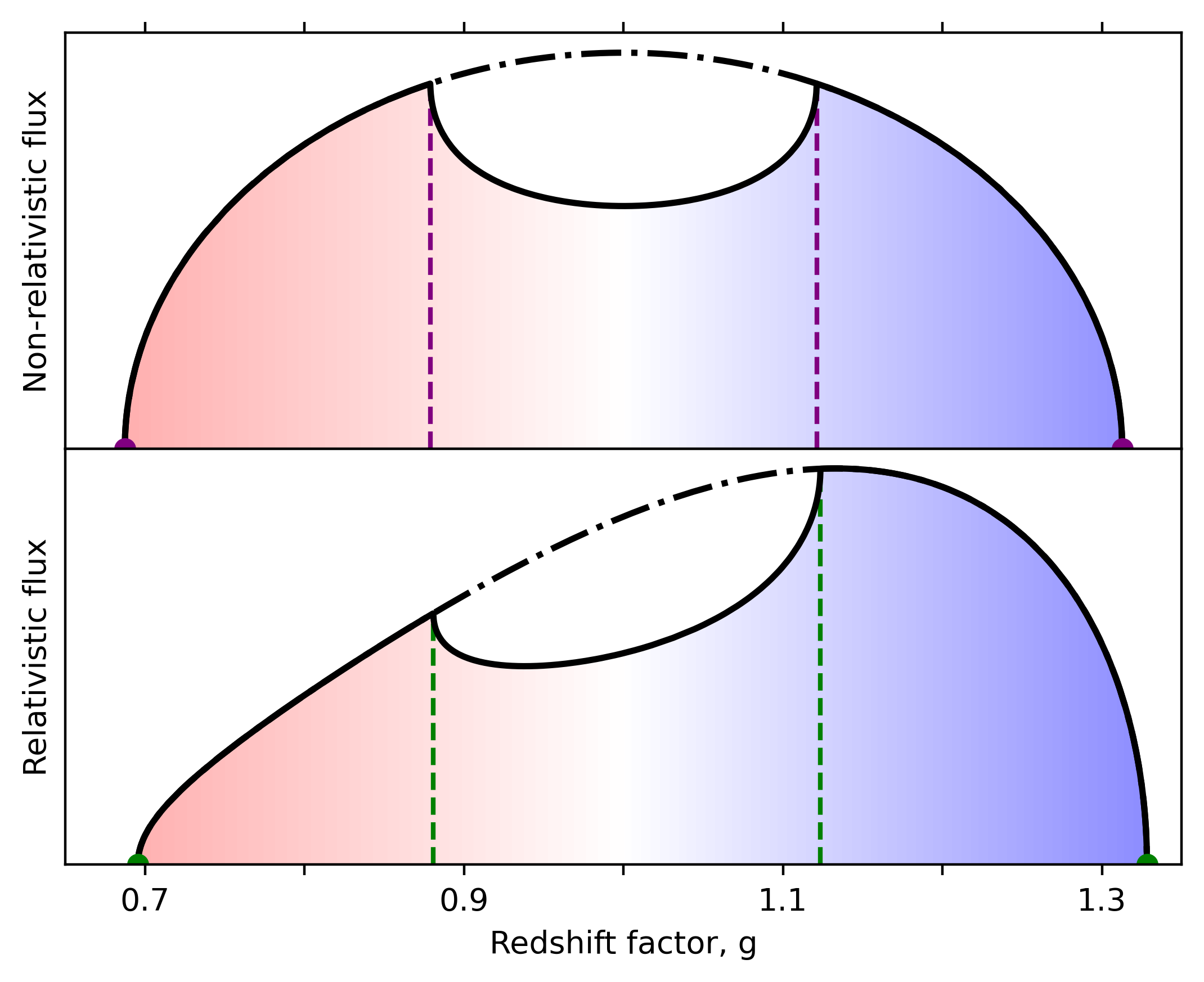}\\
    \includegraphics[width=\linewidth]{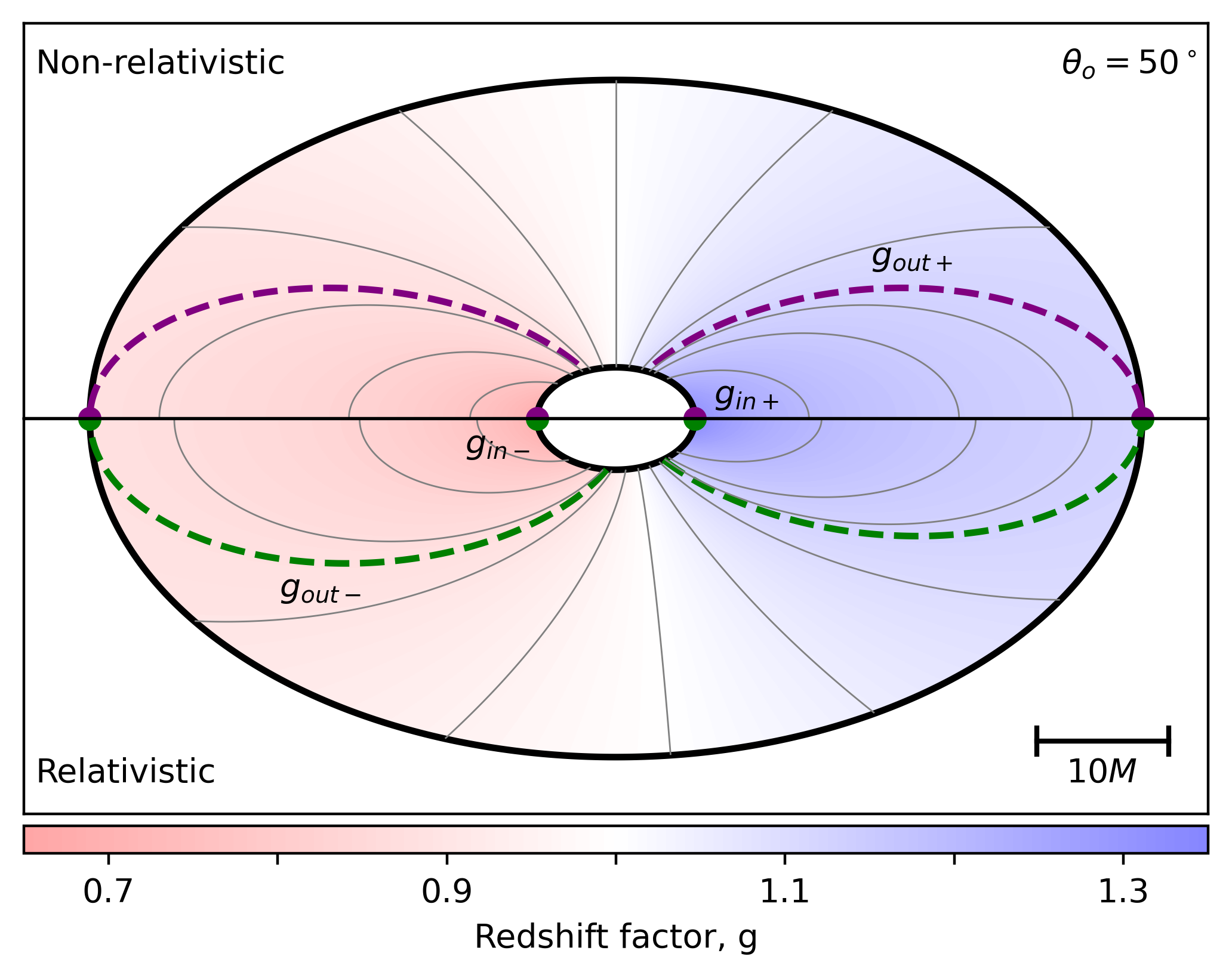}
    \caption{\textbf{Top}: Line profile of a disk of particles traveling on Keplarian orbits, with an inner radius $r_{\rm in}=6M$ and outer radius $r_{\rm out}=40M$, when considering non-relativistic (top therein) and relativistic (bottom therein) Doppler shifts. The dash-dot curves in each panel show the line profile in the limit where the outer radius goes to infinity. \textbf{Bottom}: The corresponding disk on the observer's screen, shaded by the value of the redshift factor. We show only half the disk in each case, as the disk image is symmetric about the horizontal axis. In both panels, we highlight the critical redshift values given by the extremal redshift factors of the inner and outer disk radii $\cu{g_{{\rm in}\pm},g_{{\rm out}\pm}}$ using purple (non-relativistic) and green (relativistic) dashed lines, while showing generic redshift factor curves as solid gray lines.}
\label{fig:Newtonian}
\end{figure}

In this section, we review the line profile of a disk in flat space and compare it to the line profile of the BH-disk system (explored in the body of this work). In flat space, a disk viewed at an angle $\theta_{\rm o}$ from the axis of symmetry has source radius and angle given by
\begin{subequations}
\begin{align}
    r_{\rm s}&=\sqrt{\alpha^2+\beta^2 \sec^2\theta_{\rm o}},\\
    \phi_{\rm s}&=\arctan(\alpha,\beta\sec\theta_{\rm o}),
\end{align}    
\end{subequations}
where $(\alpha,\beta)$ correspond to the Cartesian coordinates of the observer's screen. According to Newtonian gravity, particles in the disk move on circular orbits with velocity of magnitude $v_{\rm s}={r_{\rm s}}^{-1/2}$ and velocity projected along the line of sight $v_{\parallel}=\mp v_{\rm s}\cos\phi_{\rm s}\sin\theta_{\rm o}$, where the minus/plus sign corresponds to a disk that rotates clockwise/counter-clockwise. The non-relativistic and relativistic redshift (Doppler shift) factors are
\begin{align}
    g^{\rm nr}=1-v_{\parallel},\quad  g^{\rm r}=\frac{\sqrt{1-v_{\rm s}^2}}{1-v_{\parallel}},
\end{align}
respectively. The bottom panel of Fig.~\ref{fig:Newtonian} shows an example of these redshifts for a disk with inner and outer radii $r_{\rm in}=6M$ and $r_{\rm out}=40M$, respectively. 

In both cases, we can invert $(g,r_{\rm s})$ and write the line profile explicitly. The non-relativistic line profile is
\begin{subequations}
\label{eq:NonRelativisticLineProfile}
\begin{align}
    F_g^{\rm nr}=&\frac{1}{r_{\rm o}}\int_{r_{\rm in}}^{r_+(g)} \ed r \abs{\frac{\partial\alpha\partial\beta}{\partial g \partial r_{\rm s}}} I_{\rm s}(g,r_{\rm s}),\\
    \abs{\frac{\partial\alpha\partial\beta}{\partial g \partial r_{\rm s}}}=&\frac{r_{\rm s}^{3/2}\cos\theta_{\rm o}}{\sqrt{\sin^2\theta_{\rm o}-r(g-1)^2}},\\
    r_+(g)=&\min\pa{r_{\rm out},\frac{\sin^2\theta_o}{(g-1)^2}},
\end{align}   
\end{subequations}
which is symmetric about $g=1$, and thus exhibits the (familiar) symmetric double peak (as shown on the top panel of Fig.~\ref{fig:Newtonian}).  The ``$g^3$'' factor is exclude from Eq.~\ref{eq:NonRelativisticLineProfile} due to the lack of relativistic beaming. On the other hand, taking into account special relativistic effects, the line profile is
\begin{subequations}
\begin{align}
    F_g^{\rm r}=&\frac{g^3}{r_{\rm o}}\int_{r_{\rm in}}^{r_+(g)} \ed r \abs{\frac{\partial\alpha\partial\beta}{\partial g \partial r_{\rm s}}}  I_{\rm s}(g,r_{\rm s}),\\
    \abs{\frac{\partial\alpha\partial\beta}{\partial g \partial r_{\rm s}}}=&\frac{\sqrt{2}r_{\rm s}\cos\theta_{\rm o}}{g}\sqrt{\frac{r_{\rm s}-1}{R}},\\
     R=&2-2r_{\rm s}+4g\sqrt{r(r-1)}\nonumber\\
     & +g^2(1-2r-\cos(2\theta_{\rm o})),\nonumber\\
    r_+(g)=&\min\pa{r_{\rm out},\frac{{1+g^2 (S-\cos^2\theta_{\rm o})}}{(g-1)^2}},\\
    S=&\sin\theta_{\rm o}\pa{2\sqrt{1-g^2\cos^2\theta_{\rm o}}+g^2\sin\theta_{\rm o}},\nonumber
\end{align}   
\end{subequations}
which corresponds to a pair of skewed peaks (as shown on the bottom case of the top panel of Fig.~\ref{fig:Newtonian}). 

Each radius has extremal redshift factors
\begin{align}
    g_{{s}\pm}^{\rm nr}=1\mp\frac{\sin\theta_{\rm o}}{\sqrt{r_{\rm s}}}, \quad g_{{s}\pm}^{\rm r}=\frac{\sqrt{r_{\rm s}-1}}{\sqrt{r_{\rm s}}\pm\sin\theta_{\rm o}}.
\end{align}
These functions are monotonic in the source radius, so the MOR and MOB \eqref{eq:MORMOB} are always sourced by the inner disk radius. The critical redshift values of the flat space disk are $g_{\rm in-}< g_{\rm out-}<g_{\rm out+}< g_{\rm in+}$.

Considering the classification developed in Sec.~\ref{sec:LineProfileMorpology}, the flat space disk always has redshift factor configuration III~finite ($\CutD~\vertvert~\CutD$) when finite, and III~$\infty$ in the limit that the outer radius goes to infinity when viewed off the axis of symmetry. Thus, the flat space disk is most akin to the BH-disk system when viewed far from the spin axis. As in the BH-disk case, moving the observer away from/towards the axis of symmetry increases/decreases the broadening. But, in contrast to the BH-disk case, moving the observer towards the spin axis does not cause the source radius of the disk MOB to increase. Lastly, while the BH-disk line profile has a finite width when viewed down the axis of symmetry, the flat space disk line profile shrinks to a single spike at $g=1$.

\section{Velocity Prescriptions for the Particles in the Disk}
\label{sec:DiskModel}

In this section, we review the various velocity profiles for equatorial accretion disks around the Kerr BHs studied in this paper.

Working in Boyer--Lindquist coordinates, the geometry of the Kerr spacetime is uniquely specified in terms of the BH's mass $M$ and angular momentum $J=aM$, where $a\in\br{0,M}$ is the BH's spin parameter. The event horizon is located at 
$r_\mathrm{h}=M+\sqrt{M^2-a^2}$.

In the Kerr geometry, a particle is described by four conserved quantities: the particle's mass $m$, energy $\mathcal{E}$, energy-rescaled angular momentum $\lambda$, and Carter constant $\eta$. All the disk particles we consider are confined to the equatorial plane and thus have $\eta=0$. Further, the trajectory of photons is determined by only $(\lambda,
\eta)$, irrespective of the photon energy $\mathcal{E}$.

Assuming the observer is far from the BH ($r_o\to\infty$) with zero angular momentum ($\lambda_o=0$), Bardeen's  observer's screen coordinates, ($\alpha,\beta$), map onto the conserved quantities of an observed photon as~\cite{Bardeen1973}
\begin{subequations}
\begin{align}
    \alpha&=\mp_a\lambda/\sin\theta_{\rm o},\\
    \beta=&\pm_\beta \sqrt{\eta+a^2\cos^2{\theta_o}-\alpha^2\cot^2{\theta_o}},
\end{align}  
\end{subequations}
where $\mp_a=-1$ ($\mp_a=1$) when the BH rotates counter-clockwise (clockwise)\footnote{All resolved images in this work feature the BH rotating clockwise.}. 

The observed intensity $I_{\rm o}$ depends on the emission location in the disk (radius $r_{\rm s}$ and angle $\phi_{\rm s}$); the redshift factor of the observed photons $g$, which measures the ratio of the photon energy as measured by the observer and the emitter; and the surface normal $n_{\rm s}$ as measured in the frame of the emitter.
The source radius $r_{\rm s}(\alpha,\beta)$ and angle $\phi_{\rm s}(\alpha,\beta)$ can be found by ray tracing the photon trajectory backwards from the observer to the source, via the geodesic equations. Once a four-velocity $u_{\rm s}$ for the disk is prescribed, one can determine the redshift factor $g(\alpha,\beta)=(u_{\rm o} \cdot  k)/(u_{\rm s}\cdot k)=E_{\rm o}/E_{\rm s}$, where $k$ is the photon's four-momentum, and the emission surface normal, which is given by $n_{\rm s}(\alpha,\beta)=\arccos(\eta g)/r_{\rm s}$.

\subsection{The Standard Disk Model}
\label{sec:StandardDisk}

In the Standard disk model, particles travel on circular orbits. For prograde orbits, the conserved energy and angular momentum are given by~\cite{Bardeen1973,Cunningham1975}
\begin{subequations}
\begin{align}
    \label{eq:OrbiterE}
	{\ovcirc E}&=\frac{m\pa{r^{3/2}-2M\sqrt{r}+ a\sqrt{M}}}{\sigma(r)},\\
    \label{eq:OrbiterL}
	{\ovcirc \lambda}&= \frac {\sqrt{M}\pa{r^2+a^2- 2a\sqrt{Mr}}}{{r^{3/2}-2M\sqrt{r}+ a\sqrt{M}}},
\end{align}
\end{subequations}
with $\sigma=\sqrt{r^3-3Mr^2+2a\sqrt{M}r^{3/2}}$.
Circular orbits are stable for radii at and above the innermost stable circular orbit (ISCO) radius~\cite{Bardeen1973}
\begin{subequations}
\label{eq:ISCO}
\begin{align}
	\risco&=M\pa{3+Z_2-\sqrt{\pa{3-Z_1}\pa{3+Z_1+2Z_2}}},\\
	Z_1&=1+\sqrt[3]{1-a_*^2}\pa{\sqrt[3]{1+a_*}+\sqrt[3]{1-a_*\,}},\\
	Z_2&=\sqrt{3a_*^2+Z_1^2},
\end{align}
\end{subequations}
where $a_*\equiv a/M$.
The redshift factor of a photon emitted from the equatorial disk at radius $r_{\rm s}$ and received by a zero angular momentum orbiter at $r_{\rm o}=\infty$ is~\cite{Cunningham1975}
\begin{align}
    g&=\dfrac{\sigma(r_{\rm s})}{r_{\rm s}^{3/2}+\sqrt{M}\pa{a-\lambda}}.
    \label{eq:StandardDiskG}
\end{align}

\subsubsection{The Redshift Factor in the Large Radius Approximation}
\label{app:LargeRApprox}

Far from the BH, the contours of constant radius $r_{\rm s}$ form ellipses on the observer's screen with major and minor axes $\alpha=r_{\rm s}$ and $\beta=r_{\rm s} \cos\theta_{\rm o}$, respectively.
The redshift factor \eqref{eq:StandardDiskG} at fixed source radius is monotonic in $\alpha$, so the maximum and minimum redshift factor at fixed radii $g_{\rm s\pm}(r_{\rm s})$ occur at $\varphi=\arctan(\beta/\alpha)=\cu{0,\pi}$. In the limit of large source radius, these are
\begin{align}
    g_{\rm s\pm}&\approx1\pm \frac{x_{\rm o}\sqrt{M}}{\sqrt{r_{\rm s}}}-\frac{M(3-2 x_{\rm o}^2)}{2r_{\rm s}}+\O{\br{\frac{M}{r_{\rm s}}}^{\frac{3}{2}}},
    \label{eq:LargeRG}
\end{align}
where $x_{\rm o}=\sin\theta_o$. This approximation is manifestly spin independent as spin effects are subleading far from the BH. The function $g_{\rm s+}(r_{\rm s})$ has a maximum value $\gffp$ occurring at radius $\rffp$,
\begin{align}
    \gffp=\frac{6-3x_{\rm o}^2}{6-4x_{\rm o}^2}, \quad \rffp=\frac{M(3-2x_{\rm o}^2)^2}{x_{\rm o}^2}.
    \label{eq:LargeRMOB}
\end{align}
These expressions provide an approximation of the FFP value and its location when the FFP radius is large. This approximation works well for low viewing inclinations. On the other hand, since the FFP source radius shrinks as the observer moves away from the spin axis, the approximation deteriorates. For example, for $x_{\rm o}\lesssim1/2$ where $(\gffp,\rffp)\sim(1.05,25M)$, the percent error of $\gffp$ is $\lesssim 0.24\%$.

The approximation given in Eq.~\eqref{eq:LargeRG} for the redshift factor can be inverted to obtain the corresponding radius, giving 
\begin{subequations}
\begin{align}
    r_{1,2}&=\frac{M(Q 
    \pm P)}{2(g-1)^2},\\
    Q&=3(1-g)+(2g-1)x_{\rm o}^2,\\
    P&=x_{\rm o}\sqrt{2\br{6(1-g)+(4g-3)x_{\rm o}^2}}.
\end{align}   
\end{subequations}
This inversion allows us to  define the functions $r_{\rm s}^{\pm}(g)$, which give the source radii with maximum/maximum redshift factors $g$. Under this approximation, $r_{\rm s}^{-}=r_1$, and $r_{\rm s}^{+}$ is given by $r_1$ ($r_2$) if the source radii above (below) the FFP source radius in Eq.~\eqref{eq:LargeRMOB}. 

While the FFP approximation is only valid when the observer's inclination is near the BH's spin axis, the approximations $r_{\rm s}^{\pm}$ are still valid for any inclinations whenever they return large values. For example, restricting to $r_{\rm s}^{\pm}(g)\gtrsim25M$ limits the applicability range of the redshift factors to the range $0.78\lesssim g\lesssim 1.18$.

\subsection{Cunningham's Disk Model}
\label{sec:CunninghamDisk}
In the Cunningham disk model, particles at radii above the ISCO $r_{\rm s}\geq\risco$ still travel on stable circular orbits; but emission from particles plunging from the ISCO $r<\risco$ is assumed to be negligible. Cunningham prescribes that particles interior to the ISCO travel on geodesics while maintaining the conserved quantities associated with an ISCO orbiter \cite{Cunningham1975}. Therefore, particles in the plunging region have conserved quantities
\begin{align}
\label{eq:CunninghamDiskConservedQuantities}
    \pa{E_{\rm s},\lambda_s}=\pa{{\sqrt{1-\dfrac{2M}{3\risco}}},{\ovcirc{\lambda}}(\risco)}.
\end{align}

The redshift factor of a photon emitted from the plunging region at radius $r_{\rm s}$ and received by a zero angular momentum orbiter at $r_{\rm o}=\infty$ is
\begin{subequations}
\label{eq:Redshift}
\begin{align}
    g&=\dfrac{r_{\rm s}^2\Delta(r_{\rm s})}{E_{\rm s}A\pm_r B},\\
    A&=r_{\rm s}^4+a^2 r_{\rm s}\pa{r_{\rm s}+2 M}+\lambda_s\lambda(r_{\rm s}^2+2Mr_{\rm s}) \nonumber\\
    &\ -2Mar_{\rm s}(\lambda_s+\lambda)\\
    B&=r_{\rm s}^2\sqrt{\mathcal{R}(r_{\rm s})} \sqrt{\frac{2M\pa{\risco-r_{\rm s}}^3}{r_{\rm s}^3 \risco}},
\end{align}
\end{subequations}
where $\Delta=r_{\rm s}^2+a^2-2Mr_{\rm s}$ is a metric function of the Kerr geometry, and $\mathcal{R}=\br{\pa{r_{\rm s}^2+a^2}-a\lambda}^2-\Delta(r)\br{\eta+\pa{\lambda-a}^2}$ is the photon radial potential.

\subsection{Non-Keplerian Orbiters}
\label{App:NonKeplerian}

As in Refs.~\cite{Penna:2013zga,Cardenas-Avendano:2022csp}, non-Keplerian orbits can be phenomenologically introduced by multiplying the Keplerian specific angular momentum (Eq.~\ref{eq:OrbiterL}) with a ``Keplerianity'' parameter $\xi$. When this parameter is less (greater) than unity, the particles will move on non-geodesic, time-like, equatorial sub(super)-Keplerian orbits, with angular momentum ${\hat \lambda}=\xi {\ovcirc \lambda}$. Thus, $\xi$ is the ratio of the fluid angular momentum to the Keplerian value. 

One can generalize the particles in the disk to move on generic equatorial orbits with a four-velocity~\citep{Cardenas-Avendano:2022csp}
\begin{align}
	\label{eq:GeneralFourVelocity}
    u=u^t\left(\partial_t-\iota\partial_r+\Omega\partial_\phi\right),
\end{align}
where the angular and radial velocities are defined, respectively, as
\begin{align}
    \label{eq:OmegaAndIotaGeneral}
    \Omega=\frac{u^\phi}{u^t},\quad
    \iota=-\frac{u^r}{u^t}, 
\end{align}
and $u^\mu$ denote the contravariant components of the four-velocity. The  parameterized model presented in Ref.~\cite{Cardenas-Avendano:2022csp} relates $\Omega$ and $\iota$ to their values in Cunningham's model, using two parameters $\left(\beta_r,\beta_\phi\right)$, which range from $0$ to $1$. The explicit functional form of this general four-velocity is rather long and can be found in Appendix B of \cite{Cardenas-Avendano:2022csp}. Varying $\beta_r$ ($\beta_\phi$) from zero to unity smoothly interpolates the radial (azimuthal) velocity from that corresponding to radial infall from rest at infinity to the sub/super-Keplerian value. In this study, we assigned both parameters a value of one. However, in the public version of the code, these parameters are adjustable. Future research will explore the impact of varying these parameters.

\bibliography{Morphology.bib}

\providecommand{\href}[2]{#2}\begingroup\raggedright\begin{thebibliography}{10}

\bibitem{LISA:2022kgy}
{LISA} Collaboration, K.~G. Arun {\em et~al.}, ``{New horizons for fundamental
  physics with LISA},''
  \href{http://dx.doi.org/10.1007/s41114-022-00036-9}{{\em Living Rev. Rel.}
  {\bfseries 25} no.~1, (2022) 4},
  \href{http://arxiv.org/abs/2205.01597}{{\ttfamily arXiv:2205.01597 [gr-qc]}}.

\bibitem{Purrer:2015nkh}
M.~P\"urrer, M.~Hannam, and F.~Ohme, ``{Can we measure individual black-hole
  spins from gravitational-wave observations?},''
  \href{http://dx.doi.org/10.1103/PhysRevD.93.084042}{{\em Phys. Rev. D}
  {\bfseries 93} no.~8, (2016) 084042},
  \href{http://arxiv.org/abs/1512.04955}{{\ttfamily arXiv:1512.04955 [gr-qc]}}.

\bibitem{LIGOScientific:2020kqk}
{LIGO Scientific, Virgo} Collaboration, R.~Abbott {\em et~al.}, ``{Population
  Properties of Compact Objects from the Second LIGO-Virgo Gravitational-Wave
  Transient Catalog},'' \href{http://dx.doi.org/10.3847/2041-8213/abe949}{{\em
  Astrophys. J. Lett.} {\bfseries 913} no.~1, (2021) L7},
  \href{http://arxiv.org/abs/2010.14533}{{\ttfamily arXiv:2010.14533
  [astro-ph.HE]}}.

\bibitem{KAGRA:2021duu}
{KAGRA, VIRGO, LIGO Scientific} Collaboration, R.~Abbott {\em et~al.},
  ``{Population of Merging Compact Binaries Inferred Using Gravitational Waves
  through GWTC-3},'' \href{http://dx.doi.org/10.1103/PhysRevX.13.011048}{{\em
  Phys. Rev. X} {\bfseries 13} no.~1, (2023) 011048},
  \href{http://arxiv.org/abs/2111.03634}{{\ttfamily arXiv:2111.03634
  [astro-ph.HE]}}.

\bibitem{EventHorizonTelescope:2019pgp}
{Event Horizon Telescope} Collaboration, K.~Akiyama {\em et~al.}, ``{First M87
  Event Horizon Telescope Results. V. Physical Origin of the Asymmetric
  Ring},'' \href{http://dx.doi.org/10.3847/2041-8213/ab0f43}{{\em Astrophys. J.
  Lett.} {\bfseries 875} no.~1, (2019) L5},
  \href{http://arxiv.org/abs/1906.11242}{{\ttfamily arXiv:1906.11242
  [astro-ph.GA]}}.

\bibitem{Nemmen:2019idv}
R.~Nemmen, ``{The Spin of M87*},''
  \href{http://dx.doi.org/10.3847/2041-8213/ab2fd3}{{\em Astrophys. J. Lett.}
  {\bfseries 880} no.~2, (2019) L26},
  \href{http://arxiv.org/abs/1905.02143}{{\ttfamily arXiv:1905.02143
  [astro-ph.HE]}}.

\bibitem{EventHorizonTelescope:2022exc}
{Event Horizon Telescope} Collaboration, K.~Akiyama {\em et~al.}, ``{First
  Sagittarius A* Event Horizon Telescope Results. IV. Variability, Morphology,
  and Black Hole Mass},''
  \href{http://dx.doi.org/10.3847/2041-8213/ac6736}{{\em Astrophys. J. Lett.}
  {\bfseries 930} no.~2, (2022) L15},
  \href{http://arxiv.org/abs/2311.08697}{{\ttfamily arXiv:2311.08697
  [astro-ph.HE]}}.

\bibitem{Johnson:2024ttr}
M.~D. Johnson {\em et~al.}, ``{The Black Hole Explorer: motivation and
  vision},'' \href{http://dx.doi.org/10.1117/12.3019835}{{\em Proc. SPIE Int.
  Soc. Opt. Eng.} {\bfseries 13092} (2024) 130922D},
  \href{http://arxiv.org/abs/2406.12917}{{\ttfamily arXiv:2406.12917
  [astro-ph.IM]}}.

\bibitem{Lupsasca:2024xhq}
A.~Lupsasca, A.~C\'ardenas-Avenda\~no, D.~C.~M. Palumbo, M.~D. Johnson, S.~E.
  Gralla, D.~P. Marrone, P.~Galison, P.~Tiede, and L.~Keeble, ``{The Black Hole
  Explorer: photon ring science, detection, and shape measurement},''
  \href{http://dx.doi.org/10.1117/12.3019437}{{\em Proc. SPIE Int. Soc. Opt.
  Eng.} {\bfseries 13092} (2024) 130926Q},
  \href{http://arxiv.org/abs/2406.09498}{{\ttfamily arXiv:2406.09498 [gr-qc]}}.

\bibitem{Zhang:1997dy}
S.~N. Zhang, W.~Cui, and W.~Chen, ``{Black hole spin in X-ray binaries:
  Observational consequences},'' \href{http://dx.doi.org/10.1086/310705}{{\em
  Astrophys. J. Lett.} {\bfseries 482} (1997) L155},
  \href{http://arxiv.org/abs/astro-ph/9704072}{{\ttfamily
  arXiv:astro-ph/9704072}}.

\bibitem{Fabian:2000nu}
A.~C. Fabian, K.~Iwasawa, C.~S. Reynolds, and A.~J. Young, ``{Broad iron lines
  in active galactic nuclei},'' \href{http://dx.doi.org/10.1086/316610}{{\em
  Publ. Astron. Soc. Pac.} {\bfseries 112} (2000) 1145},
  \href{http://arxiv.org/abs/astro-ph/0004366}{{\ttfamily
  arXiv:astro-ph/0004366}}.

\bibitem{Brenneman:2013oba}
L.~Brenneman, ``{Measuring Supermassive Black Hole Spins in Active Galactic
  Nuclei},'' \href{http://arxiv.org/abs/1309.6334}{{\ttfamily arXiv:1309.6334
  [astro-ph.HE]}}.

\bibitem{Reynolds2019}
C.~S. {Reynolds}, ``{Observing black holes spin},''
  \href{http://dx.doi.org/10.1038/s41550-018-0665-z}{{\em Nature Astronomy}
  {\bfseries 3} (Jan., 2019) 41--47},
  \href{http://arxiv.org/abs/1903.11704}{{\ttfamily arXiv:1903.11704
  [astro-ph.HE]}}.

\bibitem{Reynolds:2020jwt}
C.~S. Reynolds, ``{Observational Constraints on Black Hole Spin},''
  \href{http://dx.doi.org/10.1146/annurev-astro-112420-035022}{{\em Ann. Rev.
  Astron. Astrophys.} {\bfseries 59} (2021) 117--154},
  \href{http://arxiv.org/abs/2011.08948}{{\ttfamily arXiv:2011.08948
  [astro-ph.HE]}}.

\bibitem{Fabian:1989ej}
A.~C. Fabian, M.~J. Rees, L.~Stella, and N.~E. White, ``{X-ray fluorescence
  from the inner disc in Cygnus X-1},''
  \href{http://dx.doi.org/10.1093/mnras/238.3.729}{{\em Mon. Not. Roy. Astron.
  Soc.} {\bfseries 238} (1989) 729--736}.

\bibitem{Tanaka:1995en}
Y.~Tanaka {\em et~al.}, ``{Gravitationally Redshifted Emission Implying an
  Accretion Disk and Massive Black Hole in the Active Galaxy MCG:-6-30-15},''
  \href{http://dx.doi.org/10.1038/375659a0}{{\em Nature} {\bfseries 375} (1995)
  659}.

\bibitem{Krolik:2002ae}
J.~H. Krolik and J.~F. Hawley, ``{Where is the inner edge of an accretion disk
  around a black hole?},'' \href{http://dx.doi.org/10.1086/340760}{{\em
  Astrophys. J.} {\bfseries 573} (2002) 754},
  \href{http://arxiv.org/abs/astro-ph/0203289}{{\ttfamily
  arXiv:astro-ph/0203289}}.

\bibitem{2004ApJS..153..205D}
M.~{Dov{\v{c}}iak}, V.~{Karas}, and T.~{Yaqoob}, ``{An Extended Scheme for
  Fitting X-Ray Data with Accretion Disk Spectra in the Strong Gravity
  Regime},'' \href{http://dx.doi.org/10.1086/421115}{{\em \apjs} {\bfseries
  153} no.~1, (July, 2004) 205--221},
  \href{http://arxiv.org/abs/astro-ph/0403541}{{\ttfamily
  arXiv:astro-ph/0403541 [astro-ph]}}.

\bibitem{Zdziarski:2023zuh}
A.~A. Zdziarski, S.~Banerjee, S.~Chand, G.~Dewangan, R.~Misra, M.~Szanecki, and
  A.~Niedzwiecki, ``{Black Hole Spin Measurements in LMC X-1 and Cyg X-1 Are
  Highly Model Dependent},''
  \href{http://dx.doi.org/10.3847/1538-4357/ad1b60}{{\em Astrophys. J.}
  {\bfseries 962} no.~2, (2024) 101},
  \href{http://arxiv.org/abs/2308.06167}{{\ttfamily arXiv:2308.06167
  [astro-ph.HE]}}.

\bibitem{Novikov1973}
I.~D. {Novikov} and K.~S. {Thorne}, ``{Astrophysics of black holes.},'' in {\em
  Black Holes (Les Astres Occlus)}, pp.~343--450.
\newblock Jan., 1973.

\bibitem{Lasota:2024lcl}
J.-P. Lasota and M.~Abramowicz, ``{The stress at the ISCO of black-hole
  accretion discs is not a free parameter},''
  \href{http://arxiv.org/abs/2410.06200}{{\ttfamily arXiv:2410.06200
  [astro-ph.HE]}}.

\bibitem{Iwasawa:1996uh}
K.~Iwasawa {\em et~al.}, ``{The Variable iron k emission line in
  MCG-6-30-15},'' \href{http://dx.doi.org/10.1093/mnras/282.3.1038}{{\em Mon.
  Not. Roy. Astron. Soc.} {\bfseries 282} (1996) 1038--1048},
  \href{http://arxiv.org/abs/astro-ph/9606103}{{\ttfamily
  arXiv:astro-ph/9606103}}.

\bibitem{Brenneman2006}
L.~W. {Brenneman} and C.~S. {Reynolds}, ``{Constraining Black Hole Spin via
  X-Ray Spectroscopy},'' \href{http://dx.doi.org/10.1086/508146}{{\em \apj}
  {\bfseries 652} no.~2, (Dec., 2006) 1028--1043},
  \href{http://arxiv.org/abs/astro-ph/0608502}{{\ttfamily
  arXiv:astro-ph/0608502 [astro-ph]}}.

\bibitem{1988ApJ...332..646A}
M.~A. {Abramowicz}, B.~{Czerny}, J.~P. {Lasota}, and E.~{Szuszkiewicz}, ``{Slim
  Accretion Disks},'' \href{http://dx.doi.org/10.1086/166683}{{\em \apj}
  {\bfseries 332} (Sept., 1988) 646}.

\bibitem{2009ApJS..183..171S}
A.~{Sadowski}, ``{Slim Disks Around Kerr Black Holes Revisited},''
  \href{http://dx.doi.org/10.1088/0067-0049/183/2/171}{{\em \apjs} {\bfseries
  183} no.~2, (Aug., 2009) 171--178},
  \href{http://arxiv.org/abs/0906.0355}{{\ttfamily arXiv:0906.0355
  [astro-ph.HE]}}.

\bibitem{Abdikamalov:2020oci}
A.~B. Abdikamalov, D.~Ayzenberg, C.~Bambi, T.~Dauser, J.~A. Garcia,
  S.~Nampalliwar, A.~Tripathi, and M.~Zhou, ``{Testing the Kerr black hole
  hypothesis using X-ray reflection spectroscopy and a thin disk model with
  finite thickness},'' \href{http://dx.doi.org/10.3847/1538-4357/aba625}{{\em
  Astrophys. J.} {\bfseries 899} no.~1, (2020) 80},
  \href{http://arxiv.org/abs/2003.09663}{{\ttfamily arXiv:2003.09663
  [astro-ph.HE]}}.

\bibitem{Penna:2013zga}
R.~F. Penna, A.~Kulkarni, and R.~Narayan, ``{A new equilibrium torus solution
  and GRMHD initial conditions},''
  \href{http://dx.doi.org/10.1051/0004-6361/201219666}{{\em Astron. Astrophys.}
  {\bfseries 559} (2013) A116},
  \href{http://arxiv.org/abs/1309.3680}{{\ttfamily arXiv:1309.3680
  [astro-ph.HE]}}.

\bibitem{Cardenas-Avendano:2022csp}
A.~C\'ardenas-Avenda\~no, A.~Lupsasca, and H.~Zhu, ``{Adaptive analytical ray
  tracing of black hole photon rings},''
  \href{http://dx.doi.org/10.1103/PhysRevD.107.043030}{{\em Phys. Rev. D}
  {\bfseries 107} no.~4, (2023) 043030},
  \href{http://arxiv.org/abs/2211.07469}{{\ttfamily arXiv:2211.07469 [gr-qc]}}.

\bibitem{Cunningham1975}
C.~T. {Cunningham}, ``{The effects of redshifts and focusing on the spectrum of
  an accretion disk around a Kerr black hole.},''
  \href{http://dx.doi.org/10.1086/154033}{{\em \apj} {\bfseries 202} (Dec.,
  1975) 788--802}.

\bibitem{eXTP:2018anb}
{eXTP} Collaboration, S.-N. Zhang {\em et~al.}, ``{The enhanced X-ray Timing
  and Polarimetry mission\textemdash{}eXTP},''
  \href{http://dx.doi.org/10.1007/s11433-018-9309-2}{{\em Sci. China Phys.
  Mech. Astron.} {\bfseries 62} no.~2, (2019) 29502},
  \href{http://arxiv.org/abs/1812.04020}{{\ttfamily arXiv:1812.04020
  [astro-ph.IM]}}.

\bibitem{Barret:2016ett}
D.~Barret {\em et~al.}, ``{The Athena X-ray Integral Field Unit (X-IFU)},''
  \href{http://dx.doi.org/10.1117/12.2232432}{{\em Proc. SPIE Int. Soc. Opt.
  Eng.} {\bfseries 9905} (2016) 99052F},
  \href{http://arxiv.org/abs/1608.08105}{{\ttfamily arXiv:1608.08105
  [astro-ph.IM]}}.

\bibitem{2020SPIE11444E..22T}
M.~Tashiro {\em et~al.}, \href{http://dx.doi.org/10.1117/12.2565812}{``{Status
  of x-ray imaging and spectroscopy mission (XRISM)},''} in {\em Space
  Telescopes and Instrumentation 2020: Ultraviolet to Gamma Ray}, J.-W.~A. {den
  Herder}, S.~{Nikzad}, and K.~{Nakazawa}, eds., vol.~11444 of {\em Society of
  Photo-Optical Instrumentation Engineers (SPIE) Conference Series},
  p.~1144422.
\newblock Dec., 2020.

\bibitem{Reynolds:2023vvf}
C.~S. Reynolds {\em et~al.}, ``{Overview of the advanced x-ray imaging
  satellite (AXIS)},'' \href{http://dx.doi.org/10.1117/12.2677468}{{\em Proc.
  SPIE Int. Soc. Opt. Eng.} {\bfseries 12678} (2023) 126781E},
  \href{http://arxiv.org/abs/2311.00780}{{\ttfamily arXiv:2311.00780
  [astro-ph.IM]}}.

\bibitem{Bambi2024}
C.~Bambi, {\em Testing Gravity with Black Hole X-Ray Data},
  \href{http://dx.doi.org/10.1007/978-981-97-2871-8_5}{pp.~149--182}.
\newblock Springer Nature Singapore, Singapore, 2024.
\newblock \href{http://arxiv.org/abs/2210.05322}{{\ttfamily arXiv:2210.05322
  [gr-qc]}}.
\newblock \url{https://doi.org/10.1007/978-981-97-2871-8_5}.

\bibitem{Gates2020}
D.~E.~A. {Gates}, S.~{Hadar}, and A.~{Lupsasca}, ``{Maximum observable
  blueshift from circular equatorial Kerr orbiters},''
  \href{http://dx.doi.org/10.1103/PhysRevD.102.104041}{{\em \prd} {\bfseries
  102} no.~10, (Nov., 2020) 104041},
  \href{http://arxiv.org/abs/2009.03310}{{\ttfamily arXiv:2009.03310 [gr-qc]}}.

\bibitem{2009ApJ...707L..87T}
J.~A. {Tomsick}, K.~{Yamaoka}, S.~{Corbel}, P.~{Kaaret}, E.~{Kalemci}, and
  S.~{Migliari}, ``{Truncation of the Inner Accretion Disk Around a Black Hole
  at Low Luminosity},''
  \href{http://dx.doi.org/10.1088/0004-637X/707/1/L87}{{\em \apjl} {\bfseries
  707} no.~1, (Dec., 2009) L87--L91},
  \href{http://arxiv.org/abs/0911.2240}{{\ttfamily arXiv:0911.2240
  [astro-ph.HE]}}.

\bibitem{2010A&A...521A..15A}
M.~A. {Abramowicz}, M.~{Jaroszy{\'n}ski}, S.~{Kato}, J.~P. {Lasota},
  A.~{R{\'o}{\.z}a{\'n}ska}, and A.~{S{\k{a}}dowski}, ``{Leaving the innermost
  stable circular orbit: the inner edge of a black-hole accretion disk at
  various luminosities},''
  \href{http://dx.doi.org/10.1051/0004-6361/201014467}{{\em \aap} {\bfseries
  521} (Oct., 2010) A15}, \href{http://arxiv.org/abs/1003.3887}{{\ttfamily
  arXiv:1003.3887 [astro-ph.HE]}}.

\bibitem{Fabian:2014tda}
A.~C. Fabian, M.~L. Parker, D.~R. Wilkins, J.~M. Miller, E.~Kara, C.~S.
  Reynolds, and T.~Dauser, ``{On the determination of the spin and disc
  truncation of accreting black holes using X-ray reflection},''
  \href{http://dx.doi.org/10.1093/mnras/stu045}{{\em Mon. Not. Roy. Astron.
  Soc.} {\bfseries 439} no.~3, (2014) 2307--2313},
  \href{http://arxiv.org/abs/1401.1615}{{\ttfamily arXiv:1401.1615
  [astro-ph.HE]}}.

\bibitem{Aldi:2016ntn}
G.~F. Aldi and V.~Bozza, ``{Relativistic iron lines in accretion disks: the
  contribution of higher order images in the strong deflection limit},''
  \href{http://dx.doi.org/10.1088/1475-7516/2017/02/033}{{\em JCAP} {\bfseries
  02} (2017) 033}, \href{http://arxiv.org/abs/1607.05365}{{\ttfamily
  arXiv:1607.05365 [astro-ph.HE]}}.

\bibitem{Reynolds:2007rx}
C.~S. Reynolds and A.~C. Fabian, ``{Broad iron K-alpha emission lines as a
  diagnostic of black hole spin},''
  \href{http://dx.doi.org/10.1086/527344}{{\em Astrophys. J.} {\bfseries 675}
  (2008) 1048}, \href{http://arxiv.org/abs/0711.4158}{{\ttfamily
  arXiv:0711.4158 [astro-ph]}}.

\bibitem{Reynolds:2013qqa}
C.~S. Reynolds, ``{Measuring Black Hole Spin using X-ray Reflection
  Spectroscopy},'' \href{http://dx.doi.org/10.1007/s11214-013-0006-6}{{\em
  Space Sci. Rev.} {\bfseries 183} no.~1-4, (2014) 277--294},
  \href{http://arxiv.org/abs/1302.3260}{{\ttfamily arXiv:1302.3260
  [astro-ph.HE]}}.

\bibitem{Bambi:2020jpe}
C.~Bambi {\em et~al.}, ``{Towards Precision Measurements of Accreting Black
  Holes Using X-Ray Reflection Spectroscopy},''
  \href{http://dx.doi.org/10.1007/s11214-021-00841-8}{{\em Space Sci. Rev.}
  {\bfseries 217} no.~5, (2021) 65},
  \href{http://arxiv.org/abs/2011.04792}{{\ttfamily arXiv:2011.04792
  [astro-ph.HE]}}.

\bibitem{Wilkins:2012zm}
D.~R. Wilkins and A.~C. Fabian, ``{Understanding X-ray reflection emissivity
  profiles in AGN: Locating the X-ray source},''
  \href{http://dx.doi.org/10.1111/j.1365-2966.2012.21308.x}{{\em Mon. Not. Roy.
  Astron. Soc.} {\bfseries 424} (2012) 1284},
  \href{http://arxiv.org/abs/1205.3179}{{\ttfamily arXiv:1205.3179
  [astro-ph.HE]}}.

\bibitem{Dauser:2013xv}
T.~Dauser, J.~Garcia, J.~Wilms, M.~Bock, L.~W. Brenneman, M.~Falanga,
  K.~Fukumura, and C.~S. Reynolds, ``{Irradiation of an Accretion Disc by a
  Jet: General Properties and Implications for Spin Measurements of Black
  Holes},'' \href{http://dx.doi.org/10.1093/mnras/sts710}{{\em Mon. Not. Roy.
  Astron. Soc.} {\bfseries 430} (2013) 1694},
  \href{http://arxiv.org/abs/1301.4922}{{\ttfamily arXiv:1301.4922
  [astro-ph.HE]}}.

\bibitem{Gonzalez:2017gzu}
A.~G. Gonzalez, D.~R. Wilkins, and L.~C. Gallo, ``{Probing the geometry and
  motion of AGN coronae through accretion disc emissivity profiles},''
  \href{http://dx.doi.org/10.1093/mnras/stx2080}{{\em Mon. Not. Roy. Astron.
  Soc.} {\bfseries 472} no.~2, (2017) 1932--1945},
  \href{http://arxiv.org/abs/1708.03205}{{\ttfamily arXiv:1708.03205
  [astro-ph.HE]}}.

\bibitem{Zhang:2024ahe}
W.~Zhang, M.~Dov\v{c}iak, M.~Bursa, J.~Svoboda, and V.~Karas, ``{Inferring the
  iron K emissivity profiles of accretion discs irradiated by extended
  coronae},'' \href{http://dx.doi.org/10.1093/mnras/stae1714}{{\em Mon. Not.
  Roy. Astron. Soc.} {\bfseries 532} no.~4, (2024) 3786--3796},
  \href{http://arxiv.org/abs/2407.08336}{{\ttfamily arXiv:2407.08336
  [astro-ph.HE]}}.

\bibitem{Wu:2007sk}
S.-M. Wu and T.-G. Wang, ``{Iron Line Profiles from Relativistic Thick
  Accretion Disk},'' {\em ASP Conf. Ser.} {\bfseries 373} (2007) 119--120,
  \href{http://arxiv.org/abs/0704.2460}{{\ttfamily arXiv:0704.2460
  [astro-ph]}}.

\bibitem{Draghis:2022ngm}
P.~A. Draghis, J.~M. Miller, A.~Zoghbi, M.~Reynolds, E.~Costantini, L.~C.
  Gallo, and J.~A. Tomsick, ``{A Systematic View of Ten New Black Hole
  Spins},'' \href{http://dx.doi.org/10.3847/1538-4357/acafe7}{{\em Astrophys.
  J.} {\bfseries 946} no.~1, (2023) 19},
  \href{http://arxiv.org/abs/2210.02479}{{\ttfamily arXiv:2210.02479
  [astro-ph.HE]}}.

\bibitem{Draghis:2023vzj}
P.~A. Draghis, J.~M. Miller, E.~Costantini, L.~C. Gallo, M.~Reynolds, J.~A.
  Tomsick, and A.~Zoghbi, ``{Systematically Revisiting All NuSTAR Spins of
  Black Holes in X-Ray Binaries},''
  \href{http://dx.doi.org/10.3847/1538-4357/ad43ea}{{\em Astrophys. J.}
  {\bfseries 969} no.~1, (2024) 40},
  \href{http://arxiv.org/abs/2311.16225}{{\ttfamily arXiv:2311.16225
  [astro-ph.HE]}}.

\bibitem{Mummery:2024mrq}
A.~Mummery, A.~Ingram, S.~Davis, and A.~Fabian, ``{Continuum emission from
  within the plunging region of black hole discs},''
  \href{http://dx.doi.org/10.1093/mnras/stae1160}{{\em Mon. Not. Roy. Astron.
  Soc.} {\bfseries 531} no.~1, (2024) 366--386},
  \href{http://arxiv.org/abs/2405.09175}{{\ttfamily arXiv:2405.09175
  [astro-ph.HE]}}.

\bibitem{Dovciak:2003jym}
M.~Dovciak, V.~Karas, and T.~Yaqoob, ``{An Extended scheme for fitting x-ray
  data with accretion disk spectra in the strong gravity regime},''
  \href{http://dx.doi.org/10.1086/421115}{{\em Astrophys. J. Suppl.} {\bfseries
  153} (2003) 205--221},
  \href{http://arxiv.org/abs/astro-ph/0403541}{{\ttfamily
  arXiv:astro-ph/0403541}}.

\bibitem{Wilkins:2020pgu}
D.~R. Wilkins, C.~S. Reynolds, and A.~C. Fabian, ``{Venturing beyond the ISCO:
  Detecting X-ray emission from the plunging regions around black holes},''
  \href{http://dx.doi.org/10.1093/mnras/staa628}{{\em Mon. Not. Roy. Astron.
  Soc.} {\bfseries 493} no.~4, (2020) 5532--5550},
  \href{http://arxiv.org/abs/2003.00019}{{\ttfamily arXiv:2003.00019
  [astro-ph.HE]}}.

\bibitem{Cardenas-Avendano:2020xtw}
A.~Cardenas-Avendano, M.~Zhou, and C.~Bambi, ``{Modeling uncertainties in X-ray
  reflection spectroscopy measurements II: Impact of the radiation from the
  plunging region},'' \href{http://dx.doi.org/10.1103/PhysRevD.101.123014}{{\em
  Phys. Rev. D} {\bfseries 101} no.~12, (2020) 123014},
  \href{http://arxiv.org/abs/2005.06719}{{\ttfamily arXiv:2005.06719
  [astro-ph.HE]}}.

\bibitem{Reynolds:1997ek}
C.~S. Reynolds and M.~C. Begelman, ``{Iron fluorescence from within the
  innermost stable orbit of black hole accretion disks},''
  \href{http://dx.doi.org/10.1086/304703}{{\em Astrophys. J.} {\bfseries 488}
  (1997) 109}, \href{http://arxiv.org/abs/astro-ph/9705136}{{\ttfamily
  arXiv:astro-ph/9705136}}.

\bibitem{Young:1998ha}
A.~J. Young, R.~R. Ross, and A.~C. Fabian, ``{Iron line profiles including
  emission from within the innermost stable orbit of a black hole accretion
  disc},'' \href{http://dx.doi.org/10.1046/j.1365-8711.1998.02058.x}{{\em Mon.
  Not. Roy. Astron. Soc.} {\bfseries 300} (1998) 11},
  \href{http://arxiv.org/abs/astro-ph/9808089}{{\ttfamily
  arXiv:astro-ph/9808089}}.

\bibitem{Garcia:2013oma}
J.~Garcia, T.~Dauser, C.~S. Reynolds, T.~R. Kallman, J.~E. McClintock,
  J.~Wilms, and W.~Eikmann, ``{X-ray reflected spectra from accretion disk
  models. III. A complete grid of ionized reflection calculations},''
  \href{http://dx.doi.org/10.1088/0004-637X/768/2/146}{{\em Astrophys. J.}
  {\bfseries 768} (2013) 146}, \href{http://arxiv.org/abs/1303.2112}{{\ttfamily
  arXiv:1303.2112 [astro-ph.HE]}}.

\bibitem{Dong:2023bbd}
J.~Dong, G.~Mastroserio, J.~A. Garc\i{}a, A.~Ingram, E.~Nathan, and R.~Connors,
  ``{X-ray Reflection from the Plunging Region of Black Hole Accretion
  Disks},'' \href{http://arxiv.org/abs/2312.09210}{{\ttfamily arXiv:2312.09210
  [astro-ph.HE]}}.

\bibitem{Bambi:2016sac}
C.~Bambi, A.~Cardenas-Avendano, T.~Dauser, J.~A. Garcia, and S.~Nampalliwar,
  ``{Testing the Kerr black hole hypothesis using X-ray reflection
  spectroscopy},'' \href{http://dx.doi.org/10.3847/1538-4357/aa74c0}{{\em
  Astrophys. J.} {\bfseries 842} no.~2, (2017) 76},
  \href{http://arxiv.org/abs/1607.00596}{{\ttfamily arXiv:1607.00596 [gr-qc]}}.

\bibitem{Cardenas-Avendano:2019pec}
A.~Cardenas-Avendano, J.~Godfrey, N.~Yunes, and A.~Lohfink, ``{Experimental
  Relativity with Accretion Disk Observations},''
  \href{http://dx.doi.org/10.1103/PhysRevD.100.024039}{{\em Phys. Rev. D}
  {\bfseries 100} no.~2, (2019) 024039},
  \href{http://arxiv.org/abs/1903.04356}{{\ttfamily arXiv:1903.04356 [gr-qc]}}.

\bibitem{Bauer:2021atk}
A.~M. Bauer, A.~C\'ardenas-Avenda\~no, C.~F. Gammie, and N.~Yunes, ``{Spherical
  Accretion in Alternative Theories of Gravity},''
  \href{http://dx.doi.org/10.3847/1538-4357/ac3a03}{{\em Astrophys. J.}
  {\bfseries 925} no.~2, (2022) 119},
  \href{http://arxiv.org/abs/2111.02178}{{\ttfamily arXiv:2111.02178 [gr-qc]}}.

\bibitem{GitHub}
\url{https://github.com/iAART/LineAART.}

\bibitem{Bardeen1973}
J.~M. {Bardeen}, ``{Timelike and null geodesics in the Kerr metric.},'' in {\em
  Black Holes (Les Astres Occlus)}, C.~{Dewitt} and B.~S. {Dewitt}, eds.,
  pp.~215--239.
\newblock Gordon and Breach Science Publishers, Jan, 1973.

\end{thebibliography}\endgroup
\bibliographystyle{utphys2}

\newpage
\newpage

\end{document}